.

# 東南大學

# 博士学位论文

## 实用语音情感识别若干关键技术研究

专 业 名 称：信息与通信工程

研究生姓名 ： ____黄程韦____

导 师 姓 名：____赵 力____

# RESEARCH ON SEVERAL KEY TECHNOLOGIES IN PRACTICAL SPEECH EMOTION RECOGNITION

A Dissertation Submitted to

Southeast University

For the Academic Degree of Doctor of

Engineering

School of Information Science and Engineering

Southeast University

2013

# 东 南 大 学 学 位 论 文 独 创 性 声 明

本人声明所呈交的学位论文是我个人在导师指导下进行的研究工作及取得的研究成果。尽我所知，除了文中特别加以标注和致谢的地方外，论文中不包含其他人已经发表或撰写过的研究成果，也不包含为获得东南大学或其它教育机构的学位或证书而使用过的材料。与我一同工作的同志对本研究所做的任何贡献均已在论文中作了明确的说明并表示了谢意。

研究生签名：___________________ 日 期：___________

# 东 南 大 学 学 位 论 文 使 用 授 权 声 明

东南大学、中国科学技术信息研究所、国家图书馆有权保留本人所送交学位论文的复印件和电子文档，可以采用影印、缩印或其他复制手段保存论文。本人电子文档的内容和纸质论文的内容相一致。除在保密期内的保密论文外，允许论文被查阅和借阅，可以公布（包括刊登）论文的全部或部分内容。论文的公布（包括刊登）授权东南大学研究生院办理。

研 究 生 签 名 ____________ 导 师 签 名 ____________

日　　　期 ____________

# 摘要

语音情感识别技术能够从语音信号中识别说话人的心理情感状态，具有重要的理论意义和应用价值。以往的研究主要围绕基本情感类别进行，不能满足实际应用的需要。本文研究了实用语音情感识别技术，包括烦躁、自信和疲倦等三种与认知过程有关的情感。

高自然度的情感语料的获取是研究的基础。本文中通过以下途径诱发实用语音情感数据：认知作业、计算机游戏、噪声刺激、睡眠剥夺和观看影视片断等。通过人工标注和筛选，建立了实用语音情感数据库。通过增加心电情感信号，建立了语音和心电的双模态数据库。

语音情感特征分析是重要的一个环节。本文中提取了 481 个静态统计特征，包括了音质特征和韵律特征。对烦躁、自信和疲倦等实用语音情感进行了特征分析，对谐波噪声比特征在实用语音情感上的分布特点进行了研究。通过特征降维与识别实验验证了本文中所提取的特征的有效性。

在获取的高自然度的情感语料上，研究了基于高斯混合模型的实用语音情感识别系统。第一、在充足的数据量下获得了理想的识别性能。第二、针对高斯混合模型的不足，在小样本条件下结合两类分类器组进行了系统的性能改进。第三、考虑连续情感语音中的上下文关系，提出了嵌入马尔科夫网络的高斯混合模型，对情感的连续性进行建模，提高了识别率。

在建立了性能较为完善的高斯混合模型语音情感识别系统后，进一步研究了系统鲁棒性的问题。第一、研究了噪声对语音情感识别系统的影响，首次将基于听觉掩蔽效应的降噪算法应用到实用语音情感识别中，获得了理想的效果。第二、针对实际条件下复杂的未知情感类型，提出了一种可拒判的情感识别方法，增强了系统对未知情感样本的兼容性。第三、针对大量非特定说话人带来的识别困难，提出了一种基于说话人特征聚类的情感特征规整化方法，提高了系统对非特定说话人的鲁棒性。第四、通过增加心电信号通道，首次进行了语音信号与心电信号融合的双模态情感识别，提高了系统的识别率和抗毁性。

本文中的语音情感识别系统能够扩展应用到跨语言和耳语音情感识别领域中。本文分析了跨语言语音情感识别中面临的困难，研究了在汉语和德语上通用性较高的特征，进行了跨语言的情感识别。实验结果显示，本文中的生气情感模型在两种语言中具有较强的共通性。本文还分析了耳语音情感识别的特点和难点，提取了有效的耳语音情感特征，并且将本文中提出的嵌入马尔科夫网络的高斯混合模型在耳语音情感识别中进行了验证，获得了理想的识别效果。

最后，本文对实用语音情感识别技术的应用进行了分析和讨论，对语音情感计算技术的发展进行了展望。

**关键词：情感识别；噪声；非特定说话人；高斯混合模型；马尔科夫网络**

# ABSTRACT

Speech emotion recognition technology is designed for recognizing the speaker's emotional states from speech signals, which is of both theoretical value and practical value. In the past researches the basic emotion types were the main concerns, which, however, might not satisfy the real world requirements. In this dissertation the practical speech emotion recognition technology is studied, including several cognitive related emotion types, namely fidgetiness, confidence and tiredness.

The high quality of naturalistic emotional speech data is the basis of this research. In this dissertation the following techniques are used for inducing practical emotional speech: cognitive task, computer game, noise stimulation, sleep deprivation and movie clips. By human annotation and selection the practical speech emotion database is achieved. By adding the electrocardiogram signals the bi-modal emotional database is achieved.

The emotional speech features analysis is an important step. In this dissertation 481 static features are extracted, including voice quality features and prosodic features. Feature analysis is carried out on practical speech emotions like fidgetiness, confidence and tiredness. Harmonic-to-noise ratio is studied for its distribution characters in the practical speech emotions. Through feature reduction and recognition experiments the effectiveness of the features proposed in this dissertation is justified.

With the high-quality naturalistic emotional speech data achieved, the practical speech emotion recognition system is studied based on Gaussian mixture model. First, promising recognition results are achieved under sufficient training data. Second, considering the drawbacks of Gaussian mixture model, two-class classifier set is adopted for performance improvement under the small sample case. Third, considering the context information in continuous emotional speech, a Gaussian mixture model embedded with Markov networks is proposed. The recognition rate is improved by modeling the continuality of emotions.

Upon establishing a satisfactory speech emotion recognition system based on Gaussian mixture model, a further study is carried out for system robustness analysis. First, noise influence on the speech emotion recognition system is studied. Noise reduction algorithm based on auditory masking properties is fist introduced to the practical speech emotion recognition, and achieved satisfactory results. Second, to deal with the complicated unknown emotion types under real situation, an emotion recognition method with rejection ability is proposed, which enhanced the system compatibility against unknown emotion samples. Third, coping with the difficulties brought by a large number of unknown speakers, an emotional feature normalization method based on speaker-sensitive feature clustering is proposed. Speaker-independent robustness of the system is improved. Fourth, by adding the electrocardiogram channel, a bi-modal emotion recognition system based on speech signals and electrocardiogram signals is first introduced. The emotion recognition rate and the system survivability are improved.

The speech emotion recognition methods studied in this dissertation may be extended into the cross-language speech emotion recognition and the whispered speech emotion recognition. The difficulties in cross-language speech emotion recognition are analyzed. The general features across the Chinese and the German speech data are studied, and cross-language test is carried out. The experimental results show that the angry emotion model in this dissertation has a relatively strong generality in both languages. The difficulties and characters in the whispered speech emotion recognition are also studied. Effective whispered speech emotional features are extracted and the proposed Gaussian mixture model embedded with Markov networks is tested on the whispered speech emotion data. The experimental results show a promising improvement.

At the end of this dissertation the applications of the practical speech emotion recognition technique are discussed and the development of affective computing in speech signals is prospected.

# 目录

# 图索引

# 表索引

# 缩略语对照表

| | | |
|---|---|---|
| AIBO | Artificial Intelligence Robot | 人工智能机器人（索尼机器宠物） |
| ANN | Artificial Neural Network | 人工神经网络 |
| BP | Back Propagation | 反向传播 |
| CMS | Cepstral Mean Subtraction | 倒谱均值减法 |
| DES | Danish Emotional Speech | 丹麦情感语音 |
| ECG | Electrocardiogram | 心电图 |
| ELRA | European Language Resources Association | 欧洲语言资源协会 |
| EM | Expectation-Maximization | 期望最大化 |
| FDR | Fisher Discriminant Ratio | Fisher 判别系数 |
| GMM | Gaussian Mixture Model | 高斯混合模型 |
| HIV | Human Immunodeficiency Virus | 人类免疫缺陷病毒 |
| HMM | Hidden Markov models | 隐马尔科夫模型 |
| HNR | Harmonic-to-Noise Ratio | 谐波噪声比 |
| HRV | Heart rate variability | 心率变异性 |
| IMO | International Maritime Organization | 国际海事组织 |
| KNN | K-Nearest Neighbor | K 近邻 |
| LDA | Linear Discriminant Analysis | 线性判别分析 |
| LDC | The Language Data Consortium | 语言数据联盟 |
| LLD | Low Level Descriptor | 低层描述符 |
| LPC | Linear Predictive Coding | 线性预测编码 |
| MFCC | Mel-Frequency Cepstral Coefficients | 镁尔倒谱参数 |
| MIT | Massachusetts Institute of Technology | 麻省理工学院 |
| OGI | Oregon Graduate Institute | 俄勒冈研究生院 |
| PCA | Principal Components Analysis | 主成份分析 |
| PCM | Pulse Code Modulation | 脉冲编码调制 |
| QPBO | Quadratic Pseudo-Boolean Optimization | 正交伪布尔优化 |
| RBF | Radial Basis Function | 径向基核函数 |
| SFS | Sequential Forward Selection | 序列前向选择 |
| SFM | Spectral Flatness Measure | 谱平坦测度 |
| SMCP | Standard Marine Communication Phrases | 标准航海通信用语 |
| SN | Speaker Normalization | 说话人规整化 |
| SUSAS | Speech under Simulated and Actual Stress | 模拟与真实压力下的语音 |
| SVM | Support Vector Machine | 支持向量机 |
| TEO | Teager Energy Operator | Teager 能量 |
| VAM | The Vera am Mittag | Vera 午间脱口秀 |
| WAV | Waveform Audio Format | 波形音频格式 |

# 第一章 绪论

在本文中我们介绍了实用语音情感识别的研究工作。语音是人类交流信息最自然和便捷的手段之一。语音中除了语义信息外，还包含了丰富的情感信息，传统的语音识别仅限于语义的识别，而忽略了情感信息的识别。语音情感识别就是要从语音信号中提取出说话人情感状态的信息，它是一个交叉研究领域，涉及到模式识别、信号处理、声学、语音学，以及心理学等多个学科领域。语音情感识别是情感计算的一个子课题，同时它与社会信号处理、人工心理等研究领域有非常密切的联系。

在人工智能中，情感计算被认为是赋予计算机更高的、全面的智能的一个关键途径。MIT（Massachusetts Institute of Technology）的 Picard 教授在 1997 年发表了《情感计算》[1]一书，标志着情感计算学科的正式建立。研究者们对情感计算的可能的功用做了很多有意思的设想。Picard 在她的这本著作中，设想了一种叫做"情感镜子"的智能仪器。我们在日常生活中看到的普通的镜子，是依据牛顿光学的原理来反射出照镜子的人的容貌和影像。镜子的发明可以使我们看到自己，认识到自己的容貌。而所谓的情感镜子，可以使"照"镜子的人定量的或者定性的观察到自己的情感状态，为自己的行为表现打分。可以设想使用情感镜子并不一定需要利用光线和图像来作为媒介，也可以借助声波或者电生理参数等，甚至是通过脑电波来识别人的情感。这样的一种智能仪器可以产生很多有趣的应用，例如演员可以用它来训练自己的演技，客服人员可以用它来提醒自己的工作态度，如果制成儿童玩具，还会对培养儿童的情感智力起到很大的帮助。

1985 年，Marvin Minsky 在《思维的社会》（The Society of Mind）[2]一书中，深入探讨了情感与智能之间的关系。情感计算领域的追随者们普遍认为，如果我们可以创造一种类似于人类智能水平的智能机器人，那么它很有可能同时具备了人类的情感智能。研究的意义并不仅在于情感能力可以使得未来的人工智能机器增添绚丽的光彩，而主要是在于情感智能很可能是人工智能中的一个不可或缺的环节。"没有情感能力的智能是否可能出现？"[2]。在目前来看，在自然界的历史当中，这样的智能是没有存在过的。

除了情感计算外，同语音情感识别技术关系密切的还有另一个新兴学科——社会信号处理。随着以人为中心的计算技术的进一步发展，在社会行为学和管理科学等领域逐渐采用了先

进的测量和计算手段。在这样的技术背景下，诞生了一门新的学科：社会信号处理（Social Signal Processing），它从信息处理的角度研究人类情绪感知能力和社会交往能力，对情感计算在社会生活中的应用开辟了新的前景。2009 年，Vinciarelli 在《Image and Vision Computing》上发表了一篇总结性的论文[3]，对相关领域的最新进展进行了总结，提供了社会信号处理学科的研究框架。

如果说情感计算是将人类情感行为中总结出的规律，用到自然界中去创造出"人工智能"，那么社会信号处理的着眼点则是将自然科学当中产生的计算工具，为探索人类社会活动的规律所服务。语音情感识别技术，则是在这样的技术背景下受到了研究者的重视，它既能够促进认知科学的基础性理论研究，又能够在广泛的社会生活领域发挥应用价值，还能间接的推动语音信号处理和机器学习等相关学科的发展。

了解了语音情感识别技术诞生的学科背景，下面我们就来简要介绍一下语音情感识别技术的研究历史、发展现状和面临的挑战等。

## 1.1 研究历史简述

早期的研究者们从心理学的角度对语音信号中的情感信息进行研究。1962 年，Friedhoff 等人研究了人类语音当中情感的传递，他们的研究成果在《Nature》上获得了发表[4]，指出了对人类情感进行客观测量和评价的可能性。1984 年，Scherer 等人详细研究了人类声音中能够反映出情感变化的一系列特征[5, 6]，他们的研究特别分析了基音频率的变化与情感变化之间的关系，这为后来的计算机自动识别语音情感提供了重要的理论基础。1996 年，卡内基梅隆大学的 Dellaert 等人[7]采用几种模式识别的算法，从工程学的角度开发了一个较早的语音情感自动识别系统。随后，Thomas Huang 等人研究了双模态的情感识别系统，融合了表情和语音进行自动情感识别[8]。随着计算机科学的迅速发展，以及自然人机交互的需求，越来越多的研究者投身于语音情感识别研究。诸多研究成果，受到了同行的关注[9-15]。日本的 Nicholson[15]，慕尼黑工业大学的 Schuller[10, 16-18]，北爱尔兰 Queens 大学的 Cowie[11]等人，发表了一系列重要的研究成果，推动了这个领域的发展。

日本文部省在九十年代发布的关于语音情感识别的研究计划，是最早受到政府支持的情感计算的研究计划之一。Nicholson 等人采用了人工神经网络的方法，对语音情感进行自动的识别[15]。结果显示，几种基本的情感得到了正确的区分。德国的 Schuller 等人在一系列著名的国际会议上，发表了语音情感识别的诸多研究成果[10, 16, 17]，其中包括了对建立一个合理的

情感识别框架的诸多构想。例如：采用语音识别中获得成功的隐马尔科夫模型；采用数量非常庞大的、通过多种函数统计获得的声学特征；引入情感关键词等语义特征来拓宽特征空间所反映的情感信息维度；以及交叉数据库上的情感识别研究等等。北爱尔兰的 Cowie 等人[11]较早的总结了语音情感识别领域的发展状况，并且倡导采用高自然度的数据来进行研究，说明了自然语料在这个领域中的关键作用。还有很多学者在语音情感识别领域取得了令人瞩目的研究成果，我们在本文中的各个章节内再按照章节的主题进行详细的介绍。

## 1.2 研究现状分析

近年来的研究进展可以大致分为以下几个方面：一、情感特征的寻找；二、建模算法的研究；三、自然情感数据库的建立；四、环境自适应的方法，如上下文信息、跨语言、跨文化，和性别差异等，这一类方法着重关注情感模型的适应能力。下面我们对这四个方面分别讨论目前存在的主要问题，和可能的解决途径。

### 一、情感特征

在语音信号中，情感的表达并不总是十分明显的。不同的情感可以具有相似的特征[9]，不同的说话人可以存在较大的个体差异，这对情感特征的研究造成了较大的困难，甚至是对人耳听辨也造成了很大的困难。在语音情感的自动识别中一个极为重要的问题就是寻找情感在特征空间中的一个清晰的表达，也就是需要进行情感特征的提取和优化。

早期的研究者，主要是来自于心理学领域[5, 19]，他们对情感语音的发音特点做了总结，特别是发现了基音频率、发音时间持续长度等特征受到情感变化的影响较大。在心理学实验中，通过人耳听辨来进行语音情感识别，能够找到不同情感类型的语料对应的不同听觉特性。听辨实验的结果可以为计算机实验提供方向性的指导，然而通过听觉实验总结出来的发音特点，往往还不能直接用于计算机实验。因此，来自工程领域的研究者们[7, 8, 12, 20]借鉴了自动语音识别等领域的研究经验，采用镁尔倒谱系数、共振峰频率、基音频率、过零率等声学参数进行情感特征的统计分析。采用统计模式识别的研究框架，对采集的情感语料数据进行训练和识别，在特定的数据库上实现了语音信号中的情感识别。

进一步的研究发现，在不同数据库上，情感特征的研究结果差别很大。由于对语音特征产生较大影响的，是文本的变化，也就是音位信息的干扰。因此，为了获得鲁棒性较高的情感特征，研究者们从特征构造的角度进行研究，希望能够消除音位信息的影响，寻找与文本无关的情感特征。目前，效果较好的方法是将一系列的统计函数施加在基本的语音参数之上[21]。例

如，针对一条语料，求取其基音、共振峰等参数的最大值、最小值、均值、差分等。由于这样的统计分析是在整条语句之上进行的，它能够反映出一段时间之上的语音参数的变化规律，而消除了短时特征对音位敏感的弱点。

然而，目前的特征分析结果大多是针对基本情感类别进行的，对实际当中具有实用意义的新的情感类型的研究非常少。为了识别更多具有实用意义的情感类型，我们需要提供充足的基本语音特征，寻找合适的特征优化方法。

二、情感建模

情感建模是语音情感识别中的一个基本问题。我们可以采用心理学理论中的情感类别模型或者维度空间模型对情感进行建模。情感类别模型将情感区分为互不相同的类别，是一种离散模型。维度空间模型将情感视为多维空间中的连续变量，是一种连续模型[22]。

早期的研究，几乎都采用了情感的类别模型。这种理论认为，人类的情感可以分为基本情感和复合情感，复合情感是由基本情感按照一定比例混合构成。在工程实际中，我们将各种情感类别看成是相互独立的离散类别，对情感语料的识别，就是给每个样本"贴上"某一个情感类别"标签"。采用这种模型较为直观，与模式分类的思想最为贴近。但是，不能够细腻的刻画情感的微小变化，对于一些似是而非的情感样本，往往难以准确的给出识别结果。特别是当遇到训练数据当中没有出现过的未知情感类别时，分类器往往就无法正常工作。因此，在采用情感的类别模型时，有必要结合"可拒判"的方法来处理超出训练数据范围的未知情感类别。

近年来，部分研究者采用了维度空间模型，对情感的几个维度（例如唤醒维、效价维）进行定量的分析[14]。通过回归分析等方法，获得情感语料对应的唤醒度和效价度，用连续的矢量来描述语音信号中的情感信息。这种方法的优点是可以定量的进行分析，理论上，能够对情感的微小变化进行识别。然而在实际当中，情感维度的分析方法与人们在日常生活中的习惯相差较远，所获得的识别结果并不具有直接的应用价值。通过维度分析获得的唤醒度和效价度数值，仍然需要依靠情感类别模型映射到离散的情感类别上，才能说明所识别的是"何种"情感。

在情感的心理模型框架下，我们可以寻找合适的统计学习算法来建立情感的数学模型。很多常用的分类算法，都曾被用于语音情感的分类。例如，隐马尔科夫模型、支持向量机、K近邻分类器、人工神经网络以及高斯混合模型等。其中，隐马尔科夫模型是借鉴了自动语音识别中

的研究经验，它适合与短时的动态特征相结合。但由于以帧为单位的动态特征对文本的变化较为敏感，因此它不利于建立与文本无关的情感识别系统。支持向量机的优点是适合小样本条件下的模型学习，在早期的语音情感研究中，由于语料获取的困难较大，一些著名的数据库如柏林情感语音库等[23]，仅包含数量较少的情感语料。在这些规模较小的数据库上，支持向量机是具有先天的优势的。在一些实际应用当中，我们可能需要定制某种特殊类型的情感的识别系统，在难以获得充足训练语料的场合，可以考虑采用支持向量机方法。人工神经网络在九十年代的情感识别文献中出现的频率较高，日本的研究者在这方面做了一些初步的情感识别研究[15]，人工神经网络可以应用于人脸识别和语音情感识别，然而识别率仍然有待提高。高斯混合模型，是近年来在语音情感识别中取得较高识别性能的一种方法。由于高斯混合模型对数据的拟合能力较强，因此可能在语种识别、说话人识别、语音情感识别等"个体差异"较大的数据类型上的学习能力较强。然而高斯混合模型的缺点也是受制于训练数据，成功建立情感模型需要有充足的训练数据。对于小样本条件下的高斯混合模型的改进研究，具有一定的理论意义和实际价值。

三、情感数据库

语音情感模型的训练依赖于高质量的数据库，情感数据库的质量在建立情感识别系统的过程当中起到了至关重要的作用[24]。2005年，柏林工业大学的Burkhardt等人，建立了柏林情感语音数据库[23]。这是一个包含了七种基本情感类别的表演数据库，由专业的男女演员通过朗读的方式表现指定的情感。柏林库的数据规模虽然不大，文本内容也比较有限，但是制作规范、语料质量较高，且是较早建立和免费发布的情感语音数据库。大量的语音情感识别研究都在柏林库上做了实验验证，因此是影响力较大的一个重要数据库。

英国北爱尔兰Queen's大学的Cowie等人较早的认识到表演语料的不足之处，指出了情感识别研究中自然语料的重要性和必要性。在表演数据库上获得的一些研究结果，如情感的最佳特征组，基于各种学习算法的情感的统计模型，在实际应用当中并不能很好的识别真实的情感。造成这种性能瓶颈的原因，正是表演情感与真实情感的差异。在表演语料中的情感，往往带有夸张的成分，通常其唤醒度会高于自然语料中的情感。在表演语料基础上建立的统计模型，与现实生活中人们的真实情感有一定的差别。

2009年的Interspeech会议上，Schuller等人[18]举办了语音情感识别的测评比赛（Emotion Challenge），并且指定AIBO数据库为比赛的专用数据库。该数据库是通过索尼的机器狗智能玩具采集的自然语料，在儿童与机器狗对话的过程中录制和建立了儿童的情感语音数

据库。目前，情感语料的自然度问题受到了广泛重视。如何采集自然度高的语料，也就是如何使得训练数据真实可信，以及在此基础上建立高质量的数据库是研究的重点和难点之一。一些研究者们采用了来自实验心理学的诸多诱发手段，如通过计算机游戏、睡眠剥夺与隔绝实验等。另外一些研究者们从自然的人机语音交互中录制情感数据，或是从电视脱口秀节目中选择录制情感语料。然而，在提高语料的自然度的同时，也带来了很多其它的问题。主要是一系列的不可控因素，如文本内容的随机性、说话人数据的不平衡，以及目标情感的个性化差异等。为解决这些自然语料中不可控因素干扰的问题，一种可能的途径是近年来出现的一系列"环境自适应"的方法。

四、环境自适应

情感的表达方式与很多环境因素有关，如年龄、性别、语境、语种和文化背景等[25, 26]。因此语音情感识别比起其它的模式识别问题有特殊的困难。环境自适应的方法可能是语音情感识别进入实际应用的一个关键途径。

环境自适应方法的核心是提高系统的自适应能力。根据当前语料中所包含的丰富的信息，对识别系统进行适当的调整，以适合当前语料的特点或是当前说话人的个性特征。语料当中包含了情感信息以外的丰富的信息，这些信息都可以加以利用，以提高情感识别系统的性能，包括以下方面[16, 26-29]：说话人的性别、上下文的情感语境、环境噪声的干扰程度、语种、文化差异、个性差异，以及所处的应用环境中的特殊用户需求。

对于说话人的性别的识别，可以提高情感系统的识别率。由于男性与女性的基音频率有较大的差别，表达情感的方式也不同，因此，情感特征的分析可以根据说话人性别的不同分别进行，以获得更准确的结果。王治平等[30]分析了情感语音中的性别信息，采用了基于性别差异化的情感识别器。

在实际中，情感语料往往是持续较长时间的连续语音数据。在一段较长时间内，情感的变化是需要符合一定的规律的，对这种情感连续变化的规律进行研究和建模，可以辅助语音情感识别。例如，情感的维度空间理论认为情感是连续的变量，也就是说情感在短时间内发生突变的概率较小，持续渐变的可能性较大。因此，我们可以采用马尔科夫随机场等优化工具[31]，对时域上相邻的前后情感识别结果进行配准，以使得情感识别算法的输出结果，更加符合上下文的情感语境。

上下文信息的利用，还可以通过识别情感关键词来进行[17]。从语料的语义层面分析，融合情感关键词的识别系统能够获得更广泛的情感信息。特别是控制维度的情感特征很难从声学参数当中获得，有的研究者认为，控制维度可能同语义特征的关联性更大。融合了词法分析的语音情感识别系统，可以获得更稳定的性能，当声学特征不能很好的作为情感识别的特征时，情感关键词可以提供合理的参考信息。

噪声是各种电子信息系统中面临的一个重要干扰因素。目前，对语音情感识别系统的抗噪声性能的研究还非常少[16]。如何根据环境中的噪声情况，选择合适的降噪算法、噪声鲁棒的情感特征，是非常重要的一个课题。在诸多实际应用领域，噪声的干扰非常严重，例如在车载电子中，来自发动机的噪声是制约自动语音识别系统性能的瓶颈。对环境噪声影响的研究，是提高实际中语音情感识别系统性能的重要途径。

情感语音中的情感信息的表达，还受到语言和文化的影响。语种对情感特征的影响是比较大的，在不同语言的数据库上，往往会得到不同的最佳情感特征组。目前，跨数据库的语音情感识别研究还非常欠缺，一个优秀的识别系统应该能够对不同语种的数据进行自适应的处理。从心理学的研究结果来看，来自不同民族的被试人员对情感的感知和理解是基本一致的[25]，也就是说人类情感的表达和理解在全球范围内、各民族之间是有很多的共通点的。因此，寻找这些共同的语音情感特征，对建立一个通用的语音情感识别系统、实现语种的自适应，是很有意义的。

## 1.3 今后的发展方向

上面的四点从研究方法的角度讨论了语音情感识别的研究现状，从研究课题的意义和社会价值的角度，我们还可以看到本领域有一些新的发展动向。主要是情感识别与认知科学之间的互相影响和相互促进，以及语音情感识别技术对实际需求的响应。

认知科学的基本理论一直是指导语音情感识别研究的潜在的基础。然而，一些特殊类别的情感的检测，能够形成帮助认知科学与其它学科领域（如行为学和教育学等）发展的工具。例如烦躁等负面情感的出现，会对认知能力有一定程度的威胁。再如微软公司研发的"情感手镯"，能够用来评估教育现场（如课堂上）的学生的学习状态。因此，目前语音情感识别领域的研究，有明显的向交叉学科发展的趋势。一个重要的发展方向是对特殊情感类别的自动识别的研究，从以往关注情感的识别准确度，转向关注所识别的情感的实用价值。

实用技术的发展往往会受到实际需求的影响，语音情感识别技术也不例外。随着语音情感

识别研究的深入，研究者们提出了很多颇具前景的实际应用。在车载电子中，Jones 与 Jonsson [32]提出对驾驶员的情感状态进行识别，并作出相应的响应以保证驾驶的安全。在他们的研究中，对枯燥、愉快、惊讶、生气和悲伤进行了跟踪监测。Clavel[33]研究了恐惧类别的极端情感的识别，在一个基于音频的监控系统中，对可能发生的危险情感进行了探测。Ang [34]研究了韵律特征对识别人机对话中的负面情绪的作用，有利于建立一个和谐的人机环境。

可以看到，情感状态的识别与在此基础上的心理评估具有很高的实际应用价值[35]。特别是在航空、航天、航海等军事领域中，长时间的、单调的、高强度的任务，会使得相关人员面临严酷的生理以及心理考验，引发某些负面的情绪。探讨这些情绪对工作能力的作用及其机制和影响因素，具有非常重要的应用价值，可以研究提高个体认知和工作效率的方法、避免影响认知和工作能力的因素。

我们注意到，认知心理学的研究成果显示，负面情绪对认知能力有一定的影响[36]。Pereira等人[37]的研究表明，不愉快的情绪会影响人对视觉目标的检测能力。对负面情绪的研究不仅在心理学中具有重要的意义，而且在人机交互中也同样有重要的作用：在车载电子中，可以自动监测驾驶员的情感状态能够降低事故的发生；在教育技术中监视学生的情绪状态，以提高教学质量；特别是可以与无线传感器网络技术相结合，开发出可佩带的情绪检测装置，在医疗器械市场中有较广阔的前景[38]。

因此，在本文中，我们探讨了"实用语音情感识别"的研究意义，希望能起到抛砖引玉的效果，促进情感计算领域中实用技术的发展。所谓"实用语音情感识别"，包括了两层含义，一是研究具有实用价值的情感类型的识别方法，二是研究和解决实际条件下语音情感识别系统可能面临的困难和挑战。

## 1.4 本文的工作

本文从应用环境中的实际需求出发，研究了具有特殊实用价值的语音情感类别的识别技术，着重研究了几种与认知过程有关的实用语音情感的识别，包括烦躁、自信和疲倦等。在高自然度的情感语料基础上，建立了基于高斯混合模型的实用语音情感识别系统。

本文的研究内容围绕以下三个研究问题进行：

1）如何识别与认知过程有关的实用情感类别，

2）如何提高系统的鲁棒性，

3）如何扩展应用到新的情感识别问题中。

本文的主要工作概述如下：

研究了与认知过程有关的三种实用语音情感：烦躁、疲倦和自信。通过计算机游戏、认知作业等实验进行了高自然度的情感语料的诱发和采集，建立了实用语音情感的数据库，为后续的研究提供数据基础。研究了语音情感的声学特征，分析了实用语音情感的谐波噪声比等特征，构造了481个情感语音的全局统计特征，选用适合的特征降维方法，获得了烦躁、疲倦、自信、喜悦、中性、生气、恐惧、悲伤和惊讶等九种情感的较好的识别效果。

对基本的GMM分类算法进行了研究，考虑了三种条件下的GMM算法的性能和改进：充足样本下的GMM分类器、高自然度的小样本集、连续语音中的情感识别。在小样本条件下，对高斯混合模型进行了改进：采用了两类分类器组的方法提高了实用语音情感识别性能。在连续情感语音中，结合上下文信息，提出了嵌入马尔科夫网络的高斯混合模型识别方法，提升了连续情感语音中的识别性能。

进一步研究了实用语音情感识别系统的鲁棒性问题，提出了四个方面的改进：（1）研究了噪声对情感语音识别的影响，首次将基于人耳听觉掩蔽效应的降噪方法应用于语音情感识别，提高了情感识别系统的抗噪声性能；（2）针对未知类型的情感，提出了基于模糊熵的可拒判的识别方法，扩展了系统的应用范围，提高了对未知数据的适应能力；（3）提出了一种基于说话人聚类的情感特征归一化方法，在反映说话人信息的说话人特征空间进行模糊聚类，利用聚类标签在反映情感信息的情感特征空间中进行特征归一化处理，以降低大量说话人对情感模型的干扰，实现非特定说话人的情感识别；（4）首次融合了语音信号与心电信号进行双模态的实用情感识别。

对本文中建立的实用语音情感识别方法进行了扩展应用：（1）扩展应用到不同语种的情感语料上，研究了在德语和汉语数据上的跨语音情感识别的问题，建立了生气情感的通用模型，分析了两种语言中共同的情感声学特征。（2）扩展应用到畸变语音（耳语音）情感语料上，比较研究了畸变语音与正常音下情感识别的特点，探讨了应该如何对原有的系统进行调整和合理的设置。

本文的创新点归纳为以下八个方面：

（1） 通过认知作业中的重复性四则运算、噪声刺激和睡眠剥夺等诱发手段，首次采集了烦躁、疲倦和自信等汉语实用语音情感语料，建立了实用语音情感数据库。（第二章第四小节）

（2） 在小样本条件下改进了基于 GMM 模型的实用语音情感分类器的性能。针对 GMM 模型对训练数据依赖性较强的弱点，结合两类分类器组和情感码字的相关译码，提高了 GMM 分类器在小样本条件下的识别性能。（第四章第四小节）

（3） 提出了嵌入马尔科夫网络的高斯混合模型，使得传统的 GMM 情感识别系统能够扩展用于连续语音情感识别中。通过三阶马尔科夫网络对连续语音中情感连续性进行建模，将长句和短句两个时域尺度上进行多层次的分析与融合识别，提高了实用语音情感和耳语音情感的识别性能。（第四章第五小节）

（4） 首次将基于听觉掩蔽效应的降噪算法应用在烦躁、疲倦和自信等实用语音情感识别当中，提高了噪声条件下烦躁、疲倦、自信等九种情感的识别性能，并且分析了不同噪声条件对语音情感识别系统的影响。（第五章第二小节）

（5） 提出了一种基于模糊熵的可据判的识别方法。由于真实情感的复杂性，实际条件下会出现的未知情感类别，未知情感类别的输入会导致系统的错误判决，降低系统的稳定性。基于模糊熵阈值的可据判的高斯混合模型分类器能够有效的对未知情感类别进行据判，从而提高系统的稳定性能。（第五章第三小节）

（6） 提出了一种基于说话人聚类的情感特征归一化方法，在反映说话人信息的说话人特征空间进行模糊聚类，利用聚类标签在反映情感信息的情感特征空间中进行特征归一化处理，以降低大量说话人对情感模型的干扰。针对大量未知说话人的场合，提高了非特定说话人的语音情感识别性能。（第五章第四小节）

（7） 首次融合了语音信号与心电信号进行双模态的实用情感识别。同时利用心电信号和语音信号两种通道的信息，对烦躁等实用语音情感进行了较好的识别。与单模态系统比较，其情感识别性能得到了明显的提高。（第五章第五小节）

（8） 首次研究了在德语和汉语数据上的不匹配的跨语言情感识别问题：在德语数据上学习情感模型，在汉语数据上进行验证；在汉语数据上学习情感模型，在德语数据上进行验证。建立了生气情感的通用模型，分析了两种语言中共同的情感声学特征。（第六章第一小节）

本文的第二章至第七章的内容安排如下：

第二章是情感的基本理论，介绍了目前心理学中关于情感与认知的一些结论，并且从诱发

实验的角度探讨了情感数据采集的问题，为下面的研究提供研究数据基础。第三章进行了实用语音情感的特征分析，在基本声学参数之上构造了可用于情感识别的统计特征，并且比较了几种特征压缩方法。第四章研究了高斯混合模型用于实用语音情感识别的若干问题，在不同的样本条件下进行了算法的分析和改进。第五章研究了系统的鲁棒性，探讨了噪声鲁棒性、情感兼容性、大量说话人条件下的鲁棒性，以及多通道抗毁性等问题。第六章将我们建立的情感识别方法扩展应用到耳语音和多语种的情况下，研究了在特殊应用场合系统参数的调整，探讨了其中一系列的制约因素，以及分析了相应的性能表现。在最后一章中，对未来情感计算技术的应用进行了展望，并总结了全文的工作。

# 第二章 情感理论与情感诱发实验

## 2.1 情感的心理学理论

情感识别研究需要以心理学的理论为指导，首先我们需要定义研究的对象——人类情感。然而"情感是什么？"这一个由来已久的问题，一直没有一个统一的答案。目前情感研究中的一个主要的问题是，缺乏对情感的一个一致的定义以及对不同情感类型的一个定性的划分。

### 2.1.1 基本情感论

基本情感论认为，人类的复杂的情感是由若干种有限的基本情感构成的，基本情感按照一定的比例混合构成各种复合情感。基本情感论认为情感可以用离散的类别模型来描述，目前大部分的情感识别系统，都是建立在这一理论体系之上的。随之而来的情感识别的研究途径即是：将模式分类中的分类算法应用到情感类别的划分中。那么基本情感包括哪些呢，不同的研究者对基本情感有不同的定义。Plutchik 认为[39]，基本情感包括："接纳"、"生气"、"期望"、"厌恶"、"喜悦"、"恐惧"、"悲伤"和"惊讶"。Ekman 与 Davidson[40]认为，基本情感包括："生气"、"厌恶"、"恐惧"、"喜悦"、悲伤和"惊讶"。James[41]则认为，基本情感包括："恐惧"、"悲伤"、"爱"和"愤怒"。

我们可以看到，在心理学领域对基本情感类别的定义还没有一个统一的结论，然而在语音情感识别的文献中，较多的研究者采用的是六种基本情感状态："喜悦"、"生气"、"惊讶"、"悲伤"、"恐惧"和"中性"。近年来，有不少研究者对基本情感类别的识别方法进行了研究，取得了一定的研究成果。相应的语音情感数据库也是建立在这些基本情感类型的表演数据上的，例如，在柏林数据库上的平均识别率可以达到百分之八十以上[28]。虽然目前对这些常见的情感类型的研究文献较多，但是对一些具有实际意义的特殊情感类别研究的还很少，特别是烦躁等负面情绪在一些人机系统中具有重要的实用价值，值得我们关注。

基本情感论认为情感可以被看作是一系列离散的类别，从工程技术的角度，对情感建模时可以采用相应的情感类别模型。情感的类别模型的优点是贴近实际应用、符合人们对情感的常识性理解。在大多数谈论到情感的场合，我们关注的是情感属于何种类别，例如，在人机交互中，我们关心用户目前是喜悦还是烦躁，而并不关心烦躁情感可能由哪些基本情感组成，也不

关心用户的情感中激活程度是什么级别，或是情感的控制维度处在什么位置。因此，在实际应用中，情感的类别模型能够满足目前的需求，而且更符合模式分类的思想，是一个较好的情感模型。

## 2.1.2 维度空间论

情感的维度空间论认为人类所有的情感都是由几个维度空间所组成的，特定的情感状态只能代表一个从亲近到退缩或者是从快乐到痛苦的连续空间中的位置，不同情感之间不是独立的，而是连续的，可以实现逐渐的、平稳的转变，不同情感之间的相似性和差异性是根据彼此在维度空间中的距离来显示的。与这种情感理论对应的情感识别方法，最直接的是机器学习中的回归分析（Regression），上文中我们已指出，基本情感类别论对应的情感识别方法则是分类（Classification）算法。

最近 20 多年来，最广为接受和得到较多实际应用的维度模型，是下面两个维度组成的二维空间：

（1）效价度（Valence）或者快乐度 （Hedonic tone），其理论基础是正负情感的分离激活，这得到了许多研究的证明，主要体现为情感主体的情绪感受，是对情感和主体关系的一种度量[42]；

（2）唤醒度（Arousal）或者激活度（Activation），指与情感状态相联系的机体能量激活的程度，是对情感的内在能量的一种度量[42]。

维度空间模型为研究者们提供了一个方便的研究和表示情感的工具。Smith 和 Ike 对日本人的五种基本情感进行了初步的研究[26]，Taylor 等人着重对人脸面部表情进行了研究[43]，金学成等人也对基本情感在唤醒度/效价度空间的分布进行了研究[42]。在本文中，通过我们采集的情感数据和人耳听辨分析，本文中所要研究的几种情感在效价度/唤醒度二维空间中所处的大致位置如图 2.1 所示，与前人的研究结果一致[44, 45]。

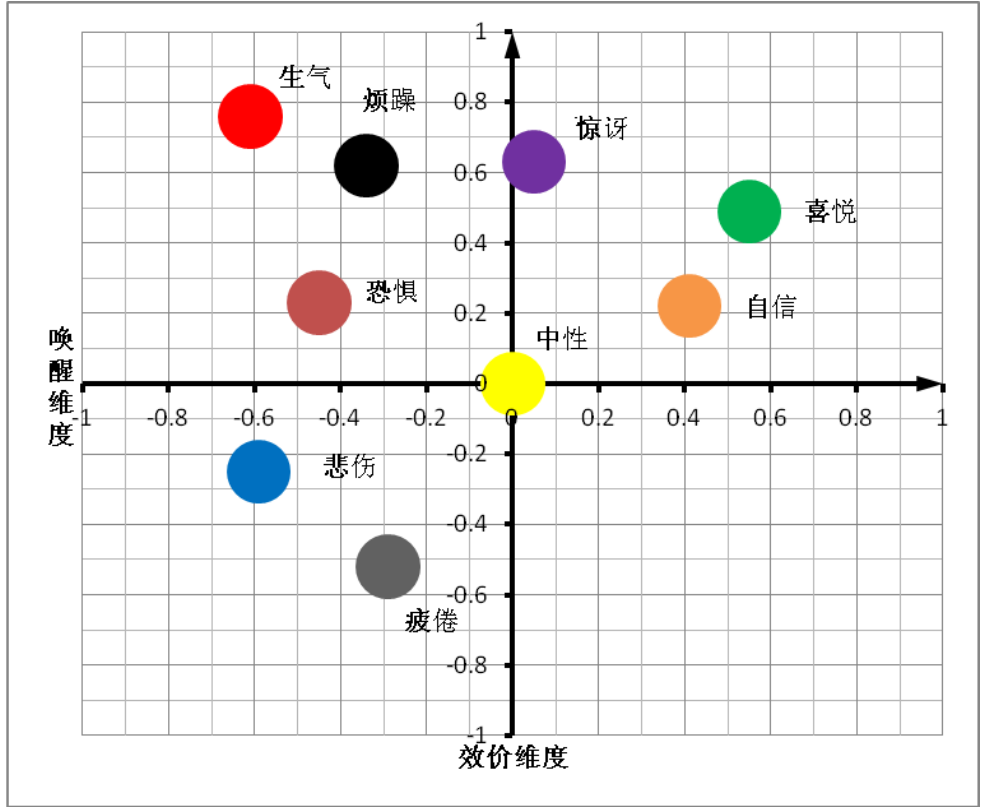

**图 2.1 情感的维度空间分布**

## 2.2 与认知过程有关的三种实用情感

语音情感识别就是要使得计算机能够对用户的情感状态做出一个实时的判断，从而在人机交互当中能够做出相应的响应，提供更好的服务。除了自然人机交互领域，语音情感识别的另一个重要的应用领域是心理状态的自动评估。例如，在话务中心，通过对客服人员的谈话录音数据的自动识别，可以评价服务质量。在教育现场，对学生负面情绪的检测有利于评估教学效果。在航空、航海和航天等军事领域，长时间、高强度的任务会引发烦躁等一系列危险的情感，对烦躁情感的监测，具有重要的实际意义。

因此，对实际当中的特殊情感类别的识别研究，具有重要的实用价值。以往的基本情感类别的识别研究，不能很好的满足实际需要。本文中，我们研究了"烦躁"、"自信"和"疲倦"三种与认知过程有关的实用语音情感。希望能够起到抛砖引玉的效果，促进语音情感识别技术在实际当中的应用。

烦躁的含义通常是指"心中烦闷不安，急躁易怒，甚则手足动作及行为举止躁动不宁的表现"[46]。烦躁在唤醒度上处于正端，在效价度上处于负端。采用反映效价维度特征的音质参数，一般能够将其与喜悦等情感进行区分。

烦躁情感如果不及时处理，可能引发烦躁症等精神疾病。例如，在长时间的载人航天环境中，狭小的轨道舱和高强度的精神压力都可能引发烦躁情感。对烦躁情感的实时监测，以及相应的心理干预，是保证这一类任务顺利完成的重要环节。

自信情感也具有重要实用价值，在军事通信场合，战斗机飞行员说话的语气和方式往往与普通人不同。自信情感在飞行员训练过程中可能出现的较多，能够反映出当前执行任务的人员处于饱满的精神状态。在此类精神状态下，与认知有关的作业的完成率也较高，一般来说出差错的概率较小。

对自信情感的检测能够有效评估人员的心理状态，能够降低人机系统中人员因素带来的潜在危险。自信处于效价度的正向、唤醒度的正向，采用短时能量、基音等韵律特征能够将其与中性、疲倦等状态进行区分。

疲倦是一种常见的负面情感，特别是在长时间的载人航行任务中，很容易引发疲倦情感。在一些重复性的作业中，很容易引发工作人员的疲倦情感，特别会引起认知作业水平的降低，出现错误多发、表现不稳定等现象。因此，在一些特殊环境下的高强度任务中，疲倦状态的检测和及时处理，是保证任务顺利完成的必要环节。

在日常生活中，长时间的使用计算机工作很容易带来一系列的疾病，危害用户的健康。因此对疲倦情感的识别，具有很大的市场应用前景。未来的个人计算机通过情感识别技术，可以对用户的疲倦程度做出一个合理的判断，进而通过播放轻松的音乐、调节舒适的背景灯光等，缓解用户的疲倦感，提醒用户合理的休息，提供更健康、更优质的和谐人机交互体验。

## 2.3 实用语音情感数据库的建立

### 2.3.1 语音情感数据库概述

语音情感数据库的建立，是研究语音情感的必需的研究基础，具有极为重要的意义。目前国际上流行的语音情感数据库有 AIBO（Artificial Intelligence Robot）语料库[47]、VAM(The Vera am Mittag)数据库[48]、丹麦语数据库[49]、柏林数据库[23]、SUSAS(Speech under Simulated and Actual Stress)数据库[50]等。

柏林数据库是一个使用较为广泛的语音情感数据库，早期的语音情感识别研究的成果在柏林库上进行了验证。它包含了生气、无聊、厌恶、恐惧、喜悦、中性和悲伤等语音情感类别，但是其中的情感数据是采用表演的方式采集的，语料的真实度得不到保证，并且数据量较少。柏林数据库中的语料是按照固定的文本进行情感渲染的表演，其文本包含了十条德语语句。十名专业演员参与了语音的录制，包括五名女性、五名男性。初期录制了大约 900 条的语料，后期经过二十个听辨人的检验，494 条语料被选出组成了柏林情感语音数据库，以保证百分之六十以上的听辨人认为这些语料表演自然，百分之八十以上的听辨人对语料的情感标注一致。

丹麦语数据库(Danish Emotional Speech, DES)由四个专业演员表演获得，包括两名男性和两名女性。情感数据中包含了五种基本情感状态：生气、喜悦、中性、悲伤和惊讶。丹麦语数据库中的语料在采集之后，经过了二十名听辨人员进行数据的校验，以保证数据的有效性。这些听辨人员的母语均为丹麦语，年龄在 18 岁至 59 岁之间。

SUSAS 数据库的全称是：Speech under Simulated and Actual Stress,即模拟与实际条件下的紧张语音数据库。SUSAS 是最早建立的自然语料数据库之一，甚至包含了部分现场噪声以增加研究的挑战性。语料库中的语言为英语，说话人数量为 32 人，文本内容包含了一部分航空指令，如"brake"（刹车）、"help"（求助）等，然而这些文本都是事先固定的，而且是长度较短的简短指令。这个数据库的录制方法对一些军事应用具有一定的参考价值，详细的数据库录制过程可以参考 Hansen 等人的报告[50]。

VAM 数据库（The Vera am Mittag German audio-visual emotional speech database）是由德语的脱口秀节目录制而成的一个公开数据库，其数据的自然度较高。VAM 数据库中的情感数据包含了情感语音和人脸表情两部分，总共包含 12 个小时的录制数据。大部分的情感数据具有情感类别情感标注，情感的标注是从唤醒度、效价度和控制度三个情感维度进行评价的。

AIBO 语料库（The FAU Aibo Emotion Corpus）是 2009 年在 Interspeech 会议上举办的 Emotion Challenge 评比中指定的语音情感数据库。情感数据的采集方式是，通过儿童与索尼的 AIBO 机器狗进行自然交互，从而进行情感数据的采集。说话人由 51 名儿童组成，年龄段为 10 岁到 13 岁，其中 30 个为女性。实验过程中，被试儿童被告知索尼的机器狗会服从他们的指挥，鼓励被试像和朋友说话一样同机器狗交谈，而实际上索尼的机器狗是通过无线装置由工作人员控制的，以达到同被试儿童更好交互的目的。语料库包含了 9.2 小时的语音数据，四万八千个左右的单词。数据录制的采样频率为 48kHz，采用 16bit 量化。该语料库中情感数据的自然度高，是目前较为流行的一个语音情感数据库。

以往的语音情感数据库，集中在对几种基本情感的研究上。通过对几种基本语音情感的研究，虽然能够在一定程度上验证识别算法的性能优劣，但是搜索到的情感特征也仅能反映基本情感类别之间的差异。仅停留在对基本情感类别的研究上，远远不能满足实际应用中的需求。

在实际的语音情感识别应用中，还面临着情感语料真实度的问题。早期的数据库往往采取表演的方式来获取情感数据，广泛使用的表演语料库导致研究成果与真实情感差异较大。根据自然程度和采集方法，情感语料可以分为自然语音、诱发语音和表演语音三类。表演语料的优点是容易采集，缺点是情感表现夸张，与实际的自然语音有一定的差别。基于表演情感语料建立情感识别系统，会带入一些先天的缺陷，这是由于用于识别模型训练的数据与实际的数据有一定的差别，导致了提取的情感特征上的差别。因此，早期基于表演语料的识别系统，它的情感模型在实验室条件下是符合样本数据的，在实验测试中也能获得较高的识别率，但是在实际条件下，系统的情感模型与真实的情感数据不能符合的很好，产生了应用中的技术瓶颈。

### 2.3.2 实用语音情感数据库的需求

面向实际应用的需求，实用语音情感数据库必须要保证语料的真实可靠，不能采用传统的表演方式采集数据。在本文中，我们将通过实验心理学中的方法来诱发实用语音情感数据，尽可能的使训练数据接近真实的情感数据，为后续的研究工作提供实验数据基础。

我们针对在长期载人航天环境以及其它类似的高强度特殊作业环境中可能面临的实际问题，选择了具有实际应用价值的语音情感，采集了"烦躁"、"自信"、"疲倦"、"喜悦"和"中性"等情感状态下的语音情感数据，建立了一个汉语音的实用语音情感数据库。下面我们介绍数据库的建立规范、情感诱发过程以及听辨校验方法。

### 2.3.3 建立过程和一般规范

国际上从事语料库大规模开发的著名机构有：美国的非营利性组织 LDC（The Language Data Consortium）、欧洲 1995 年成立的 ELRA（European Language Resources Association）以及 OGI（Oregon Graduate Institute）等。参考国内外著名语料库及其相关的规范[23, 48-51]，本文设计了实用语音情感数据库建立的流程，如图 2.2 所示，相应的规范语料采集规范[52]如表2.1所示。

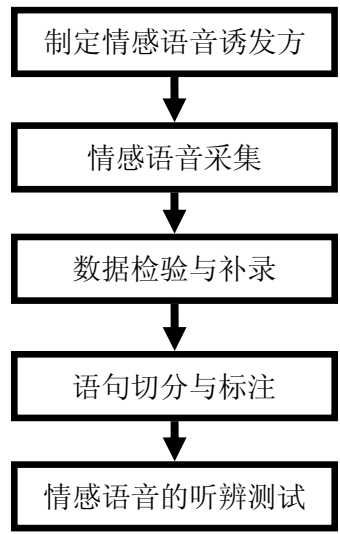

**图 2.2 实用语音情感数据库的制作过程**

**表 2.1 实用语情感语音库的制作规范**

| 规范 | 详细说明 |
|---|---|
| 发音人规范 | 描述发音人的年龄、性别、教育背景和性格特征等。 |
| 语料设计规范 | 描述语料的组织和设计内容。包括文本内容的设计、情感的选择、语料的来源场合等。 |
| 录音规范 | 描述录音环境的软硬件设备、录音声学环境等技术指标。 |
| 数据存储技术规范 | 描述采样率、编码格式、语音文件的存储格式及其技术规范。 |
| 语料库标注规范 | 情感标注内容和标注系统说明。 |
| 法律声明 | 发音人录音之后签署的有关法律条文或者声明。 |

### 2.3.4 软硬件设备环境与数据检验

录音软件采用"Adobe Audition"软件(原"Cool Edit"软件),录音时采用单声道、采样频率为 11kHz 和 48kHz 两种、16 位量化精度,录制的语音信号保存为 PCM(Pulse Code Modulation)编码的 WAV(Waveform Audio Format)格式。录音硬件设备包括:高性能计算机一台、M-audio MobilePre USB 声卡一台、大振膜电容话筒一支、监听耳机一副等。

录音过程在安静的实验室内进行，每次录音后，应进行数据的检验与补录，及时对语音文件进行人工检验，以排除录音过程中可能出现的错误。例如，查看并剔除语音中的信号过载音段、不规则噪声（例如咳嗽等）和非正常停顿造成的长时静音等。对于错误严重的录音文件，必要时进行补录。

## 2.4 情感语料的诱发方法

### 2.4.1 通过计算机游戏诱发情感语料

本节我们介绍通过计算机游戏来诱发喜悦和烦躁情感的方法，在计算机游戏创造出的虚拟的情景中诱发被试说出带有特定情感的话语，能够采集高自然度的情感数据。

根据 Scherer 等人的观点[9, 19, 53, 54]，人类声音中蕴含的情感信息，受到无意识的心理状态变化的影响（push effects），以及社会文化导致的有意识的说话习惯的控制（pull effects）。因此，实用语音情感数据库的建立需要考虑语音中情感的自然流露和有意识控制。通过实验诱发手段，诱发情感在语音中的自然流露。充分考虑录音环境中，社会文化习惯的影响，尽量排除干扰因素，确保获得充足的目标情感类别的数据。

在传统的语音情感数据库中，往往采用表演的方式来采集数据。在实际的语音通话和自然交谈中，说话人的情感对语音产生的影响，常常是不受说话人控制的，通常也不服务于有意识的交流目的，而是反映了说话人潜在的心理状态的变化[53, 54]。相反，演员能通过刻意的控制声音的变化来表演所需要的情感，这样采集的情感数据对于情感语音的合成研究是没有问题的，但是对自然情感语音的识别研究是不合适的，因为表演数据不能提供一个准确的情感模型。为了能更好的研究实际环境中的情感语音，有必要采集除表演语音以外的，较高自然度的情感数据，

针对这个问题 Johnstone 等人进行了诱发心理学实验[53, 54]，最早通过计算机游戏诱发情感的方法来采集语音情感数据。采用计算机游戏进行情感诱发实验的优势在于，通过游戏中画面和音乐的视觉、听觉刺激，能提供一个互动的、具有较强感染力的人机交互环境，能够有效的诱发出被试的正面与负面的情感。特别是在游戏胜利时，被试由于在游戏虚拟场景中的成功与满足，被诱发出喜悦等正面情感；在游戏失败时，被试在虚拟场景中受到挫折，容易引发烦躁等负面情感。

下面我们就来介绍本文中计算机游戏诱发情感的方法。在游戏场景的设置方面，为了便于烦躁、喜悦情感的诱发，我们选用了既需要耐心又具有一定挑战的计算机小游戏。游戏中被试要求用鼠标移动一个小球通过复杂的管道，在通过管道的过程中如果小球碰到管壁，则小球将爆炸，游戏失败；在规定时间内（倒计时 1 分钟）顺利通过管道后，到达终点，游戏胜利；游戏共有 5 个难度等级，以适合不同水平的被试。情感语音的诱发与录制过程如下。在被试参加游戏前，让被试平静的读出指定的文本内容，录制中性状态的语音。在每次游戏胜利后，要求被试用喜悦的语气说出指定的文本内容，录制喜悦状态的语音。在每次游戏失败后，要求被试用烦躁的语气说出指定的文本内容，录制烦躁状态的语音。为了便于对数据进行检验，在每次录制情感语音后，让被试填写情感的主观体验，记录诱发的情感类型，在实验结束后，根据被试的情感主观体验表，剔除主观体验与诱发目标情感不一致的语音数据，必要时进行适当的补录。

选择参与情感诱发实验的被试人员（发音人），发音人应具有良好的健康状况，近期无感冒，无喉部疾病，并且听力正常。发音人的选择主要考虑发音人的性别、年龄、生活背景、教育程度、职业、病理情况、听力状况、口音等。研究表明，由于生理构造的差别，男女在表达相同情感时，其声学特征有一定差异性；而不同年龄段的人群，在表达情感时同样会出现不同情况。在建库时对这些因素进行规范，可以有选择性的提高某些特定人群的情感识别率。本实验中选拔的被试人员为 5 名男性和 5 名女性，年龄在 20 岁到 30 岁之间。

在本诱发实验中采用了固定的文本内容，在目前的情感语音数据库中，有的数据库采用固定的文本内容，以便于进行情感特征的对比分析，有的数据库录制过程中不固定文本内容，获得的自然语料难度更高，对建立与文本无关的系统的要求更高。考虑到实用语音情感识别的一个主要应用领域为长期的航空、航天和航海任务所引发的负面情感的评估，本实验中的设计的文本由无情感倾向性的工作用语单词和短句组成，选自国际海事组织（IMO）发布的《标准航海通信用语》（SMCP）[55]。工作用语中的单词和短句的例子如下：

"螺旋桨"（单词）。

"保持你船左后弦受浪"（短句）。

对于固定文本的设计，主要考虑以下两个方面。第一，所选择的单词和语句必须不包含某一方面的情感倾向；第二，必须具有较高的情感自由度，对同一个单词或语句能施加各种感情进行分析比较。文本材料分为单词和短句两种类型，组成如表 2.2 所示。单词共 20 个，短句 20

句。单词着重考虑了名词、动词。分别选用名词 5 个、动词 5 个。因为形容词往往会附带感情色彩，在这里没有加以考虑。短句的选择主要按不同句型分类，着重考虑使用频率较高的陈述句，挑选部分疑问句、祈使句，如表 2.2 所示。

**表 2.2 实用语音情感数据库的文本类型**

| 类型 | 分类 | 数量 |
|------|------|------|
| 单词 | 名词 | 10 |
|      | 动词 | 10 |
| 短句 | 陈述句 | 10 |
|      | 疑问句 | 5 |
|      | 祈使句 | 5 |

### 2.4.2 通过认知作业诱发情感语料

本节中介绍另一种情感诱发的手段——通过认知作业诱发情感语料，包括烦躁、疲劳和自信等心理状态的诱发。在一个重复性的、长时间的认知作业中，我们采用噪声诱发、睡眠剥夺等手段辅助诱发负面情绪。

认知作业现场的情感识别具有重要的实际意义，特别是在航天、航空、航海等长时间的、高强度的工作环境中，对工作人员的负面情感的及时检测和调控是非常重要的一个研究课题。烦躁、疲劳和自信等心理状态对认知过程有重要的影响，是评估特殊工作人员的心理状态和认知作业水平的一个重要因素。

认知心理学的研究表明，负面情感对认知能力有影响。对情感的研究，吸引了很对不同领域的研究者，Pereira 的研究显示，负面情感会影响到对视觉目标的识别能力[37]。一个自动识别人类情感的系统会在很多领域发挥重大的作用，例如，在车载系统中可以帮助驾驶员调节烦躁情感从而避免事故；在公共场所的监视系统中，对恐惧等极端情感的检测，可以帮助识别潜在的危险情况。此外，与认知有关的实用情感的识别，在教育技术和智能人机交互等领域中，也具有广阔的应用前景。

在诱发实验中先后共六名被试参加实验，其中三名为女性。实验中要求被试进行数学四则运算测试，以模拟认知工作环境。在实验中，被试将题目和计算结果进行口头汇报，并进行录音，以获取语料数据。在实验的第一阶段，通过轻松的音乐使得被试放松情绪，进行一些较为简单容易的计算题目，以获得正面的情感语料。在实验的第二阶段，采用噪声刺激的手段来诱发负面情感（通过佩戴的耳机进行播放），采用睡眠剥夺的手段辅助诱发负面情感（如烦躁、疲

倦等），同时增加计算题目的困难程度。对于实验中对于简单的四则运算题目，被试容易做出自信的回答，对于较难的计算，被试的口头汇报中出现明显的迟疑，在实验的后半段，经过长时间的工作，被试更容易产生疲劳和烦躁的情感。认知作业结束后，我们对每一题的正确与错误进行了记录和统计。对每一段录制的语音数据进行了被试的自我评价，标注了所出现的目标情感。

通过检测与认知有关的三种实用语音情感，能够从行为特征的角度反映出被试认知能力的波动，从而客观的评估特殊工作人员的心理状态与该项工作的适合程度。因此，有必要进一步的研究负面情感对认知能力的影响。我们分析了负面情感和正面情感下的作业错误率，如图 2.3 所示。

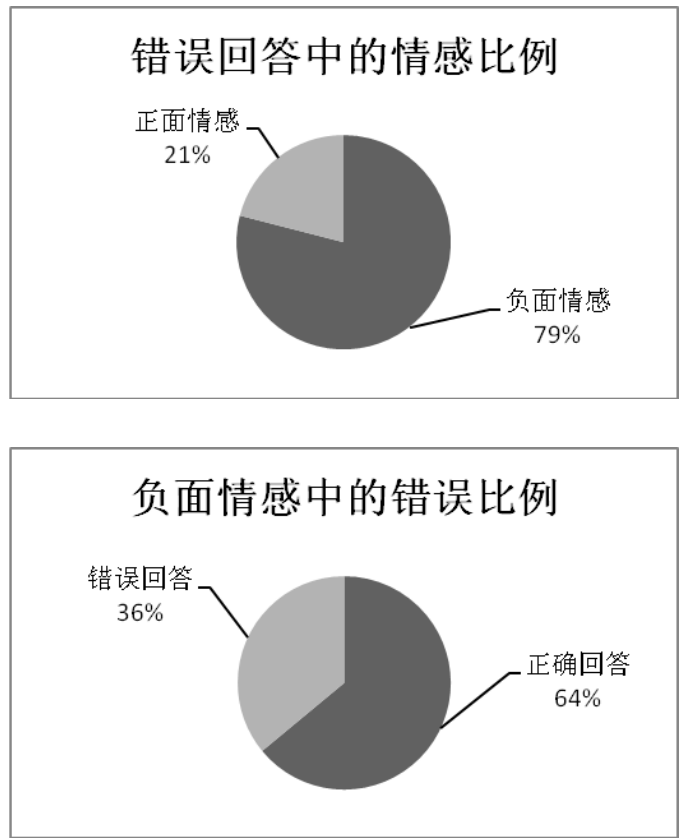

**图 2.3 负面情感和错误率之间的关系**

## 2.5 情感语料的主观评价方法

为了保证所采集的情感语料的可靠性，需要进行主观听辨评价，每条样本由 10 名未参与录音的人员进行评测。一般认为人区分信息等级的极限能力为 7±2，故可以引入九分位的比例标度

来衡量信息等级。在本文的评价方法中，我们采用标度 1、3、5、7、9 表示情感的五种强度，对应极弱，较弱，一般，较强，极强五个等级。

每条情感样本相对于每个听辨人都会产生一个评测的结果：

$$\boldsymbol{E}_{ij} = \{e_1^{ij}, e_2^{ij}, \cdots, e_K^{ij}\} \tag{2-1}$$

这里 $j$ 用来标记情感样本，$i$ 用于标记听辨人，$K$ 为情感类别数量，矢量中的不同评价因子 $e$ 代表该听辨人对于该情感语句的不同情感成分的评判值。

由于采取多人评测，为了得到第 $j$ 条情感样本的评价结果，需要将所有听辨人的测评结果进行融合，采用加权融合的准则得到该条情感样本的评判结果为：

$$\boldsymbol{E}_j = \sum_{i=1}^{M} a_i \boldsymbol{E}_{ij} \tag{2-2}$$

其中 $a_i$ 是每个听辨人的评价结果的融合权重，它代表每个听辨人的评价结果的可靠程度，有

$$\sum_{i=1}^{M} a_i = 1 \tag{2-3}$$

$M$ 为听辨人总数。融合权重对最终结果有重要的影响，其数值根据听辨人的评测质量来确定。由于在多人的评测系统中，不同听辨人的评价结果带有一定的相关性，因此我们可以从听辨结果的一致度方面来计算融合权值。

对第 $j$ 条数据，两个听辨人 $p$、$q$ 之间的相似性度量可以定义为：

$$\rho_j^{pq} = \prod_{i=1}^{K} \frac{\min\{e_i^{pj}, e_i^{qj}\}}{\max\{e_i^{pj}, e_i^{qj}\}} \tag{2-4}$$

其中 $K$ 为情感类别总数，对每次测评，两个听辨人 $p$、$q$ 之间的相似性度量为：

$$\begin{aligned}
\rho^{pq} &= \frac{1}{N} \sum_{j=1}^{N} \rho_j^{pq} \\
&= \frac{1}{N} \sum_{j=1}^{N} \prod_{i=1}^{K} \frac{\min\{e_i^{pj}, e_i^{qj}\}}{\max\{e_i^{pj}, e_i^{qj}\}}
\end{aligned} \tag{2-5}$$

其中 $N$ 为情感样本的总数。根据两人之间的相似性，可以得到一个一致度矩阵，矩阵中的每个元素代表两个听辨人之间的相互支持程度：

$$\boldsymbol{\rho} = \begin{bmatrix} 1 & \rho^{12} & \cdots & \rho^{1M} \\ \rho^{21} & 1 & \cdots & \rho^{2M} \\ \cdots & \cdots & \cdots & \cdots \\ \rho^{M1} & \rho^{M2} & \cdots & 1 \end{bmatrix} \tag{2-6}$$

据此，第 $i$ 个听辨人与其他听辨人之间的一致程度，可以通过计算平均一致度来获得：

$$\overline{\rho^i} = \frac{1}{M-1} \sum_{\substack{j=1 \\ j \neq i}}^{M} \rho^{ij} \tag{2-7}$$

以归一化之后的一致度作为每个听辨人评测结果的融合权重 $a_i$：

$$a_i = \frac{\overline{\rho^i}}{\sum_{k=1}^{M} \overline{\rho^k}} \tag{2-8}$$

将其代入（2-2）式即得到的每条情感语句的评价结果 $\boldsymbol{E}_j$：

$$\boldsymbol{E}_j = \frac{\sum_{i=1}^{M} \overline{\rho^i} \boldsymbol{E}_{ij}}{\sum_{k=1}^{M} \overline{\rho^k}} \tag{2-9}$$

根据评价结果可以对数据进行情感标注，假设 $\boldsymbol{E}_j$ 中最大的元素是 $e_p^j$，则认为该情感语句为主情感为 $p$ 的情感语料，作为最终评判的结果。

## 2.6 情感语料数据集的总结

全文中的实验数据，一共涉及了诱发数据库、表演数据库、双模态数据库三个种类。其中诱发数据中包括烦躁、疲倦和自信等与认知有关的语音情感，以及喜悦情感。表演数据中包括了三个数据库：一，基本情感类别的表演数据库[56]，二，非特定说话人数据库，三，柏林数据库[23]。我们在中文的表演数据库中选取了一部分质量较好的语料，作为实验数据集的补充。为情感训练保证足够的数据量，除了与认知有关的几种情感和喜悦情感外，进一步补充了

表演获得的其它几种基本情感，见表 2.4。另外两个表演数据库为柏林数据库和富士通非特定说话人数据库。柏林数据库，是为了研究系统的跨语言扩展性能而引入的。富士通的非特定说话人数据库是针对大量说话人的研究而采用的。双模态数据库当中，包含了语音和心电整合起来的双模态数据。我们按照时间顺序，将同一时间段的语音信号与心电信号对应保存起来，并且进行情感的标注。引入双模态数据库的目的是为了在下文中研究双通道的方式下系统鲁棒性是否能够获得提升。

本文中，主要使用下面两个测试数据集：

（1） 五种实用情感的实验数据集，如表 2.3 所示，包含烦躁、喜悦、中性、自信和疲倦等情感类别。

（2） 九种语音情感的实验数据集，如表 2.4 所示，包含烦躁、喜悦、中性、自信、疲倦、生气、悲伤、惊讶和恐惧等情感类别。

### 表 2.3 实验数据集 A（五种情感）

| 情感类型 | 烦躁 | | 喜悦 | | 中性 | | 自信 | | 疲倦 | |
|---|---|---|---|---|---|---|---|---|---|---|
| 说话人数（男/女） | 5 男/5 女 | 3 男/3 女 | 5 男/5 女 | 3 男/3 女 | 5 男/5 女 | 3 男/3 女 | 5 男/5 女 | 3 男/3 女 | 5 男/5 女 | 3 男/3 女 |
| 样本数量 | 1000 条 | 1000 条 | 1000 条 | 1000 条 | 1000 条 | 1000 条 | 1000 条 | 1000 条 | 1000 条 | 1000 条 |
| 文本种类 | 短句 | 口语 | 短句 | 口语 | 短句 | 口语 | 短句 | 口语 | 短句 | 口语 |
| 诱发方式 | 计算机游戏 | 认知作业 | 计算机游戏 | 认知作业 | 自然状态 | 自然状态 | 计算机游戏 | 认知作业 | 计算机游戏 | 认知作业 |
| 总计 | 2000 条 | | 2000 条 | | 2000 条 | | 2000 条 | | 2000 条 | |

### 表 2.4 实验数据集 B（九种情感）

| 情感类型 | 烦躁 | 喜悦 | 中性 | 自信 | 疲倦 | 生气 | 悲伤 | 惊讶 | 恐惧 |
|---|---|---|---|---|---|---|---|---|---|
| 说话人数（男/女） | 8 男/8 女 | 8 男/8 女 | 8 男/8 女 | 8 男/8 女 | 8 男/8 女 | 6 男/6 女 | 6 男/6 女 | 6 男/6 女 | 6 男/6 女 |
| 样本数量 | 2000 条 | 2000 条 | 2000 条 | 2000 条 | 2000 条 | 2000 条 | 2000 条 | 2000 条 | 2000 条 |
| 文本种类 | 短句、口语 | 短句、口语 | 短句、口语 | 短句、口语 | 短句、口语 | 短句 | 短句 | 短句 | 短句 |
| 数据来源 | 诱发 | 诱发 | 自然 | 诱发 | 诱发 | 表演 | 表演 | 表演 | 表演 |

注 1：表中增加四种基本情感：前五种情感数据与实验数据集 A 中的相同，后面四种基本情感数据是实验数据集 B 中增加的。

注 2：表中"自然"指的是中性语音在自然的状态下获得，数据来源为自然语料。

本文中大部分的实用都采用上面的两个实验数据集。在某些章节中，由于特殊的应用需求，采用了不同的测试数据集与测试方法，在具体的实验中会予以说明。

# 第三章情感的声学特征分析

## 3.1 语音情感特征研究的历史与现状

情感语音当中可以提取多种声学特征，用以反映说话人的情感行为的特点。情感特征的优劣对情感最终识别效果的好坏有非常重要的影响，如何提取和选择能有效反映情感变化的语音特征，是目前语音情感识别领域最重要的问题之一。在过去的几十年里，针对语音信号中的何种特征能有效的体现情感，研究者从心理学、语音语言学等角度出发，作了大量的研究。许多常见的语音参数都可以用来研究[13]，这些语音参数也常用于自动语音识别和说话人识别当中。例如：短时能量、过零率、有声段和无声段之比、发音持续时间、语速、基音频率、共振峰频率和带宽、镁尔倒谱参数（Mel-Frequency Cepstral Coefficients，MFCC）等等。

### 3.1.1 韵律特征与早期的研究

短时能量、发音持续时间、语速和基音频率等特征，通常被认为是韵律特征，与情感维度模型中的唤醒维度关系密切。韵律特征又被称作超音段特征，是指大于一个音位的语音单位如音节（syllable）或比音节更大的单位所表现出来的音强（intensity）、音长（length or duration）、音高（pitch）、重音（accent）、声调（tone）和语调（intonation）等语音特征参数。这一类韵律特征的变化，可能反映出情感在唤醒维度上的变换。例如，生气和中性在唤醒维度上有较大的差距，它们的短时能量特征会有明显的差异。

早期的情感特征分析主要来自心理学领域的研究，Pittam 等人的研究[1, 57, 58]表明，4kHz 以上频谱能量特征可以反映情感的变化，这一部分频段能量的增加能反映激励程度的提高，可用于区分悲伤与生气等。

Petrushin 较早的进行了语音情感特征的计算机识别[59]，一共采用了 43 个声学特征进行情感语音分析。在他的研究中，主要采用了基音频率来反映情感的变化，使用了平滑算法来获得基音频率的轮廓。除了基音特征外，Petrushin 的研究结果表明，短时能量、频谱特征、共振峰、语速、停顿以及有声段能量比例等特征对语音情感的分析也有一定的贡献。

赵力等人[13]较早的进行了汉语音的情感特征分析，主要采用了基音频率、共振峰等韵律特征和音质特征进行六种语音情感的识别，并且在此基础上进一步分析了男女性别差异对情感语音特征的影响，较早的提出了性别的规整化方法[30]。

早期的研究者们普遍采用的情感特征都是最基本的韵律特征，如基音频率、短时能量、发音持续时间、语速等。这些韵律特征确实能够反映说话人的部分情感信息，较大程度上能区分不同的基本情感类别，且参数提取的算法较为成熟。

### 3.1.2 音质特征与韵律特征的结合

随着研究的深入，另外一些研究者认为，音质特征和韵律特征相互结合才能更好的表达情感，仅有韵律特征是不够的[53，60]。音质特征主要指语音的音色和语谱方面的特征，因此也被称作是音段特征，反映发音时声门波形状的变化，其影响因素有肌肉张力、声道中央压力以及声道长度张力等。例如在发音当中是否有喘气、声音嘶哑等现象存在，都属于音质特征。从语音产生模型的角度看，音质特征主要指与声道响应相关的语谱包络特征。代表性的音质特征主要有共振峰、倒谱、LPC 系数及其衍生参数等。一般认为音质特征有可能跟情感维度模型中的效价维度的关系密切。

Alter[61]等人通过对韵律和音质之间关系的研究，发现生气和高兴状态下的发音在喘气和沙哑等方面是不同的。某些特定的元音在发音结构上的变化直接依赖于施加在上面的情感状态，而另一些元音的发音则依赖于其在句子中的位置，以及当时说话人采用的重读模式。

### 3.1.3 情感维度与情感特征的关系

不管是韵律特征还是音质特征，它们都反映了情感的维度信息，情感特征与情感维度之间并不是孤立的。Pereira 等人认为，语音信号的韵律特征与三个情感维度（效价维、唤醒维和控制维）之间的都具有一定关联性[62]，其中唤醒维和韵律特征之间具有较强的关联，在唤醒维度上距离相近的情感状态不容易用韵律特征进行区分。Gobl[60]和 Johnstone 等人[53，54]的研究同样表明语音信号中的音质特征，与情感的效价维关系密切，并且 Johnstone 等人在自然度较高的诱发语料上进行了实验验证。

金学成等人在 Pereira 的研究基础上，对共振峰等音质类特征进行了比较分析，认为此类特征与效价维的相关性较强[42]。Tato 等人[14]的研究同样表明，音质类特征对于区分唤醒

维度差异小的情感（生气和喜悦）有较好的效果，表明共振峰等音质类特征与效价维的相关性较强。

对于情感的三维模型中的控制维度，也应该能够找到对应的特征，但是在目前研究者们还没有发现有效的控制维特征。有的研究者认为音质特征也能够反映出控制维度的一些变换，有的研究者则从语义特征中去寻找控制维度的信息[17]。

### 3.1.4 近年来的研究

除了这些在语音识别和说话人识别当中常见的特征外，近年来研究者们不断提出了一些新的特征，包括：谐波噪声比特征、Teager 能量算子特征和一些谱特征等[63]。这些情感特征的有效性在部分数据库上获得了验证，谐波噪声比特征可以用在性别差异分析当中，然而其抗噪声性能可能较差；Teager 能量算子能够反映出气流中的非线性特征[64]，其在情感分析中的作用仅在少数情感语料上得到过证实；谱特征参数获得了较多研究者的关注。这些特征与效价维度的关系比较大，从在理论上看应该有利于正面情感与负面情感的区分。

近年来，在情感特征的分析过程中，研究者们开始关注到语料的真实度的问题。以为的表演语料具有一定的夸张成分，在其上获得的情感特征与实际情况可能存在一定的偏差。Truong 等人强调了情感语料在情感识别系统中的关键性作用[24]，采用了自然语音进行情感特征的分析研究，以纠正过去在表演语料上发生的一些错误。

在过去的情感特征分析中，存在的最大的问题是不同研究者之间的实验结果具有较大的差别，由于语料库的不统一，研究成果之间的可比性较差。往往在一个数据库上行之有效的特征，迁移到另一组语料上就不能获得同样的性能。因此，在今后的研究中，我们应该关注跨数据库的扩展性能的研究，对不同民族之间和不同语种之间的情感表达的差异应该受到研究者的重视。在后文中的扩展性研究章节中，我们进行了德语语料与汉语语料上的跨数据库情感识别研究，进行了初步的跨语言的情感特征分析。

此外，对于特殊人群和特殊工作环境中的情感特征的分析，具有较高的实际意义，应当受到重视。例如，高压环境下军人的情感和心理状态变化，载人航天器中狭小密闭环境引发宇航员的负面情绪，这些都是值得研究的课题。可以预期，在实际环境中引发的情感状态，其特征应该与标准数据库当中的基本情感类别的特征有所不同。因此对实用语音情感特征的研究具有较高的实际意义。

## 3.2 情感引发的基本声学参数变化

在这一节当中，我们分析基本声学参数的变化。语音情感特征的分析，可以分为两个层次：低层的描述子（Low Level Descriptor，LLD），以及用于建模识别的最优特征组或特征空间。低层描述子，即短时能量、基音频率、共振峰频率等基本的语音参数。一般还可以采用各类统计函数构造全局的静态特征，如最大值、最小值、均值、方差、一阶差分、二阶差分等等，基音轮廓的抖动等特征也包含在 LLD 特征中。在本节中，着重考察几种基本语音参数受到情感变化的影响，在后文中再进行全局特征的构造和特征的压缩降维等研究。

### 3.2.1 九种情感状态下的短时能量变化与基音轮廓变化

我们选取了有代表性的典型样本研究情感状态引起的特征变化，如图 3.1 至图 3.9 所示。为了更好的反映出情感信息带来的变化，固定了文本的内容（文本的内容为"好像快要下雨了"），这样可以排除有音位信息差别带来的特征变化。尽量采用单因素分析的方法，排除除了情感外的干扰因素，研究情感信息与声学特征之间的关联。

图 3.1 到图 3.9 中的语音样本，都出自于同一名说话人（说话人为男性，年龄在二十至三十岁之间）。固定说话人进行特征的对比分析，尽量排除由于说话人习惯和性格差异带来的影响。如此在情感样本之间总结出来的差异性特点，就不会与说话人和语义因素有关，根据单因素分析的原则，仅与情感状态有关。这样的研究方法不能消除个体差异，但是能够排除非情感的干扰因素，较"纯净"的显示出情感信息引发的声学特征的变化。

我们研究了三种对认知过程有较大影响的情感状态：烦躁、自信和疲倦，还包括了基本情感类型：喜悦、生气、惊讶、悲伤、恐惧，以及中性状态。图中的曲线使用 PRAAT 软件获得，在特征分析前，时域波形信号统一进行了标准化处理。

值得注意的是，在恐惧状态下，第一个"好"字音发生了基音缺失的情况，恐惧使得该男性被试的发音中出现了耳语音的特点。在疲倦状态下，典型样本的特征非常有特点，其基音频率较低和较平坦。在中性语音中，样本的基音轮廓同样较为平坦，尾部的"了"字音较高，相比较而言句子末吐字不连贯，基音频率发生了跳跃。悲伤情感下，结尾处的基音轮廓发生持续和平滑的下降，而喜悦情感下对应的这部分基音轮廓出现上升的凸起，同时我们看到喜悦语音的基音尾部呈现一个较明显的波浪形，起伏较大。自信和惊讶样本中，基音尾部都出现了上扬。

我们还可以看到，生气状态下短时能量的强度波动相对较小，包络较为平滑，而悲伤的短时能量的包络方差比较大。悲伤和疲倦的语音样本持续持久较久，语速较缓，而烦躁和生气的语音语速较快。

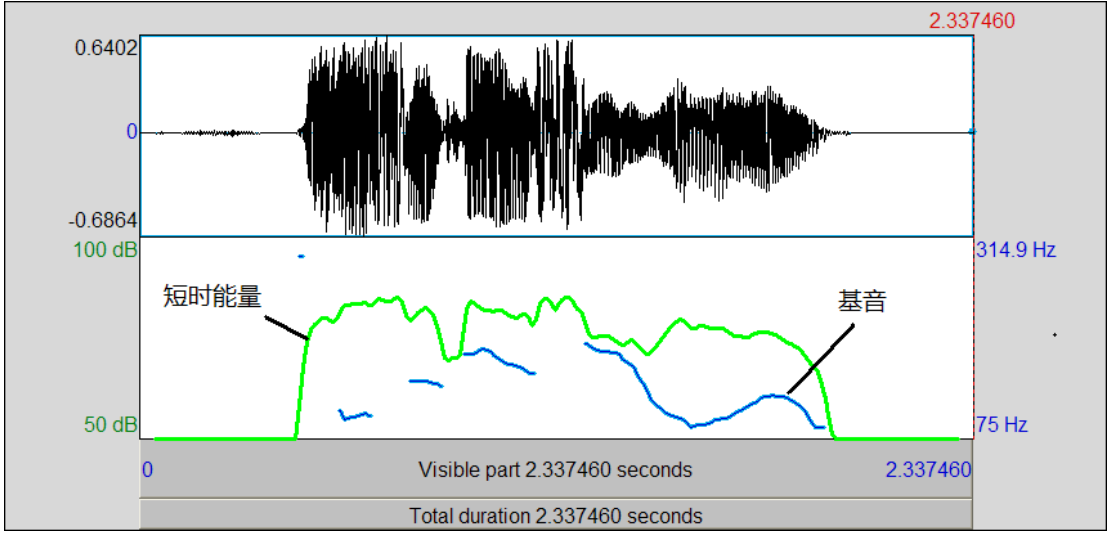

图 3.1 喜悦样本的短时能量和基音轮廓

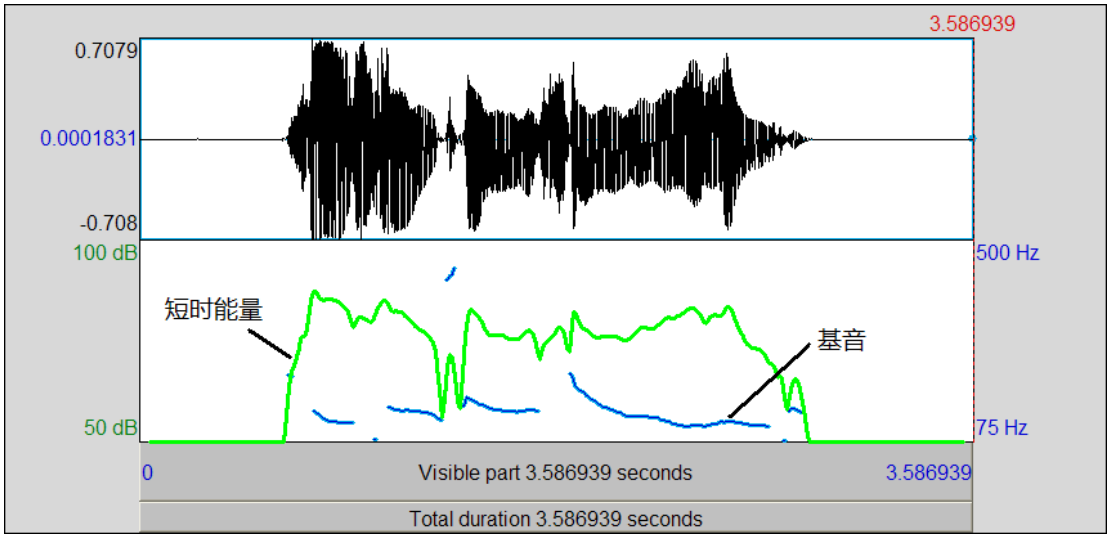

图 3.2 悲伤样本的短时能量和基音轮廓

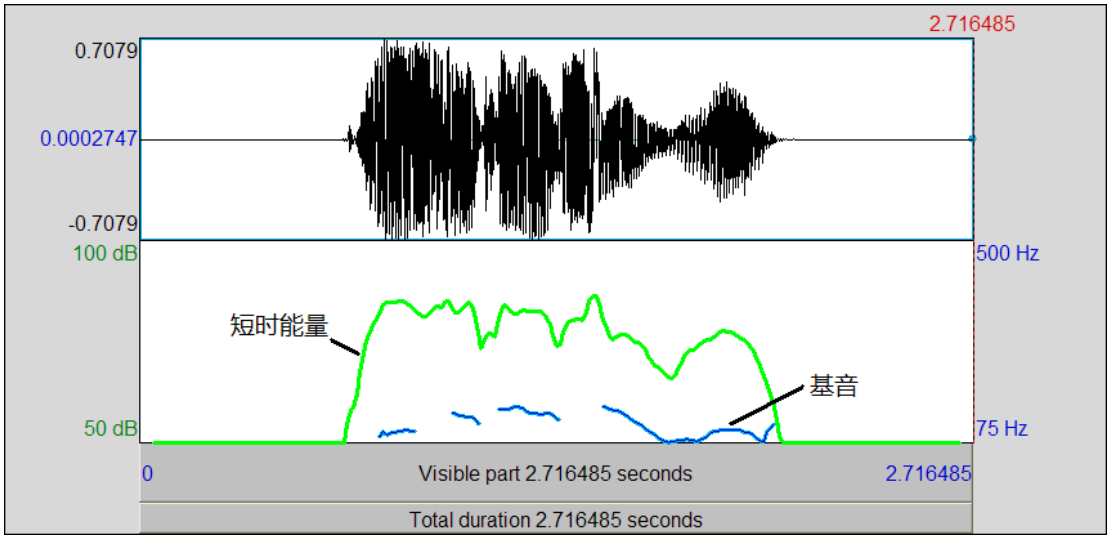

图 3.3 烦躁样本的短时能量和基音轮廓

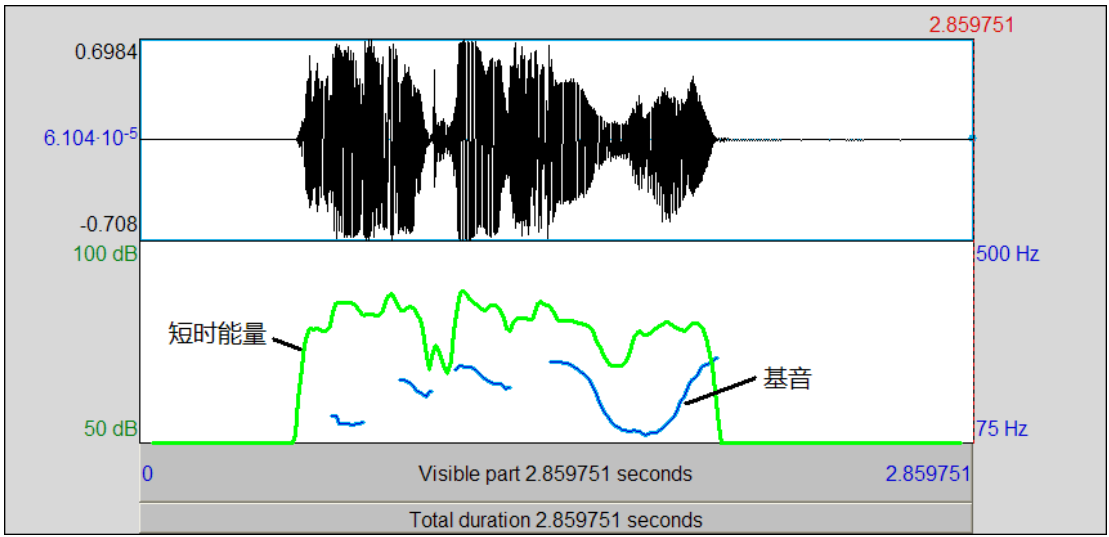

图 3.4 惊讶样本的短时能量和基音轮廓

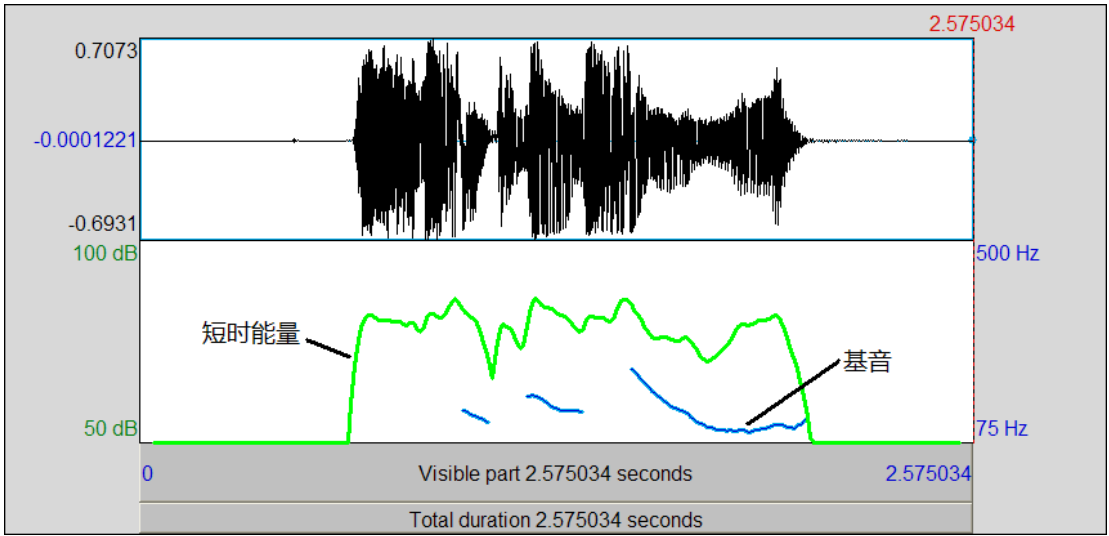

图 3.5 恐惧样本的短时能量和基音轮廓

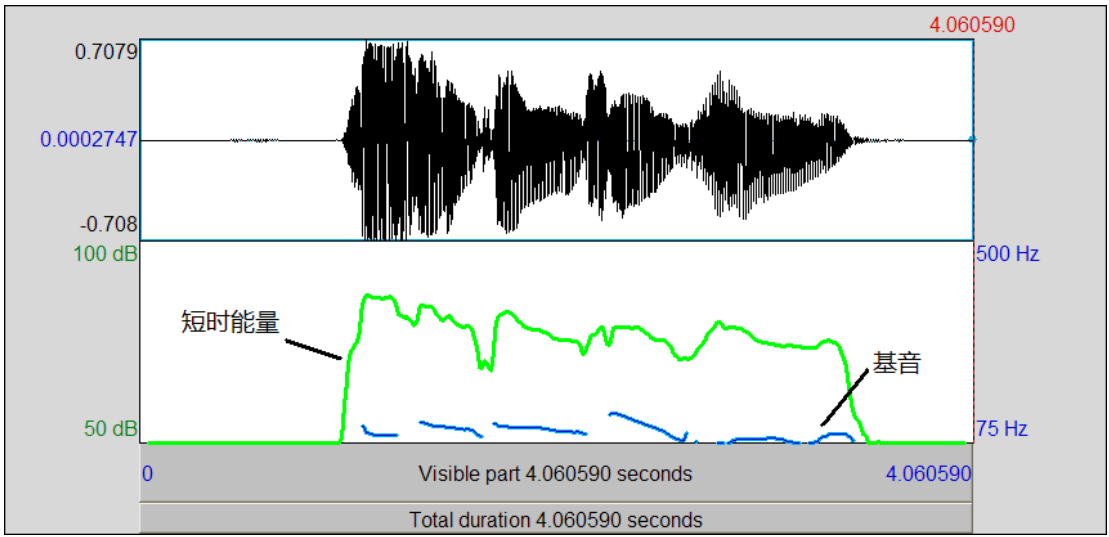

图 3.6 疲倦样本的短时能量和基音轮廓

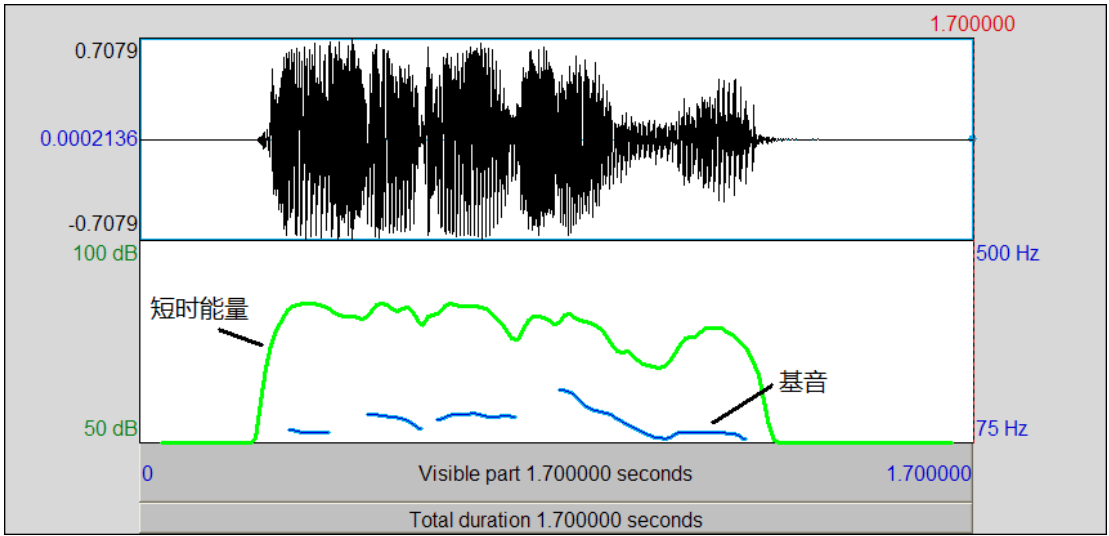

**图 3.7 生气样本的短时能量和基音轮廓**

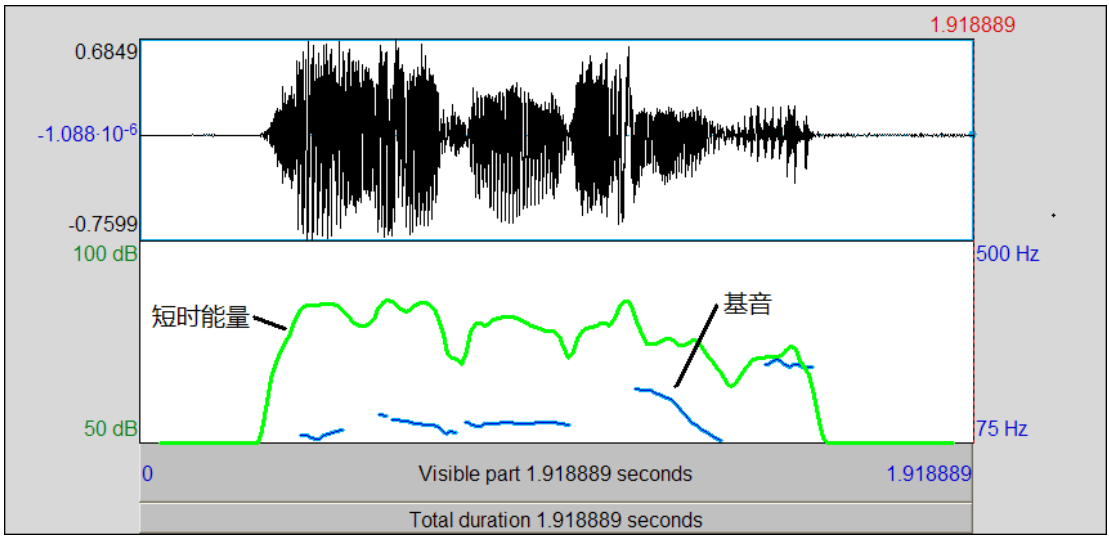

**图 3.8 中性样本的短时能量和基音轮廓**

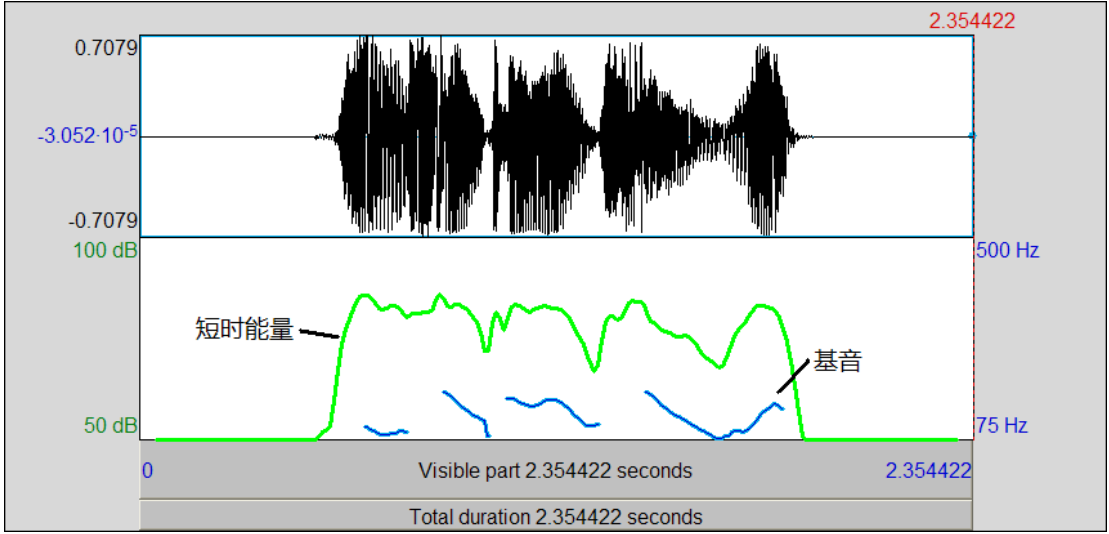

图 3.9 自信样本的短时能量和基音轮廓

## 3.2.2 实用情感状态下的基音与共振峰特征分布

上一节中我们看到，特征参数的轮廓随着情感状态的变化而变化，以帧为单位的分析方法提供了描述帧序列的动态特征。本节中，我们用整条语料上的统计均值，来研究实用情感在两种基本的语音特征，基音频率和共振峰频率上的差异。整条语句上的全局统计特征，是一种静态的特征，它受到文本语义的影响相对较小。

图 3.10 到图 3.13 为基音与共振峰特征的均值差异。从图中我们可以看到，喜悦和自信的基音均值较高，烦躁的第二共振峰均值较低，疲倦的第三共振峰均值较高。基音特征是一种韵律特征，往往与情感的唤醒度有关，从图 3.10 中可以看到，喜悦和自信的基音均值较高，而疲倦和烦躁的基音均值较低。注意到几种实用情感在情感维度空间中的位置，烦躁、喜悦和自信位于唤醒度的正向，疲倦位于唤醒度的负向。因此可以证实基音特征与唤醒维度的关系密切，两者的上升和下降变化趋势一致，并且基音频率不仅可以用于基本情感类别的区分，同样也可以用于实用语音情感的识别。

共振峰特征反映了声道的变化信息，是一种音质特征，一般认为与效价维度的情感信息有关。从图中我们可以看到，疲倦和烦躁两种负面情感在共振峰均值特征上具有自身的特点，与其他情感差别相对较大。在第二和第三共振峰特征中，疲倦处于最大值位置，烦躁处于最小值位置。我们注意到在情感的维度空间中，疲倦和烦躁都处于效价维度的负向。这说明共振峰特征上的变化与效价维度的上升和下降有关，但并不是单调函数的一致的关系。

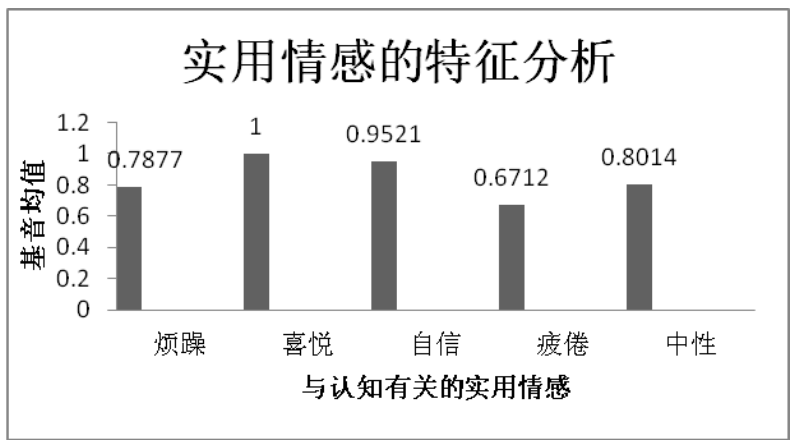

图 3.10 基音均值归一化分布

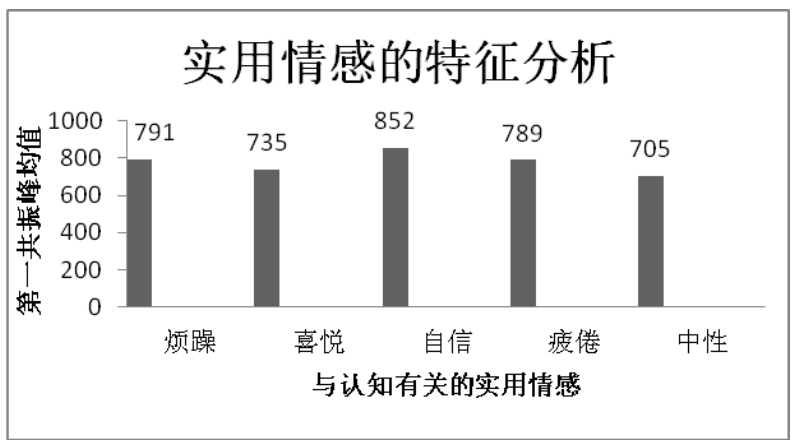

图 3.11 第一共振峰频率均值分布

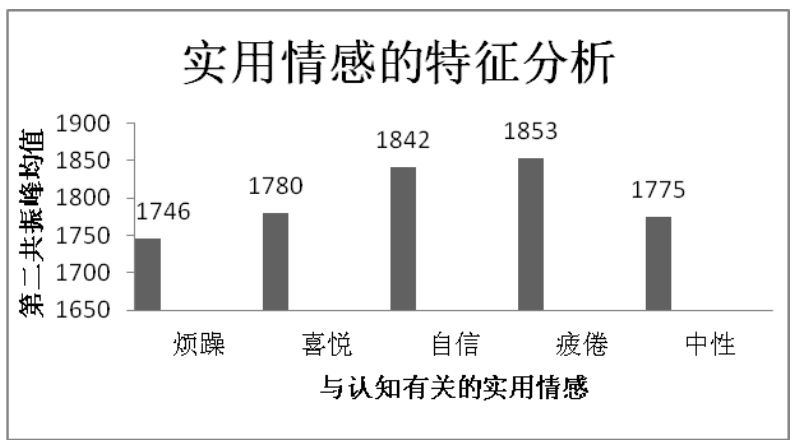

图 3.12 第二共振峰频率均值分布

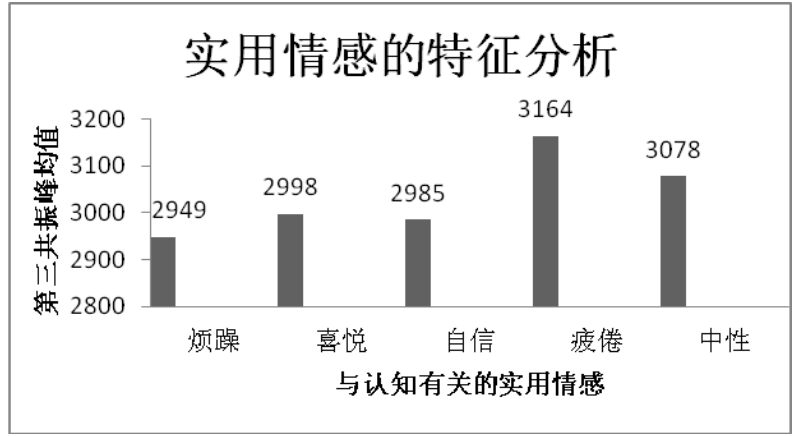

**图 3.13 第三共振峰频率均值分布**

### 3.2.3 实用语音情感的谐波噪声比特征

上一节以基音等典型的语音情感特征为例，初步显示了实用语音情感的样本分布特点。下面我们介绍一种新的情感语音特征——谐波噪声比（Harmonic-to-Noise Ratio, HNR），研究它对区分烦躁、喜悦、自信和疲倦等情感状态的作用。

谐波噪声比常用于诊断喉部疾病，常用来比较客观、定量的评价嗓音的嘶哑程度。嗓音的嘶哑与谐波噪声比是负相关的关系，73%的粗糙声可以通过谐波噪声比感知[65]。近年来Biemans等[66]将谐波噪声比作为音质特征用于评价语音的音质，可以反映出效价维度的情感信息。

在提取谐波噪声比特征时，我们假设一个稳态的元音包含两种成分：周期性的谐波分量和附加的噪声成分，假设噪声据有零均值分布。因此我们可以通过在足够多的周期内叠加的方法消除噪音。语音信号 $f(u)$ 的谐波成分 $f_a(u)$ 的能量可通过下式定义[67]：

$$H = n\sum_{u=0}^{T} f_a^2(u) \tag{3-1}$$

其中，T 为谐波周期，$n$ 为周期数，噪声分量的能量为，

$$N = \sum_{i=1}^{n}\sum_{u=0}^{T}\left[f_i(u) - f_a(u)\right]^2 \tag{3-2}$$

其中 $f_i(u)$ 为一个周期内的波形，HNR 定义为：

$$HNR = n\sum_{u=0}^{T} f_a^2(u) \bigg/ \sum_{i=1}^{n}\sum_{u=0}^{T}\Big[f_i(u)-f_a(u)\Big]^2 \qquad (3\text{-}3)$$

对五种情感，烦躁、喜悦、自信、疲倦和中性状态下的语音信号进行 HNR 提取，如图 3.14 至图 3.18 所示。图中显示了实用语音情感状态下的 HNR 特征实例，其文本内容都固定为"好像快要下雨了"，我们可以看到，HNR 的轮廓、取值范围都随着情感状态的变化而发生较为明显的变化。因而，将 HNR 用于区分本中的几种实用语音情感类别是可行的。比较图中所示的样本可以发现，疲倦状态下的 HNR 轮廓呈现出有规律的段落，在每个基音段落内 HNR 取值较为稳定，波形平坦。而在自信状态下则相反，HNR 值抖动较为严重，取值也相对较低，可见嗓音嘶哑程度上升。这可能是由于疲倦状态下该说话人的喉部肌肉群处于松弛状态，发音较为放松；而自信情感出现时，喉部紧张用力，造成声带发音中出现一定的嘶哑。

从全局的统计特征看，HNR 的均值和方差都可以区分烦躁、自信、疲倦等实用语音情感，如图 3.19 至图 3.20 所示。在均值特征上，疲倦和喜悦的取值较自信和烦躁的高；在方差特征上，疲倦和自信的取值落差较大。

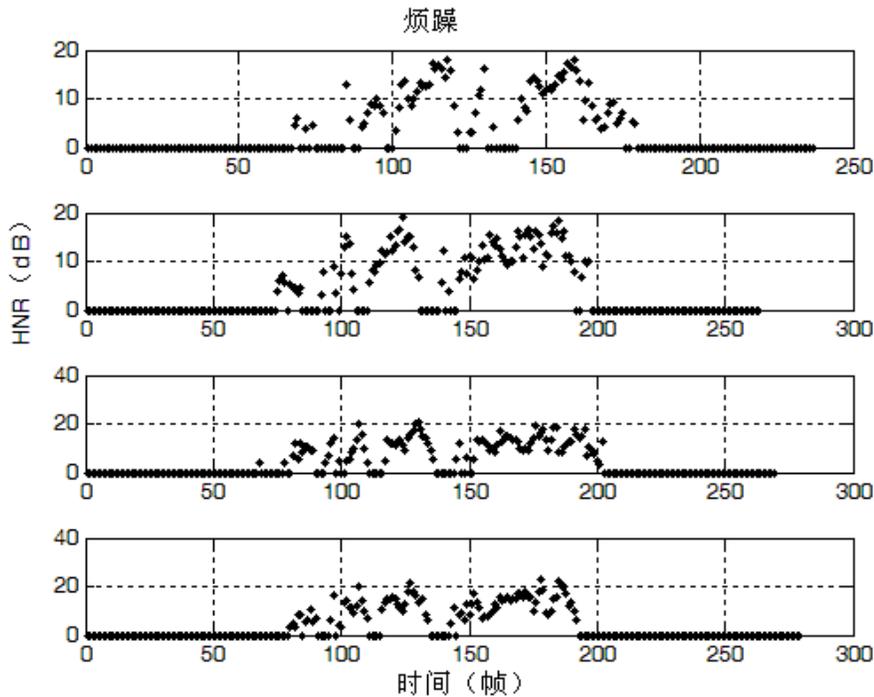

**图 3.14 烦躁样本的谐波噪声比特征实例**

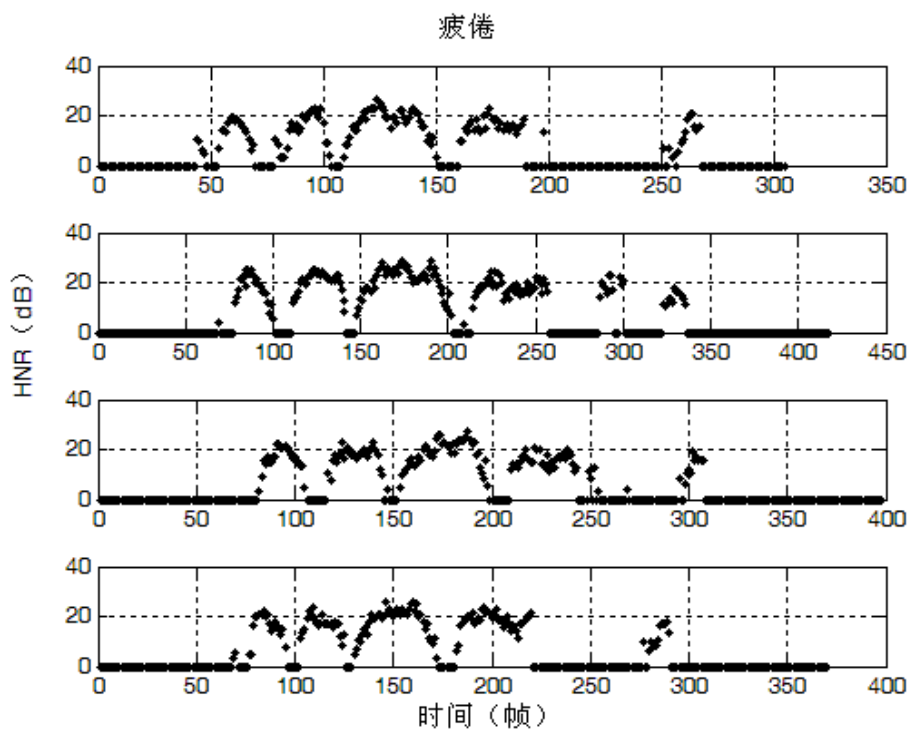

图 3.15 疲倦样本的谐波噪声比特征实例

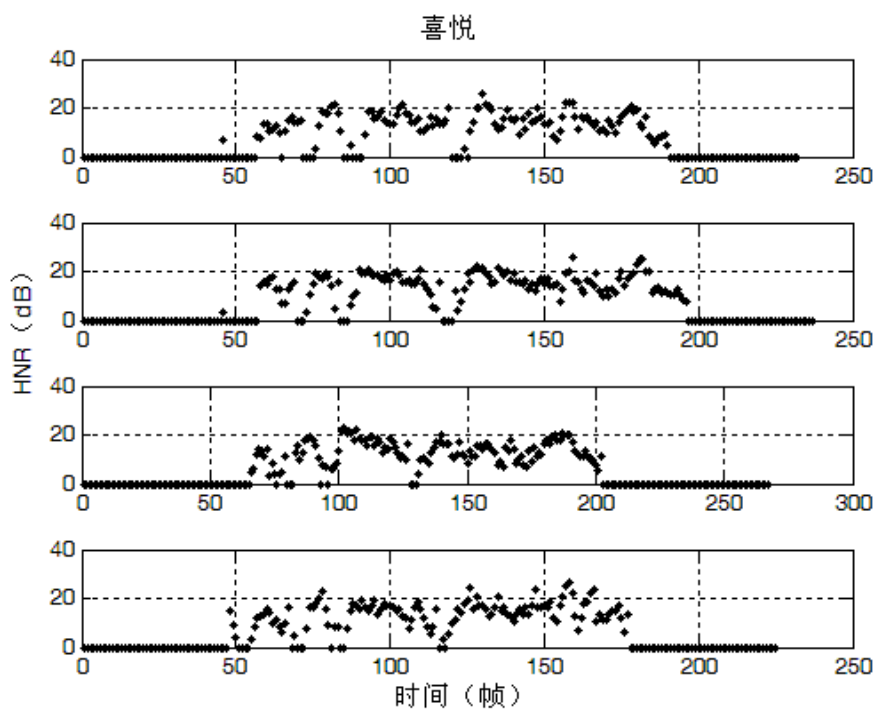

图 3.16 喜悦样本的谐波噪声比特征实例

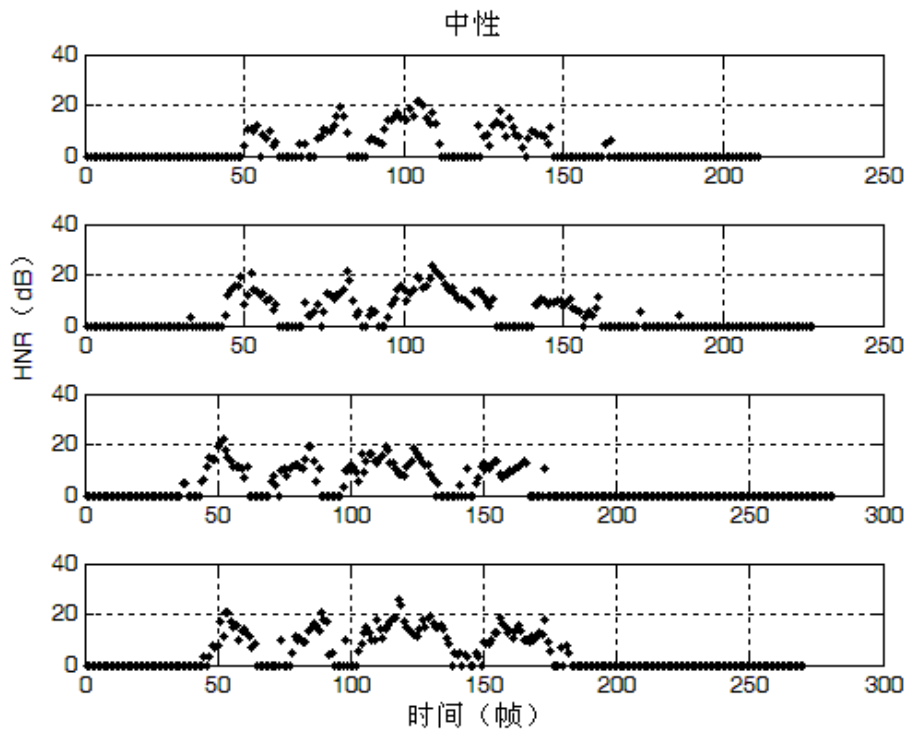

图 3.17 中性样本的谐波噪声比特征实例

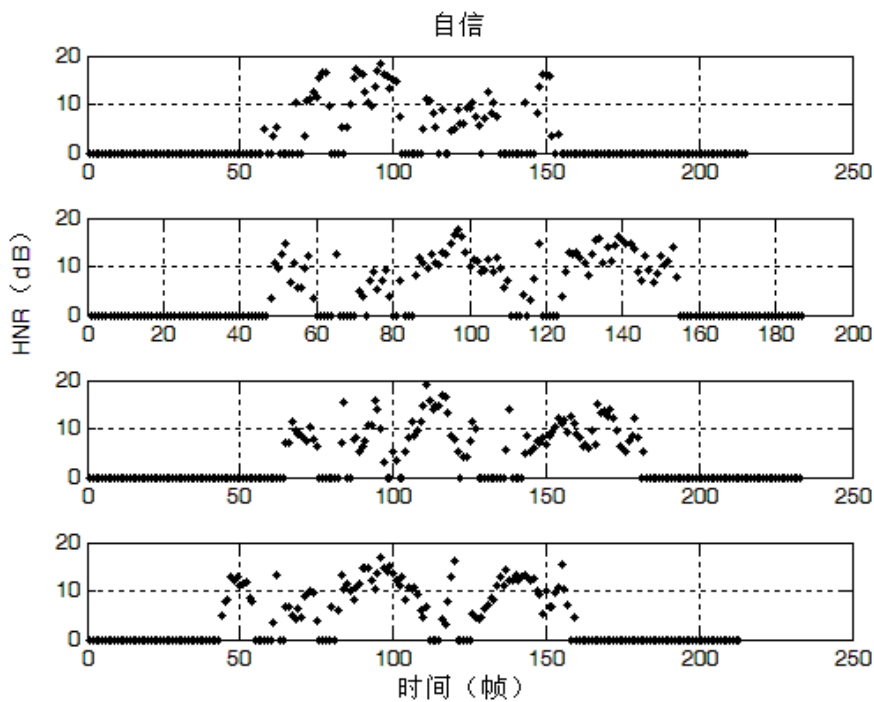

图 3.18 自信样本的谐波噪声比特征实例

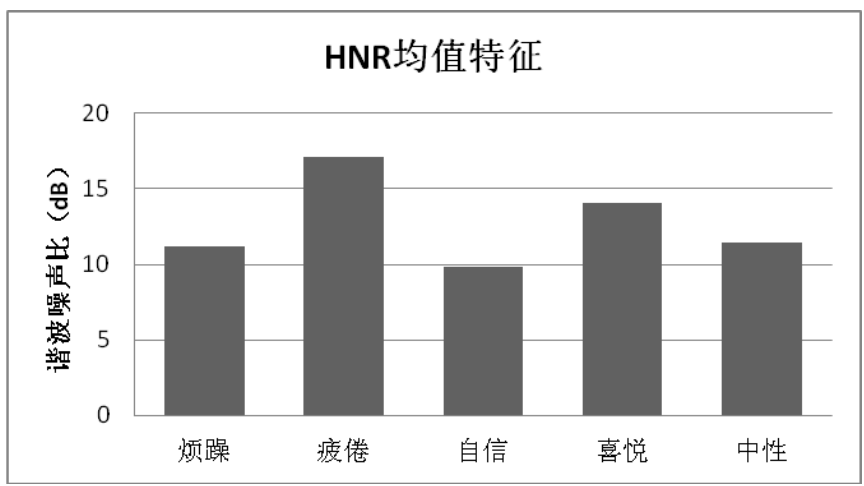

**图 3.19 实用语音情感的谐波噪声比均值分布**

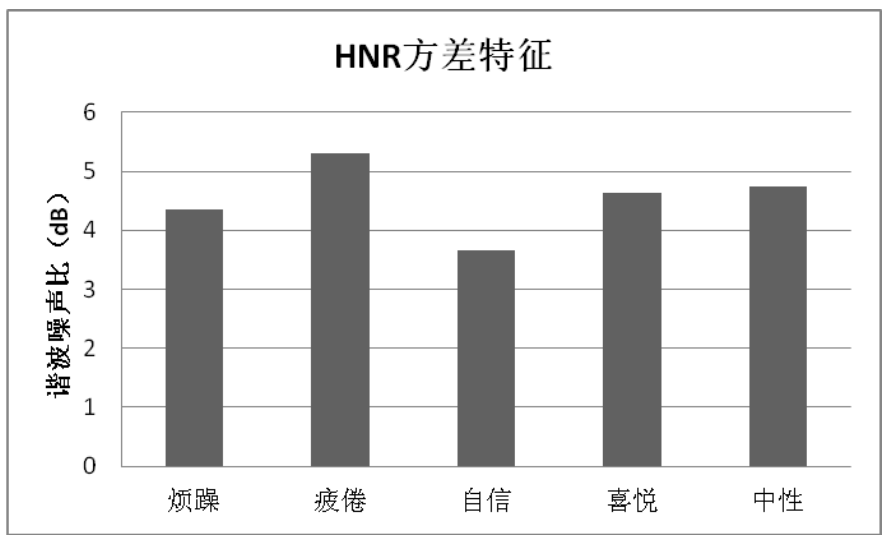

**图 3.20 实用语音情感的谐波噪声比方差分布**

本节中我们对短时能量、基音频率、共振峰频率、谐波噪声比等特征进行了实例的分析，直观的显示了不同情感状态造成的声学特征的差异。然而，仅凭少数几个特征是无法可靠的进行情感识别的。目前对语音情感的特征分析，往往采用数百甚至数千个特征，再结合特征降维方法可以获得特征空间中的有效表达。下面我们就在基本的语音参数的基础上构造大量的全局的统计特征。

## 3.3 实用语音情感的特征构造

我们参考了国内外的语音情感识别文献[9, 12, 13, 18, 33, 42, 56, 68, 69]，总结了研究者们使用的情感特征，挑选出了一些重要的特征用在实用语音情感识别中。并且我们对这些情感特征做了一些改进和补充，形成了本文中的基本声学特征组，如表3.1到表3.2所示。

用于识别和建模的特征向量一般有两种构造方法，静态统计特征和短时动态特征。动态特征对音位信息的依赖性较强，为了建立与文本无关的情感识别系统，本文中选用了静态统计特征[10, 17, 21]，如表3.1到表3.2所示。

文本的变化会对情感特征有较大的影响。情感语音当中大致包含三种信息来源，说话人信息、语义信息和情感信息。在构造情感特征和选择特征的时候，不仅需要使得特征尽可能多的反映出情感信息，也就是随着情感的变化而发生明显的变化，而且还需要尽量保持特征不受到语义变化的影响。

表中的特征包含了韵律特征与音质特征，其中一阶抖动的计算公式如下[42]：

$$\text{Jitter}^1 = \frac{\frac{1}{N-1}\sum_{i=1}^{N-1}\left|x_i - x_{i+1}\right|}{\frac{1}{N}\sum_{i=1}^{N}x_i} \times 100 \qquad (3\text{-}4)$$

二阶抖动的计算公式如下[42]：

$$\text{Jitter}^2 = \frac{\frac{1}{N-2}\sum_{i=2}^{N-1}\left|2x_i - x_{i+1} - x_{i-1}\right|}{\frac{1}{N}\sum_{i=1}^{N}x_i} \times 100 \qquad (3\text{-}5)$$

在构造频谱能量的分频段特征时，未采用 650Hz-4kHz 内的能量百分比，虽然这个频段涉及第一共振峰和几乎全部的第二共振峰，但是此频段的能量受到文本内容变化的影响较大，主要随着音位信息的变化而变化[17]。

频谱能量特征中还采用了 4kHz 以上的能量百分比，根据 Pittam 等人的研究结果显示，这一部分频段能量的增加能反映激励程度的提高，可用于区分悲伤与生气等[1, 57, 58]。

谐波噪声比特征中也同样增加了分频段的特征构造（特征 78 到特征 95）。由于谐波噪声比

会受到噪声干扰的影响，特别是在高频段中噪声的影响更加明显，因此考虑分频段构造谐波噪声比特征，以便更加细致的描述情感变化带来的信号变化。在频段的划分中，我们参考 De Krom 的方法[70]，划分为四个频段：400Hz 以下频段（包含了较低频率的谐波分量），400Hz-2000Hz 频段（大致包含了前两个共振峰的能量范围），2000Hz-5000Hz 频段（较高频率的谐波分量）。5kHz 以上频段的信号中噪声影响较严重，而且对于一些采样率较低的语料也不适用，因此没有采用。

<p style="text-align:center"><b>表 3.1 情感语音的基本声学特征构造（上）</b></p>

| 特征编号 | 特征名称 |
|---|---|
| 1-6 | 短时能量的均值、最大值、最小值、中值、范围和方差 |
| 7-12 | 短时能量一阶差分的均值、最大值、最小值、中值、范围和方差 |
| 13-18 | 短时能量二阶差分的均值、最大值、最小值、中值、范围和方差 |
| 19-24 | 基音频率的均值、最大值、最小值、中值、范围和方差 |
| 25-30 | 基音频率一阶差分的均值、最大值、最小值、中值、范围和方差 |
| 31-36 | 基音频率二阶差分的均值、最大值、最小值、中值、范围和方差 |
| 37-42 | 过零率的均值、最大值、最小值、中值、范围和方差 |
| 43-48 | 过零率一阶差分的均值、最大值、最小值、中值、范围和方差 |
| 49-54 | 过零率二阶差分的均值、最大值、最小值、中值、范围和方差 |
| 55 | 语速 |
| 56-57 | 基音频率一阶抖动、基音频率二阶抖动 |
| 58-61 | 0-250Hz 频段能量占总能量的百分比、0-650Hz 频段能量占总能量的百分比、4kHz 以上能量占总能量的百分比、短时能量抖动 |
| 62-65 | 发音帧数、不发音帧数、不发音帧数和发音帧数比、发音帧数和总帧数比 |
| 66-69 | 发音区域数、不发音区域数、发音区域数和不发音区域数之比、发音区域数和总区域数之比 |
| 70-71 | 最长发音时间、最长不发音时间 |

## 表 3.2 情感语音的基本声学特征构造（下）

| 特征编号 | 特征名称 |
|---|---|
| 72–77 | 谐波噪声比（HNR）的均值、最大值、最小值、中值、范围和方差 |
| 78–83 | 0-400Hz 频段内谐波噪声比的均值、最大值、最小值、中值、范围和方差 |
| 84–89 | 400-2000Hz 频段内谐波噪声比的均值、最大值、最小值、中值、范围和方差 |
| 90–95 | 2000-5000Hz 频段内谐波噪声比的均值、最大值、最小值、中值、范围和方差 |
| 96–101 | 第一共振峰频率（F1）的均值、最大值、最小值、中值、范围和方差 |
| 102–107 | 第二共振峰频率（F2）的均值、最大值、最小值、中值、范围和方差 |
| 108–113 | 第三共振峰频率（F3）的均值、最大值、最小值、中值、范围和方差 |
| 114–119 | 第四共振峰频率（F4）的均值、最大值、最小值、中值、范围和方差 |
| 120–125 | 第一共振峰频率一阶差分的均值、最大值、最小值、中值、范围和方差 |
| 126–131 | 第二共振峰频率一阶差分的均值、最大值、最小值、中值、范围和方差 |
| 132–137 | 第三共振峰频率一阶差分的均值、最大值、最小值、中值、范围和方差 |
| 138–143 | 第四共振峰频率一阶差分的均值、最大值、最小值、中值、范围和方差 |
| 144–149 | 第一共振峰频率二阶差分的均值、最大值、最小值、中值、范围和方差 |
| 150–155 | 第二共振峰频率二阶差分的均值、最大值、最小值、中值、范围和方差 |
| 156–161 | 第三共振峰频率二阶差分的均值、最大值、最小值、中值、范围和方差 |
| 162–167 | 第四共振峰频率二阶差分的均值、最大值、最小值、中值、范围和方差 |
| 168–171 | 第一到第四共振峰频率的一阶抖动 |
| 172–175 | 第一到第四共振峰频率的二阶抖动 |
| 176–181 | 第一共振峰带宽的均值、最大值、最小值、中值、范围和方差 |
| 182–187 | 第二共振峰带宽的均值、最大值、最小值、中值、范围和方差 |
| 188–193 | 第三共振峰带宽的均值、最大值、最小值、中值、范围和方差 |
| 194–199 | 第四共振峰带宽的均值、最大值、最小值、中值、范围和方差 |
| 200–205 | 第一共振峰带宽一阶差分的均值、最大值、最小值、中值、范围和方差 |
| 206–211 | 第二共振峰带宽一阶差分的均值、最大值、最小值、中值、范围和方差 |
| 212–217 | 第三共振峰带宽一阶差分的均值、最大值、最小值、中值、范围和方差 |
| 218–223 | 第四共振峰带宽一阶差分的均值、最大值、最小值、中值、范围和方差 |
| 224–229 | 第一共振峰带宽二阶差分的均值、最大值、最小值、中值、范围和方差 |

| 230–235 | 第二共振峰带宽二阶差分的均值、最大值、最小值、中值、范围和方差 |
| 236–241 | 第三共振峰带宽二阶差分的均值、最大值、最小值、中值、范围和方差 |
| 242–247 | 第四共振峰带宽二阶差分的均值、最大值、最小值、中值、范围和方差 |
| 248–325 | 0–12阶镁尔倒谱参数（MFCC0–MFCC12）的均值、最大值、最小值、中值、范围和方差 |
| 326–403 | 0–12阶镁尔倒谱参数一阶差分的均值、最大值、最小值、中值、范围和方差 |
| 404–481 | 0–12阶镁尔倒谱参数二阶差分的均值、最大值、最小值、中值、范围和方差 |

## 3.4 特征降维算法的性能比较

### 3.4.1 特征降维概述

在上文中我们提取了大量的基本声学特征，由于受到训练样本规模的限制，特征空间维度不能过高，在本节中，我们需要进行特征降维。特征降维，在一个模式识别系统中具有重要的作用。我们知道原始的基本特征，或多或少的能够提供可利用的信息，来增加类别之间的可区分度。从信息的增加的角度来说，原始特征的数量应该是越多越好，似乎不存在一个上限。然而，在具体的算法训练当中，几乎所有的算法都会受到计算能力的限制，特征数量的增加，最终会导致"维度灾难"的问题。以高斯混合模型为例，它的概率模型的成功训练依赖于训练样本数量、高斯模型混合度、特征空间维数三者之间的平衡。如果训练样本不足，而特征空间维数过高的话，高斯混合模型的参数就不能准确的获得。

我们在这里对上文中列出的所有基本声学特征，进行特征降维的工作，即能够反映出这些特征在区分情感类别上的能力，又是后续的识别算法研究的需要。总结语音情感识别领域近年来的一些文献，研究者们主要采用了以下的一些特征降维的方法：LDA（Linear Discriminant Analysis）、PCA（Principal Components Analysis）、FDR（Fisher Discriminant Ratio）、SFS（Sequential Forward Selection）等。其中，SFS 是一种封装器方法（Wrapper），它对具体的识别算法依赖程度比较高，当使用不同的识别算法时，可能会得到差异很大的结果。我们希望通过特征分析这部分的工作，能够给出一个相对独立的特征分析方法，而不依赖于后续的识别算法。而且，本文主要从实用的角度研究一个语音情感识别系统，SFS 方法的计算量较高，在实际当中的应用相对较困难，也较少。在后文中，我们还将从几个不同的角度，对一个实用语音情感识别系统进行一系列的实验测试，各部分结论的独立性对系统的分析问题很重要。综合以上原因，我们不采用 SFS 等封装器方法进行特征降维，尽管这类方

法可能在个别数据集上获得较高的识别率，但是对本文中有系统的研究实用语音情感识别问题的贡献和意义不大。

下面我们就对这两种特征降维的方法进行简要介绍和实验验证，我们将在实用语音情感数据集上进行测试，进而获得一个合适的特征降维方法，用于对烦躁、自信和疲劳这几种与认知过程有关的情感的研究。

### 3.4.2 基于 FDR 的特征选择

特征选择可以通过对每个特征维度上情感类别的可区分性来进行评价和优选。以两类的情况为例，在第 i 个维度上，样本的类别中心 $\mu_1$ 和 $\mu_2$ 之间的距离反映出类别之间的离散程度，每个类别的方差 $\sigma_1$ 和 $\sigma_2$ 反映出在第 i 维上的类内聚合程度。FDR 可用于情感的特征选择，基于 FDR 的特征分析方法在 HIV（Human Immunodeficiency Virus）病毒数据特征研究[71]和 Clavel 进行的恐惧情感特征研究[33]中获得了较好的效果。FDR 的定义为：

$$FDR = \sum_{i}^{M} \sum_{j \neq i}^{M} \frac{(\mu_i - \mu_j)^2}{\sigma_i^2 + \sigma_j^2} \tag{3-6}$$

其中，M为类别数，i、j为类别编号，$\mu$ 为类别中心即类别的特征向量均值，$\sigma^2$ 为相应的样本方差。

再进行特征选择之前，需要对原始的各个维度的特征进行归一化处理。因为原始的特征数值往往在不同的尺度范围内，如果不进行归一化处理，数值较大的特征会对类别可分性准则有较大的影响。各维度特征的数值归一化可通过下式完成：

$$\tilde{f}_i = \frac{f_i - \alpha_i}{\beta_i} \tag{3-7}$$

其中 $\alpha_i = \min(f_i)$，$\beta_i = \max(f_i)$。需要注意的是，$\alpha_i$、$\beta_i$ 是一组常数，对不同维度的特征是不同的取值，但对于不同样本的同一个特征维度来说是相同的取值。其中的最小值、最大值运算，只是在训练数据集上进行，在测试数据上不再进行最大值、最小值的运算。用于归一化的常数 $\alpha_i$、$\beta_i$，应该对训练数据和测试数据保持一致，其具体数值没有特殊的含义，这里不再对481个特征维度一一赘述。

由计算得到的FDR数值进行排序，选出排名最高的前25个特征，如表3.3所示。

表 3.3　有效的语音情感特征

| 序号 | 特征名称 |
|------|----------|
| 1 | 基音频率的方差 |
| 2 | 短时能量一阶差分的方差 |
| 3 | 短时能量的最小值 |
| 4 | 短时能量二阶差分的均值 |
| 5 | 0-650Hz 频段能量占总能量的百分比 |
| 6 | 基音方差 |
| 7 | 0-400Hz 谐波噪声比均值 |
| 8 | 短时能量一阶差分的均值 |
| 9 | 基音频率的均值 |
| 10 | 谐波噪声比（HNR）的方差 |
| 11 | 短时能量二阶差分的最小值 |
| 12 | 短时能量二阶差分的最大值 |
| 13 | 短时能量一阶差分的最小值 |
| 14 | MFCC3 的二阶差分的最大值 |
| 15 | 短时能量的方差 |
| 16 | F3 带宽的均值 |
| 17 | 第一共振峰的最小值 |
| 18 | MFCC3 的二阶差分的范围 |
| 19 | 语速 |
| 20 | 短时能量一阶差分的最大值 |
| 21 | 基音一阶差分的最小值 |
| 22 | F3 带宽一阶差分的均值 |
| 23 | F2 带宽的均值 |
| 24 | MFCC0 的方差 |
| 25 | 0-250Hz 频段能量占总能量的百分比 |

### 3.4.3 基于 PCA 与 LDA 的特征压缩

线性判别分析（LDA）是一种常用的特征降维优化方法，通过一种特定的线性变换，将高维特征空间映射到低纬子空间上，使得投影后的类别内模式尽量聚合，类别间模式尽量的分开。同 FDR 一样我们需要定义类内和类间离散度准则。

类内离散度矩阵定义为：

$$S_{W,i} = \frac{1}{N_i} \sum_{j}^{N_i} \left( \boldsymbol{x}_j^{(i)} - \boldsymbol{m}_i \right) \left( \boldsymbol{x}_j^{(i)} - \boldsymbol{m}_i \right)^T \tag{3-8}$$

其中 i 为类别编号，j 为样本编号，$\boldsymbol{m}_i$ 为第 i 类样本的均值。c 个类别的总体类内离散度矩阵为：

$$S_W = \sum_{i=1}^{c} P_i S_{W,i} \tag{3-9}$$

其中，$P_i$ 为第 i 类的先验概率。类间离散度矩阵定义为：

$$S_B = \sum_{i=1}^{c} P_i \left( \boldsymbol{m}_i - \boldsymbol{m} \right) \left( \boldsymbol{m}_i - \boldsymbol{m} \right)^T \tag{3-10}$$

其中 $\boldsymbol{m}$ 为全体样本的均值向量，设 LDA 中的变换矩阵为 $\boldsymbol{U}$，则可以构造可分性准则函数，以使得类内离散度最小，类间离散度最大：

$$J(\boldsymbol{U}) = \frac{\mathrm{tr}\left( S_B^* \right)}{\mathrm{tr}\left( S_W^* \right)} = \frac{\sum_{i=1}^{d} \boldsymbol{u}_i^T S_B \boldsymbol{u}_i}{\sum_{i=1}^{d} \boldsymbol{u}_i^T S_W \boldsymbol{u}_i} \tag{3-11}$$

其中 $S_B^*$ 为 LDA 变换后的类间离散度矩阵，$S_W^*$ 为 LDA 变换后的类内离散度矩阵。上式即为我们要优化的代价函数，构造 Largrange 函数得到：

$$L(\boldsymbol{U}, \lambda) = \sum_{i=1}^{d} \boldsymbol{u}_i^T S_B \boldsymbol{u}_i - \sum_{i=1}^{d} \lambda_i \left( \boldsymbol{u}_i^T S_W \boldsymbol{u}_i - b_i \right) \tag{3-12}$$

$$\frac{\partial \mathrm{L}(\boldsymbol{U}, \lambda)}{\partial \boldsymbol{u}_i} = \boldsymbol{S}_B \boldsymbol{u}_i - \lambda_i \boldsymbol{S}_W \boldsymbol{u}_i = 0 \qquad (3\text{-}13)$$

所以有：

$$\boldsymbol{S}_W^{-1} \boldsymbol{S}_B \boldsymbol{u}_i = \lambda \boldsymbol{u}_i \qquad (3\text{-}14)$$

由此得到的变换矩阵可将原始特征空间降维至 $c-1$ 维的低维空间中，在特征维数较高时，LDA 的压缩性能是非常明显的。然而在实际中 LDA 的应用会受到训练数据量的限制，当原始特征维数非常高，而训练数据量不足时，会导致矩阵出现奇异值，LDA 无法正常使用。因此，在处理高维数据时，可以采用 PCA 进行第一步降维，然后再使用 LDA 降维[72]。

### 3.4.4 特征降维实验

我们通过基于高斯混合模型（Gaussian Mixture Model，GMM）的实用语音情感分类器来进行特征降维方法的评价。高斯混合模型分类器在第四章中有详细的介绍和探讨， 我们将两类特征降维算法同 GMM 分类器结合，在第二章表 2.4 中的实验数据集 B 上进行情感识别测试，数据集中包含了九种情感类型，每种情感 2000 条样本。

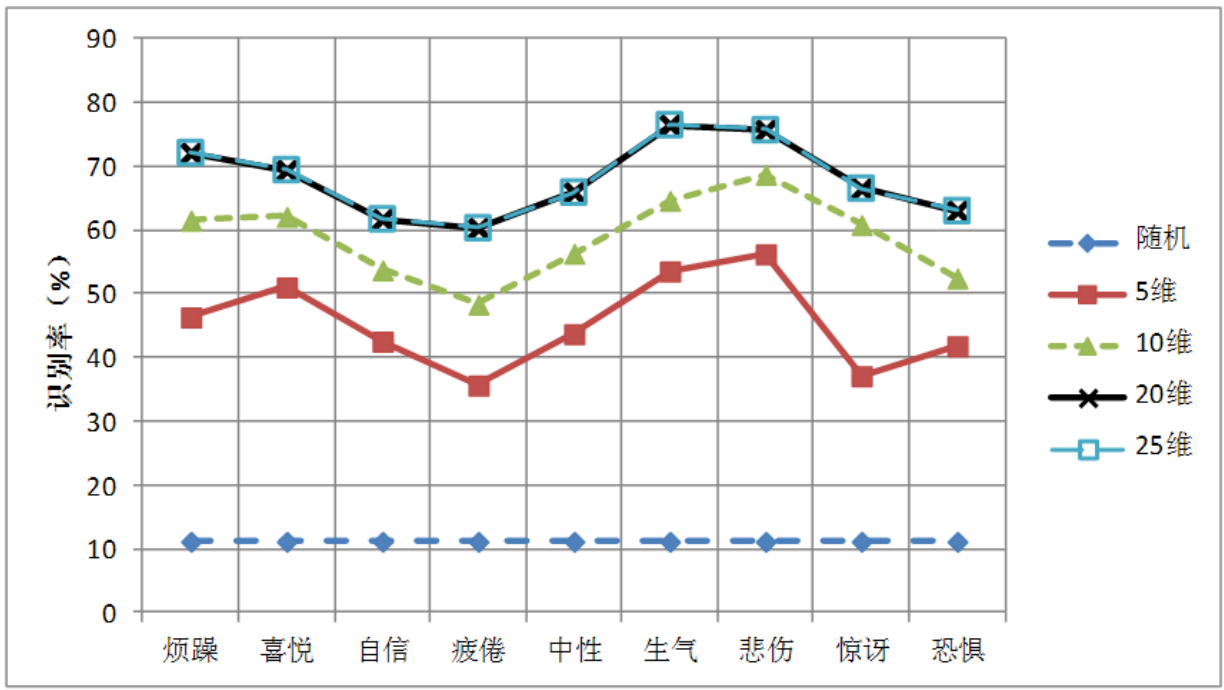

图 3.21 特征选择后的识别率结果

　　实验中的训练与测试比例为 9：1，采用交叉验证的测试方法，实验结果如图 3.21 和图 3.22 所示。图 3.21 中改变 FDR 选择出的特征维数，当维数达到 20 维和 25 维时，两条识别率曲线不再变化，发生重合。图 3.22 中改变 PCA+LDA 特征压缩方法中的 PCA 截取维数，当维数增加到 20 维和 30 维时，识别率曲线不再变化，两种条件下的识别率曲线发生重合。我们比较图中识别率的高低可以看到，PCA+LDA 的特征压缩方法的性能要略优于 FDR 方法。

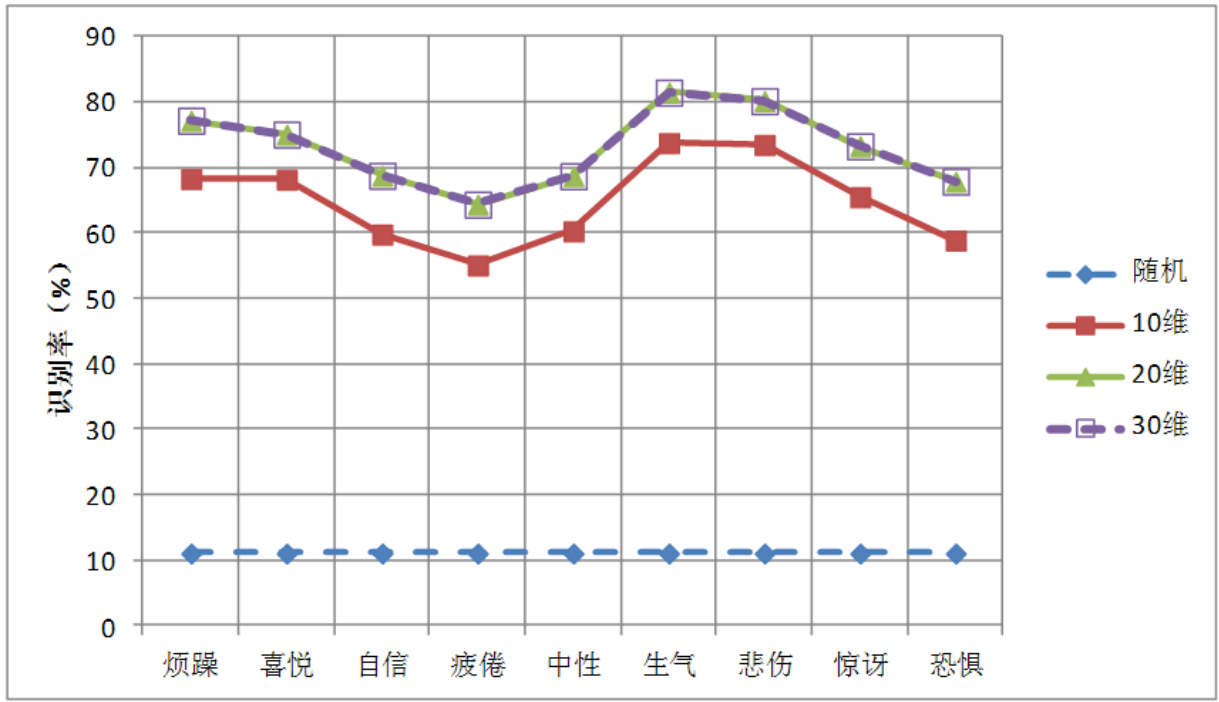

**图 3.22 特征优化后的识别率结果**

# 第四章 实用语音情感的识别算法研究

在本章里，介绍了基于高斯混合模型（Gaussian Mixture Model，GMM）的语音情感识别系统。分别考虑了在三种情况下，高斯混合模型如何应用于实用语音情感识别。第一，在充足的训练样本下，高斯混合模型应用于烦躁、自信、疲劳等实用语音情感的识别。第二，针对训练样本不足的情况，改进基于高斯混合模型的分类系统。第三，在实际应用环境中，针对连续语音信号中情感的连贯性特性，结合马尔科夫网络对上下文情感信息进行建模，以提升单一的高斯混合模型的识别效果。

**表 4.1 各种识别算法在语音情感识别应用中的特性比较**

| 识别算法 | 对语音情感数据的拟合性能 | 识别率 | 优点 | 缺点 |
|---|---|---|---|---|
| GMM[18, 73–75] | 高 | 在 AIBO 数据库、本文数据库上表现较高 | 对数据的拟合能力较高 | 对训练数据依赖性强 |
| SVM[76, 77] | 较高 | 在柏林库上表现较高 | 适合于小样本训练集 | 多类分类问题中存在不足 |
| KNN[7] | 较高 | 在柏林库上表现一般 | 易于实现，较符合语音情感数据的分布特性 | 计算量较大 |
| HMM[7, 10, 78] | 一般 | 在柏林库上表现较高 | 适合于时序序列的识别 | 受到音位信息的影响较大 |
| 决策树[79] | 一般 | 在 AIBO 数据库上表现一般 | 易于实现，适合于离散情感类别的识别 | 识别率有待提高 |
| ANN[15, 80, 81] | 较高 | 在日语情感语音上表现一般 | 逼近复杂的非线性关系 | 容易陷入局部极小特性和算法收敛速度较低的 |
| 混合蛙跳算法[80, 81] | 较高 | 在汉语语音情感数据上表现较高 | 优化能力强，有利于发现情感数据中潜在的模式 | 在迭代后期容易陷入局部最优，收敛速度较慢 |

我们简要总结了各种现有的语音情感识别算法，如表4.1所示。模式识别领域中的诸多算法都曾用于语音情感识别的研究，典型的有隐马尔科夫模型（Hidden Markov models，HMM）、高斯混合模型和支持向量机（Support Vector Machine，SVM）、人工神经网络（Artificial

Neural Network，ANN）等，表4.1中初步比较了它们各自的优缺点，以及在部分数据库上的识别性能表现。

## 4.1 高斯混合模型的基本原理

高斯混合模型，是一种拟合能力很强的统计建模工具。GMM 的主要优势在于对数据的建模能力强，理论上来说，它可以拟合任何一种概率分布函数。而 GMM 的主要缺点，也正是对数据的依赖性过高。因此在采用 GMM 建立的语音情感识别系统中，训练数据的特性会对系统性能产生很大的影响。在本章中，我们将会根据训练数据的充足与否，对 GMM 算法进行研究和改进。

高斯混合模型在说话人识别和语种识别中获得了成功的应用。就目前来说，很多研究的结果显示，GMM 在语音情感识别中是一种较合适的建模算法[18, 51, 73, 74, 82]。近年来的研究文献中，报道了不少采用 GMM 建立的语音情感识别系统。这些基于 GMM 的识别系统，相对于其它识别算法来说，获得了较好的识别率。在 2009 年，语音领域的著名的国际会议（INTERSPEECH）上，举行语音情感识别的评比（EMOTION CHALLENGE）。基于 GMM 的识别系统在总体性能上获得了该次比赛的第一。

采用何种建模算法最适合语音情感识别，一直是研究者们非常关注的问题。我们认为，在不同的情感数据库上、不同的测试环境中，不同的识别算法各有优劣，对此不能一概而论。然而，目前研究者们对自然语料非常重视，在自然语料中的情感模式较为复杂，不同的说话人、不同的性格特点、不同的上下文环境等等因素都会增加数据的复杂度。高斯混合模型对这些数据的适应能力较强，可能是多数应用场合的一种合理选择。在本章中，我们在同一个测试集上比较了不同算法的识别效果，包括高斯混合模型、K 近邻算法和支持向量机。

下面我们首先介绍高斯混合模型的基本定义。GMM 可以通过（4-1）式定义：

$$p(\boldsymbol{X}_t|\lambda) = \sum_{i=1}^{M} a_i b_i(\boldsymbol{X}_t) \tag{4-1}$$

这里 $\boldsymbol{X}$ 是语音样本的 D 维特征向量，t 为其样本序号；$b_i(\boldsymbol{X})$，$i=1,2,...,M$ 是成员密度；$a_i$，$i=1,2,...,M$ 是混合权值。每个成员密度是一 D 维变量的关于均值矢量 $\boldsymbol{U}_i$ 和协方差矩阵 $\boldsymbol{\Sigma}_i$ 的高斯函数，形式如下：

$$b_i(\boldsymbol{X}_t) = \frac{1}{(2\pi)^{D/2} |\boldsymbol{\Sigma}_i|^{1/2}} \exp\left\{-\frac{1}{2}(\boldsymbol{X}_t - \boldsymbol{U}_i)^{'}\boldsymbol{\Sigma}_i^{-1}(\boldsymbol{X}_t - \boldsymbol{U}_i)\right\} \tag{4-2}$$

其中混合权值满足条件：

$$\sum_{i=1}^{M} a_i = 1 \tag{4-3}$$

完整的高斯混和密度由所有成员密度的均值矢量、协方差矩阵和混合权值参数化。这些参数聚集一起表示为：

$$\boldsymbol{\lambda}_i = \{\boldsymbol{a}_i, \boldsymbol{U}_i, \boldsymbol{\Sigma}_i\}, \quad i = 1,2,...,M \tag{4-4}$$

根据贝叶斯判决准则，基于 GMM 的情感识别可以通过最大后验概率来获得，

$$EmotionLabel = \arg\max_{k}(p(\boldsymbol{X}_t|\boldsymbol{\lambda}_k)) \tag{4-5}$$

其中 k 为情感类别序号。

对于高斯混合模型的参数估计，可以采用 EM（Expectation-maximization）算法进行。EM 是最大期望算法，它的基本思想是从一个初始化的模型 $\overline{\lambda}$ 开始，去估计一个新的模型 $\overline{\lambda}$，使得 $p(X|\overline{\lambda}) \geq p(X|\lambda)$。这时新的模型对于下一次重复运算来说成为初始模型，该过程反复执行直到达到收敛门限。这类似于用来估计隐马尔科夫模型（HMM）参数的 Baum-Welch 重估算法。每一步的 EM 重复中，下列重估公式保证模型的似然值单调增加[83]：

混合参数的重估：

$$a_m^i = \frac{\sum_{t=1}^{T} \gamma_{tlm}^i}{\sum_{t=1}^{T}\sum_{m=1}^{M} \gamma_{tlm}^i} \tag{4-6}$$

均值矢量的重估：

$$\boldsymbol{\mu}_m^i = \frac{\sum_{t=1}^{T} \gamma_{tm}^i \boldsymbol{X}_t}{\sum_{t=1}^{T} \gamma_{tm}^i} \tag{4-7}$$

方差矩阵的重估:

$$\boldsymbol{\Sigma}_m^i = \frac{\sum_{t=1}^{T} \gamma_{tm}^i (\boldsymbol{X}_t - \boldsymbol{\mu}_m^i)(\boldsymbol{X}_t - \boldsymbol{\mu}_m^i)^{'}}{\sum_{t=1}^{T} \gamma_{tm}^i} \tag{4-8}$$

$$\gamma_{tm}^i = \frac{a_m^{i-1} N(\boldsymbol{X}_t \mid \boldsymbol{\mu}_m^{i-1}, \boldsymbol{\Sigma}_m^{i-1})}{\sum_{m=1}^{M} a_m^{i-1} N(\boldsymbol{X}_t \mid \boldsymbol{\mu}_m^{i-1}, \boldsymbol{\Sigma}_m^{i-1})} \tag{4-9}$$

GMM 各个分量的权重、均值和协方差矩阵的估计值,通过每一次迭代趋于收敛。

高斯混合模型中的混合度,在理论上只能推导出一个固定的范围,具体的取值需要在实验中确定,各高斯分量的权重可以通过 EM 算法估计得到,在 EM 算法的迭代中,要避免协方差矩阵变为奇异矩阵,保证算法的收敛性。

以权重为例,EM 算法的迭代曲线如图 4.1 所示,图中显示出了每一次迭代的收敛情况。纵坐标代表各个高斯分量的权重的数值,横坐标代表 EM 算法的迭代优化次数,不同颜色和形状的曲线代表不同的高斯分量。其中部分高斯混合分量为零,说明混合度设置偏高。初始值是由 K 均值聚类初始化得到,在迭代 35 次左右之后,算法收敛。

本节中我们对高斯混合模型应用于语音情感识别,做了一个初步的介绍,介绍了算法原理、分类方法和参数估计。在下面几节中我们通过几组实验,来研究不同条件下的算法特性和优缺点,引出本文中相应的算法改进。

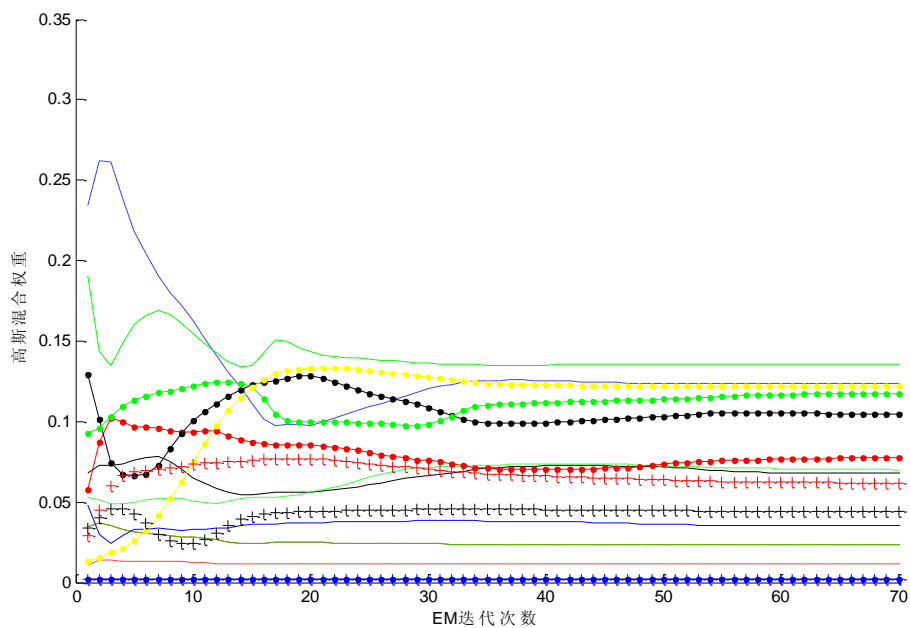

**图 4.1 高斯分量的权重迭代曲线**

## 4.2 充足样本条件下的高斯混合模型识别实验

在本小节中，我们进行了三个实验，使用 GMM 对烦躁、自信和疲倦等实用语音情感进行建模。通过实验，初步显示出 GMM 算法在这几种特殊情感上的识别性能。不同的训练数据条件下，GMM 显示出了性能的波动。通过与两种较好的语音情感识别算法的比较研究，进一步显示出了 GMM 算法性能的优点和缺陷。另一方面，实验中不仅对三种与认知过程有关的情感类型进行区分，而且进一步对包括基本情感在内的九中情感类型进行区分。显示出了实用语音情感与基本情感类型之间的可区分性。

实验中使用的数据集有两个，在第二章中介绍了本文中主要使用的这两个数据集——实验数据集 A 和实验数据集 B。实验数据集 A 包含五种情感：烦躁、喜悦、自信、疲倦和中性。实验数据集 B 中包含了九种情感：烦躁、喜悦、自信、疲倦、中性、生气、悲伤、惊讶和恐惧。详细的说话人数量和样本数量参见第二章的 2.6 小节，这里不再重复叙述。

数据集 A 中主要是与认知有关的实用语音情感，识别测试结果如表 4.2 所示。实验中采用轮换测试方法，将数据集等分成十份，采用 9：1 的训练/测试比例，交叉验证后取得平均值。

数据集 B 中增加了几种基本的语音情感，将情感类别扩展到九类，采用同样的轮换测试方法，识别测试结果如表 4.3 所示。

表 4.2 基于 GMM 的五种实用情感的识别率（%）

|  | 烦躁 | 喜悦 | 自信 | 疲倦 | 中性 |
|---|---|---|---|---|---|
| 烦躁 | 84.3 | 0 | 6.9 | 5.7 | 3.1 |
| 喜悦 | 0.3 | 81.1 | 5.1 | 3.4 | 10.1 |
| 自信 | 7.6 | 9.2 | 72.9 | 4.1 | 6.2 |
| 疲倦 | 8.6 | 2.7 | 5.2 | 68.5 | 15.0 |
| 中性 | 2.4 | 7.7 | 6.1 | 9.5 | 74.3 |

注：表格的左侧方向各行代表测试样本标注的情感类型，第二列起的各列方向代表识别判决的情感类别，本文中的其余识别率表格亦采用这样的格式——混淆矩阵（Confusion Matrix）。

表 4.3 基于 GMM 的九种语音情感的识别率（%）

|  | 烦躁 | 喜悦 | 自信 | 疲倦 | 中性 | 生气 | 悲伤 | 惊讶 | 恐惧 |
|---|---|---|---|---|---|---|---|---|---|
| 烦躁 | 77.2 | 0 | 3.0 | 3.1 | 1.4 | 2.9 | 5.8 | 2.5 | 4.1 |
| 喜悦 | 0.2 | 75.1 | 3.5 | 1.7 | 5.7 | 3.6 | 1.8 | 3.2 | 5.2 |
| 自信 | 3.8 | 4.8 | 68.8 | 2.5 | 2.9 | 8.3 | 4.5 | 3.0 | 1.4 |
| 疲倦 | 5.5 | 2.0 | 4.4 | 64.4 | 10.2 | 2.1 | 4.7 | 1.1 | 5.6 |
| 中性 | 1.5 | 6.6 | 5.1 | 7.9 | 68.7 | 1.2 | 3.5 | 2.9 | 2.6 |
| 生气 | 4.8 | 2.1 | 4.9 | 1.2 | 0.9 | 81.5 | 0 | 2.2 | 2.4 |
| 悲伤 | 2.1 | 1.1 | 2.3 | 5.5 | 3.9 | 0 | 80.2 | 0.4 | 4.5 |
| 惊讶 | 2.1 | 8.3 | 3.0 | 1.1 | 5.9 | 2.8 | 1.4 | 73.3 | 2.1 |
| 恐惧 | 3.3 | 0 | 1.6 | 8.8 | 3.2 | 3.3 | 9.0 | 2.9 | 67.9 |

高斯混合模型采用 K 均值聚类算法进行初始化，EM 算法进行参数估计，EM 算法的迭代次数上限设为 50 次，GMM 的混合度设为 32。特征降维部分采用 3.4 节中的方法，PCA 压缩中截取前 20 个维度的分量用于 LDA 降维。

从识别率中可以看到，烦躁的识别率最高，疲倦的识别率最低。说明我们选用的特征对烦躁情感有较好的体现，疲倦情感的个体差异可能较大，导致识别困难。此外，我们注意到，在第三章的特征分析中，疲倦的基音很有特点，轮廓相对平坦，然而在这里的识别测试中，识别率不是很高。这可能是由于在诱发实验中，被试会隐藏和克服自己的疲倦状态，从而导致诱发语料识别难度增大。被试的体验情感为疲倦时，听辨情感不一定表现出明显的疲倦特征。

图 4.2 中是九种情感识别率的柱状图，我们可以较为直观的看到九种情感类别都获得了较好的识别率。图 4.3 中通过比较，显示了情感类别增加后导致的识别率下降。我们可以看到，在这个实验中情感识别的过程仅仅是对样本进行非此即彼的硬性划分，这是语音情感识别系统中普遍存在的缺陷。在分类实验中，先验的认为样本数据集中有且仅有 N 个情感类别，在此假

设的基础上进行识别实际上是简化了问题。这样建立的系统，在面对少数几个情感的识别时，性能较好，在情感类别增加时，性能自然下降。为了弥补这个缺陷，我们可以考虑可据判的识别方法，使得语音情感识别系统能够处理未知类别的情感，我们在第五章中会专门讨论可据判的识别算法，以增加系统在未知情感数据出现时的鲁棒性。

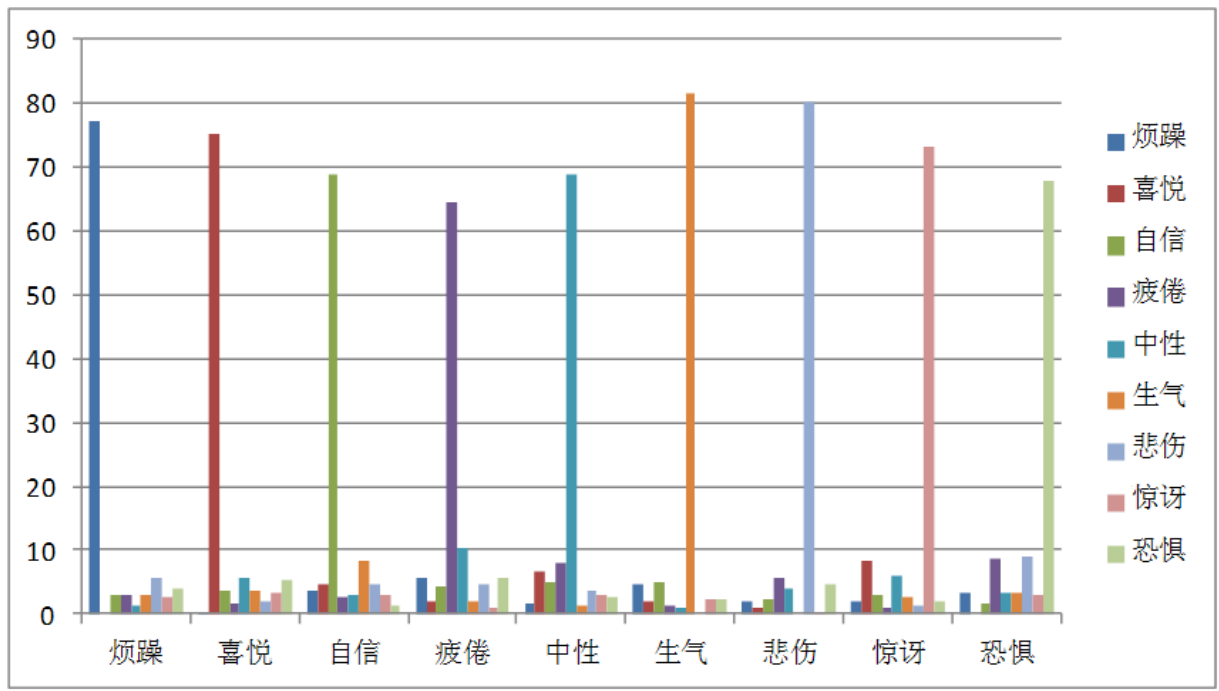

图 4.2 九种情感识别率的柱状图

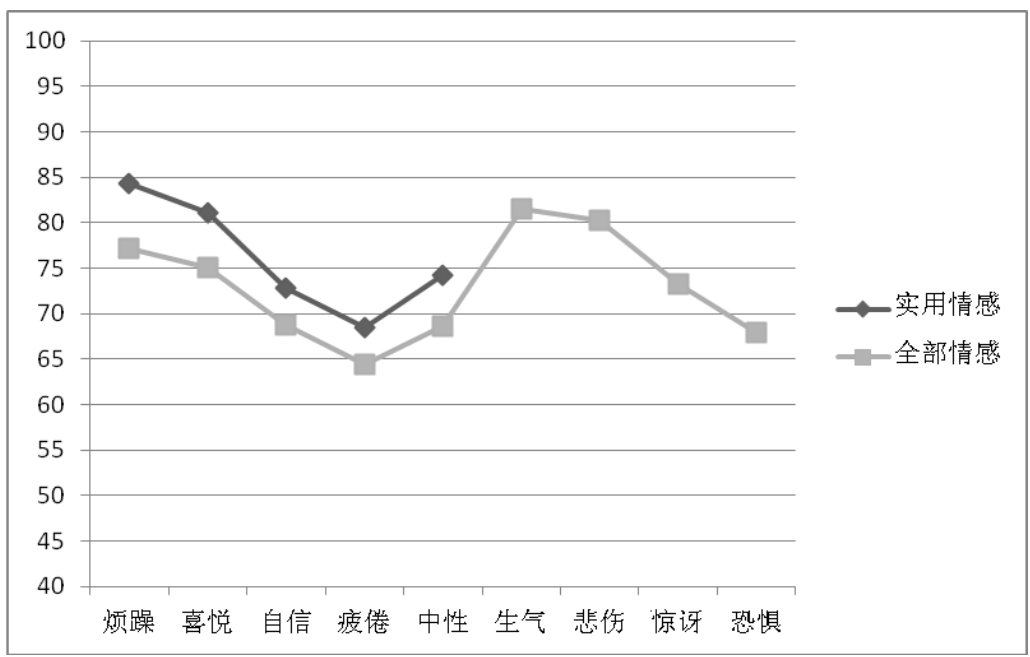

**图 4.3 基本情感与实用情感识别率的比较图**

## 4.3 支持向量机、K 近邻分类器和高斯混合模型的比较实验

诸多模式识别的算法可以应用于语音情感识别，参考近年来国内外研究文献[7, 9, 10, 13, 18, 28, 51, 73, 76, 78, 84]，在本节中我们选用了 SVM 分类器和 KNN 分类器进行识别测试，同上文中的 GMM 分类器进行比较。

### 4.3.1 支持向量机

支持向量机是由 Cortes 和 Vapnik 等人提出的一种机器学习的算法[85]，它是建立在统计学习理论和结构风险最小化的基础之上的。支持向量机在诸多模式分类应用领域中具有优势，如解决小样本问题、非线性模式识别问题以及函数拟合等。

支持向量机能够将数据样本映射到一个更高维度的特征空间里，建立一个最大间隔的超平面以达到线性可分。区分样本点的特征空间中，超平面的两边建有两个互相平行的超平面，平行超平面间的距离或差距越大，分类器的总误差越小。

在本节的实验中，SVM 的核函数选取为径向基函数（Radial Basis Function，RBF）函数，采用二叉树结构实现 SVM 的多类分类，如图 4.4 所示。在两分法分类器的树状结构中，首先识别何种情感（图中首先识别了烦躁情感），对系统的性能是有一定的影响的。在树状的分类器组结构中，误差会进行传播和积累，前面的分类错误，在后续的分类中无法纠正。例如，将烦躁样本误判为非烦躁情感后，在后续的分类器中就无法再识别出这些烦躁样本。

因此，我们考虑设计二叉树结构的两个原则是：一，首先将重要的情感区分出来，二，首先进行误判率低的情感分类。我们将烦躁情感首先进行识别，力求能够尽可能多的检测出烦躁情感，哪怕有些样本被误判为情感，其代价也比较低。疲倦和自信等实用语音情感在最后的几个分类器中进行识别，这是由于这两种情感的识别率较低，置于二叉树底端可以减少误差扩散。

关于误判和漏判的代价问题，是语音情感识别系统中必须要考虑的，我们在本章的最后 4.6 节讨论了如何评价一个情感识别系统的性能的问题。烦躁等威胁较大的负面情感，应该提高其识别率，降低其漏识别率，必要时可以牺牲其它情感类别的误判率。

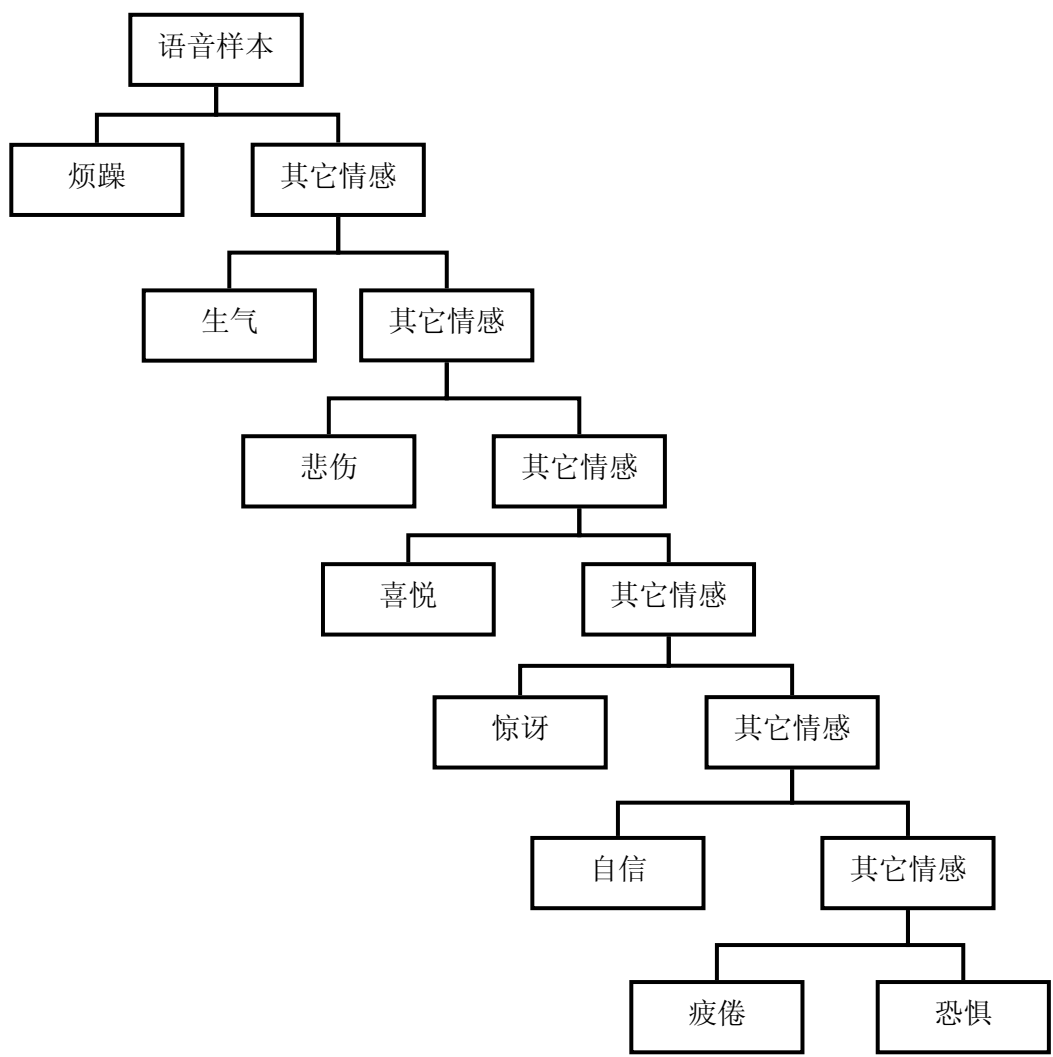

**图 4.4 二叉树结构的多分类支持向量机**

### 4.3.2 K 近邻分类器

K 近邻（k-Nearest Neighbor，KNN）分类器，采用一种较为简单直观的分类法则，其在语音情感识别应用中有较好的性能表现[7，69]。KNN 分类器的分类思想是：给定一个在特征空间中的待分类的样本，如果其附近的 K 个最邻近的样本中的大多数属于某一个类别，那么当前待分类的样本也属于这个类别。

在 KNN 分类器中，样本点附近的 K 个近邻都是已经正确分类的对象。在分类决策上只依据最邻近的一个或者几个样本的类别信息来决定待分类的样本应该归属的类别。 KNN 分类器虽然

原理上也依赖于极限定理，但在实际分类中，仅同少量的相邻样本有关，而不是靠计算类别所在特征空间区域。因此对于类别域交叉重叠较多的分类问题来说，KNN方法具有优势。

### 4.3.3 实用语音情感识别实验

我们在实验数据集 A 上，进行了识别测试。训练与测试的比例为 9：1，与上文中高斯混合模型实验一样，采用相同的轮换测试的方法。KNN分类器中K参数设置为10，SVM分类器中的核函数选择为 RBF 函数，GMM 分类器中的混合度设置为 32。实验中我们调节了各个分类器中的参数设置，使其在当前的数据集上获得最好的识别性能。表 4.4 中为 SVM、KNN 和 GMM 的识别率结果，可以看到 GMM 的总体识别率最高，其中烦躁的识别率最高，达到 77.2%。在 SVM 中，悲伤的识别率最高，而在 KNN 中生气的识别率最高，可见不同的分类算法对情感建模有较大的影响，在实际中应根据数据集的特点进行合理的选择。

**表 4.4 三种识别算法在实用语音情感中的平均识别率比较**

| 识别算法 | 平均识别率 | 各类情感的识别率（%） | | | | | | | | |
|---|---|---|---|---|---|---|---|---|---|---|
| | | 烦躁 | 喜悦 | 自信 | 疲倦 | 中性 | 生气 | 悲伤 | 惊讶 | 恐惧 |
| GMM | 73.0% | 77.2 | 75.1 | 68.8 | 64.4 | 68.7 | 81.5 | 80.2 | 73.3 | 67.9 |
| SVM | 67.1% | 69.2 | 74.5 | 62.2 | 60.2 | 59.3 | 75.7 | 77.3 | 68.4 | 56.7 |
| KNN | 63.1% | 66.1 | 70.4 | 52.3 | 59.2 | 62.5 | 76.2 | 70.1 | 62.3 | 48.9 |

KNN 分类器的核心思想是相似的样本观察应该属于相似的类别中，一般用欧式距离来度量这种相似性。在很多应用领域中，KNN 都是一种直接有效的解决途径。而 GMM 的优点在于对数据分布的拟合能力。与支持向量机等模式识别方法相比，高斯混合模型在理论上可以拟合任何概率分布模型。在大量情感数据的条件下，高斯混合模型对数据的拟合能力非常高，适合用于非特定说话人的情感识别。高斯混合模型也具有自身的缺陷，主要是对训练数据的依赖性强，当训练数据缺乏的时候，不容易获得较好的性能。在训练数据足够的情况下，如果混合度过高也会导致过学习。

在下一节中我们将研究不同的训练数据量对 GMM 的分类性能的影响，以及在小样本条件下对 GMM 分类系统的改进方法。

## 4.4 基于高斯混合模型的两类分类器组

由于采集情感数据较为困难，一般建立大规模的情感数据库困难更大。对于特殊类别的情感，特别是负面情感，被试人员往往不愿意在实验室环境下表达出来。而且特殊类别的情感在诱发时需要配合特殊的实验诱发手段，例如隔绝实验、睡眠剥夺等等。在研究某些新的实用语

音情感时，往往会遇到数据量不足的困难。特别是在研究的起步阶段，很可能会遇到某种感兴趣的情感类别的数据量缺乏的情况。

在一些特殊的应用领域，存在数据采集的困难。例如在载人航天中，对宇航员的情感和心理状态的监测是一个非常重要的环节。然而由于部分领域的特殊环境，难以获得大量的目标情感语料。因此，在这样的实际条件下，有必要探讨小样本条件下 GMM 分类器的性能。研究高斯混合模型如何在小样本条件下发挥最大的性能优势是一个非常有价值的问题。

在本节中，我们结合两类分类器组来改进高斯混合模型在小样本条件下的性能。我们将一个多类分类器分解为各个部分，即若干个两类分类器，每个两类分类器用于识别一对情感类别，对每个两类分类器进行各自的特征空间的优化，最佳区分一对情感类别对，最后通过输出融合将各个两类分类器重组成一个整体的多类分类器。采用了两类分类器组后，对不同的情感类别对，分别设计各自不同的最优的特征空间，使得待识别的情感类别两两之间都达到最佳的区分效果。

通过 LDA 变换，两类分类器组的每个分类器的特征空间维度降为一维。由于维度小，能够在数据量小的情况下也训练充分。如果采用多类分类器，则特征空间维度为 N-1，N 为类别数。需要的训练数据量则会相应增加，样本不足时训练中会出现奇异矩阵。因此，小样本条件下，适合采用两类分类器组进行高斯混合模型的训练。

### 4.4.1 多类分类问题分解为两类分类问题

为了设计一组两类分类器来代替一个多类分类器，下面考虑：n 个类别的分类问题分解为 $C_n^2$ 个两类分类问题。

设总共有 n 个类别，记为：

$$\omega_i, i = 1, 2, \cdots, n \tag{4-10}$$

待识别的样本记为 $x$，样本属于某个类别的概率分别表示为：

$$P(x \in \omega_i), i = 1, 2, \cdots, n \tag{4-11}$$

且满足：

$$\sum_{i=1}^{n} P(x \in \omega_i) = 1 \tag{4-12}$$

多类分类问题可以表示为：

$$x \in \omega_j, j = \arg\max \left\{ P(x \in \omega_i) \right\}, i = 1, 2, \cdots, n \tag{4-13}$$

这里 $\arg\max\{\ \}$ 表示取最大值的下标，并假设最大值存在。

同样，两类分类问题可以表示为：

$$x \in \omega_j, j = \arg\max\left\{P\left(x \in \omega_i\right)\right\}, i = 1, 2 \tag{4-14}$$

也可以写为：

$$x \in \omega_j, j = 3 - \arg\min\left\{P\left(x \in \omega_i\right)\right\}, i = 1, 2 \tag{4-15}$$

这里 $\arg\min\{\ \}$ 表示取最小值的下标，并假设最小值存在，即

$$P(x \in \omega_1) \neq P(x \in \omega_2) \tag{4-16}$$

记一个两类分类器的输出为：$C_{i,j}$

这里 i，j 为该两类分类器识别的两个类别的标号，且 $i \neq j$。

根据式(4-15)我们设计两类分类器的输出为：

$$C_{i,j} = \arg\min\left\{P\left(x \in \omega_k\right)\right\}, k = i, j \tag{4-17}$$

由式(4-13)可知：

$$P\left(x \in \omega_i\right) < P\left(x \in \omega_j\right) \underset{i,j \leq 2}{\rightleftharpoons} x \notin \omega_i \tag{4-18}$$

在式(4-16)的约定下，根据式(4-17)和式(4-18)有：

$$C_{i,j} \Rightarrow x \notin \omega_{C_{i,j}} \tag{4-19}$$

其中，$i$，$j$ 表示 n 个待识别类别的标号，$\Rightarrow$ 为推出符号。

式(4-18)表明，在分类正确的前提下，每一个两类分类器，可以排除多类分类问题中的一个类别（待识别样本不属于这个类别）。

由此，我们考虑将一个 n 个类别的分类器，分解为若干个两类分类器的组合，每个两类分类器记为：$C(\omega_i, \omega_j), i \neq j$，则共有 $C_n^2$ 个不同的两类分类器。

对应的分类器组的输出组成的集合为

$$C = \left\{C_{i,j} \mid 1 \leq i, j \leq n, i \neq j\right\} \tag{4-20}$$

为了证明多类分类问题的等效分解，我们需要下面的这个命题：

命题(1)："两类分类器组的 $C_n^2$ 个输出值中必然存在 $n-1$ 个互不相等的值"

这里的两类分类器组的输出由式(4-17)得到，将这 $n-1$ 个互不相同的值记为，

$$K = \left\{k_1, k_2, \cdots, k_{n-1}\right\} \tag{4-21}$$

采用数学归纳法证明上面的命题：

$n=3$ 的情况，

有三个输出值 $C_{1,2}$、$C_{1,3}$、$C_{2,3}$，假设不存在 $n-1=2$ 个互不相等的值，

则，

$$C_{1,2}=C_{1,3}=C_{2,3} \tag{4-22}$$

上式与式（4-17）矛盾

假设不成立，$n=3$ 时，必然存在 $n-1$ 个互不相等的值

考察 $n+1$ 的情况，即 $n=4$ 时，

因为 $C_{1,2}$、$C_{1,3}$、$C_{2,3}$ 中有 2 个互不相等的值，

根据式（4-17）知，

$$C_{1,2},C_{1,3},C_{2,3} \neq 4 \tag{4-23}$$

假设 $C_{1,4}$、$C_{2,4}$、$C_{3,4}$ 中有一个值为 4，则有三个互不相等的值

假设 $C_{1,4}$、$C_{2,4}$、$C_{3,4}$ 中没有一个值为 4，结合式(4-16)得，

$$C_{1,4}=1,C_{2,4}=2,C_{3,4}=3 \tag{4-24}$$

此为三个不同的值。

因此 $n=4$ 时，$C_n^2$ 个分类器输出值中同样必然有 $n-1$ 个不同的值。

根据数学归纳法，对于 $n \geq 3$，命题(1)成立。

在式(4-19)与命题(1)的基础上可以证明根据 $C_n^2$ 个两类分类器可以完成 n 个类别分类的问题。

令 $x \in \omega_x$，由于 $K \subset C$，注意到式(4-19)，有

$$\omega_x \notin \{\omega_i \mid i=k_1,k_2,\cdots,k_{n-1}\} \tag{4-25}$$

令集合

$$\Omega=\{\omega_i \mid i=1,2,\cdots,n\} \tag{4-26}$$

$$W=\{\omega_i \mid i=k_1,k_2,\cdots,k_{n-1}\} \tag{4-27}$$

所以有：

$$\omega_x \in \Omega-W \tag{4-28}$$

根据命题(1)，$W$ 中有 $n-1$ 个不同的值，

$$|W| = n-1 \qquad (4\text{-}29)$$

注意到

$$|\Omega| = n \qquad (4\text{-}30)$$

以及

$$W \subset \Omega \qquad (4\text{-}31)$$

所以有：

$$|\Omega - W| = 1 \qquad (4\text{-}32)$$

在集合 $\Omega - W$ 中只包含一个元素 $\omega_x$，即为识别的结果。

由此证明了一个 n 类的分类问题可以通过 $C_n^2$ 个两类分类问题来完成，也即表明了 $C_n^2$ 个两类分类器中，包含了足够的信息来，完成一个 n 类的多类分类器的识别工作。下面我们将给出用两类分类器组完成多类识别的具体做法。

### 4.4.2 置信度相关译码融合

本文中采用 GMM 对每种情感的概率分布进行建模，采用贝叶斯分类器来实现上文中的每个两类分类器，以后验概率代替式（4-13）到式(4-18)中样本属于某个类别的概率 $P(x \in \omega_i)$，则这组分类器组以各自的识别正确概率 $p_{i,j}$ 给出式(4-19)中的判决，此时式(4-19)变为：

$$\mathrm{C}_{i,j} \xrightarrow{\quad p_{i,j} \quad} x \notin \omega_{C_{i,j}} \qquad (4\text{-}33)$$

因此，两类分类器组给出的信息，就在这一组概率中。对分类器组的输出融合，也就是对这一组概率值的融合。

贝叶斯分类器的错误概率（或正确率）有多种计算方法，通常是计算相应的样本分布曲线的积分。这种方法是计算的平均意义上的错误概率，本文中考虑采用一种样本自适应的方法[13]，来计算对于每个样本的判决的置信度 $w_{i,j}$，用来作为上式中的概率值 $p_{i,j}$。在贝叶斯分类器中，当样本处于后验概率密度分布曲线的重叠区域时，分类器可能发生错判，发生错判的可能性可以用当前样本的后验概率的差来度量，进入分类器的样本属于不同类别的后验概率相差越大，

误判的可能性就越小。样本处于重叠区域的可能性的度量，作为每个分类器的置信度，可以用式（4-34）来得到。

$$w_{i,j} = 2 \times \frac{\left| \ln\left( p_i \right) - \ln\left( p_j \right) \right|}{\left| \ln\left( p_i \right) + \ln\left( p_j \right) \right|} \tag{4-34}$$

其中，$p_i$，$p_j$ 是后验概率密度的取值，当分类器判决越可靠时，差值越大，$w_{i,j}$ 越大，反之当 $w_{i,j}$ 越小时，说明样本距离重叠区域越近，分类可靠性越差。

得到了分类器的置信度 $w_{i,j}$，据此进行权值融合，将分类器的输出定义为，

$$C_{i,j}^{\ *} = w_{i,j} \cdot I，\quad I = \text{+1,-1} \tag{4-35}$$

其中 $I$ 是两类分类的判决，$I = +1$ 表示判断为两类分类中的第一个类别，$I = -1$ 表示判断为另一个类别。

为了实现判决融合，下面将这组分类器的输出构成一个超矢量，用相关译码的方法来进行判决，如图 4.5 所示。

在理想的情况下，判决可靠度 $p_{i,j}$ 为 1，此时得到的输出值 $C_{i,j}^{\ *} = I$，当待识别样本不属于两类分类器所能识别的两个类别时，理想中，输出值不应该给出任何信息，置为零。以此作为当前类别的码字，如表 4.5 所示。在实际情况下，输出值 $C_{i,j}^{\ *} = p_{i,j} \cdot I$，围绕在理想值（码字）的周围，可根据实际输出值与码字的距离进行译码。相关译码器的作用即是通过相关运算来衡量实际值与理想值之间的接近程度，最大的相关值对应的情感类别，即为识别结果，

$$i^* = \arg\max\left\{ r_i \right\} \tag{4-36}$$

$i^*$ 表示识别出的类别标号，$r_i$ 为相关值，通过式（4-37）得到，

$$\boldsymbol{R}^T = \boldsymbol{C}^T \cdot \boldsymbol{I}_{m \times n} \tag{4-37}$$

其中，

$$\boldsymbol{R} = \left\{ r_1, r_2, \cdots, r_n \right\} \tag{4-38}$$

$\boldsymbol{C}$ 是每个分类器输出值构成的列向量，$\boldsymbol{I}_{m \times n}$ 是表 4.5 中码字构成的矩阵，m 为分类器的个数，n 为情感类别数。

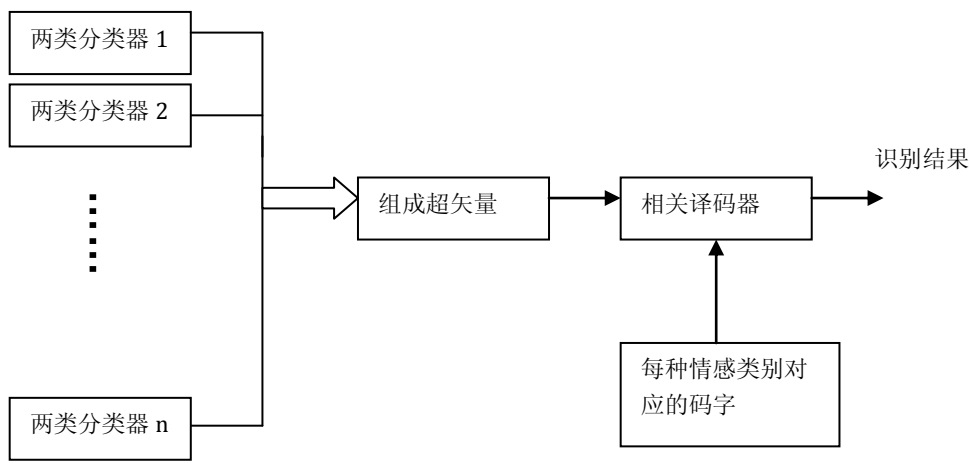

图 4.5 分类器组的判决融合框图

表 4.5 情感类别的码字

| 两类分类器 | 情感类别对应的输出值 | | | | |
|---|---|---|---|---|---|
| | 喜悦 | 烦躁 | 自信 | 疲劳 | 中性 |
| 喜悦/烦躁 | +1 | −1 | 0 | 0 | 0 |
| 喜悦/自信 | +1 | 0 | −1 | 0 | 0 |
| 喜悦/疲劳 | +1 | 0 | 0 | −1 | 0 |
| 喜悦/中性 | +1 | 0 | 0 | 0 | 1 |
| 烦躁/自信 | 0 | +1 | −1 | 0 | 0 |
| 烦躁/疲劳 | 0 | +1 | 0 | −1 | 0 |
| 烦躁/中性 | 0 | +1 | 0 | 0 | −1 |
| 自信/疲劳 | 0 | 0 | +1 | −1 | 0 |
| 自信/中性 | 0 | 0 | +1 | 0 | −1 |
| 疲劳/中性 | 0 | 0 | 0 | +1 | −1 |

### 4.4.3 分类器组的实验结果

在前面几个实用语音情感的识别实验中，测试环境是按照 9：1 的训练与测试比例来设定的。由于 GMM 算法对数据的依赖性较大，因此有必要考察改变测试比例时，系统的识别性能的变化，如图 4.6 中曲线所示。

实验中采用了实验数据集 A（详见第二章 2.6 节），包含五种情感，烦躁、喜悦、中性、自信和疲倦。随着测试比例的变化，训练数据量变的不足，因而在特征模块中也要有相应的参数调整。在训练与测试比例下降为 2：8 时，在送入 LDA 模块之前，PCA 截取的特征维数要从原本的 20 调整到 10，以防止 LDA 模块计算中出现奇异矩阵。在训练数据不足的情况下 GMM 的混合度不能设置的过高，测试比列为 2:8 时，GMM 混合度设置为 12，以保证 EM 算法的精确度。在第三章中，我们比较了 PCA 截断维数对识别率的影响，在其中我们采用了较充足的训练样本，而此处随着训练样本的减少 PCA 截断维数应该设置的较小。

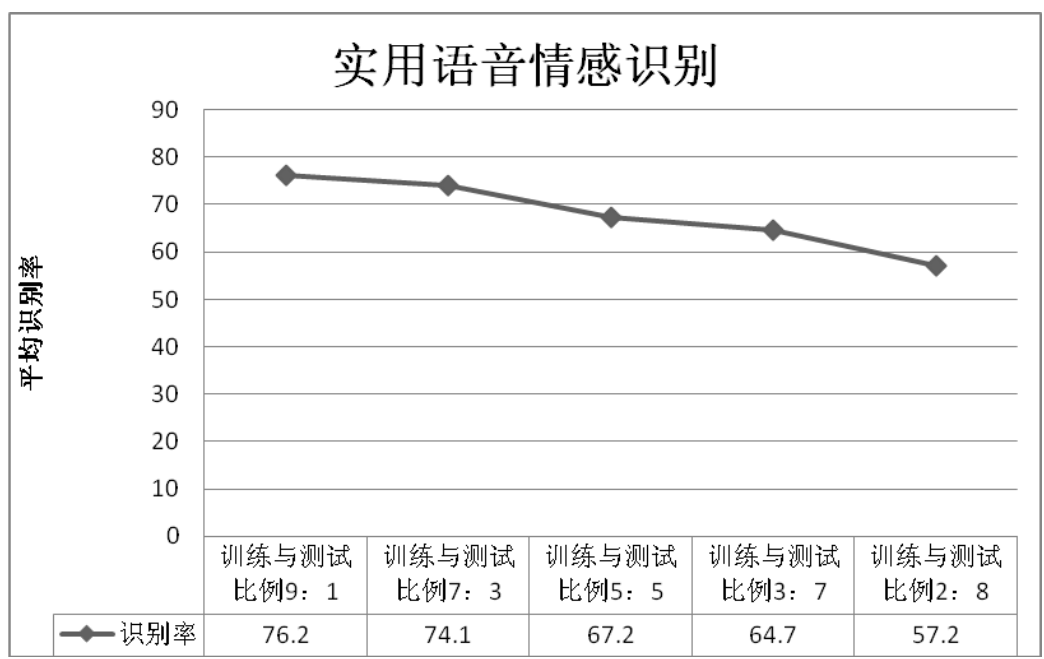

图 4.6 改变训练与测试比例后的识别结果

表 4.6 测试比例 2：8 时五种情感的识别率（%）

|  | 烦躁 | 喜悦 | 自信 | 疲倦 | 中性 |
|---|---|---|---|---|---|
| 烦躁 | **64.2** | 0.5 | 15.1 | 12.6 | 7.6 |
| 喜悦 | 0.6 | **61.4** | 12.0 | 5.4 | 20.6 |
| 自信 | 12.8 | 17.6 | **52.7** | 7.0 | 9.9 |
| 疲倦 | 13.9 | 3.2 | 8.6 | **48.3** | 26.0 |
| 中性 | 4.1 | 11.8 | 10.1 | 14.6 | **59.4** |

通过改变训练与测试比例，我们可以看到图 4.6 小样本条件下多类分类器性能的下降。测试比例为 2：8 时，五种情感的平均识别率为 57.2%，如表 4.6 所示。而训练测试比例为 9：1 的大样本条件下，平均识别率为 73%。从识别率的下降看到，有必要特别研究小样本条件下高斯混

合模型的性能改进。由于高斯混合模型对数据的依赖程度较高，它虽然能够很好的拟合底层数据样本的分布，但是在训练数据不足的情况下，性能下降明显。

表 4.7 改进的 GMM 分类器组的识别率结果（%）

|  | 烦躁 | 喜悦 | 自信 | 疲倦 | 中性 |
|---|---|---|---|---|---|
| 烦躁 | **67.4** | 0.5 | 15.1 | 11.0 | 6.0 |
| 喜悦 | 0.4 | **69.8** | 9.1 | 3.5 | 17.2 |
| 自信 | 11.1 | 15.9 | **59.1** | 6.0 | 7.9 |
| 疲倦 | 11.9 | 2.0 | 6.7 | **56.5** | 22.9 |
| 中性 | 3.2 | 8.9 | 8.1 | 11.6 | **68.2** |

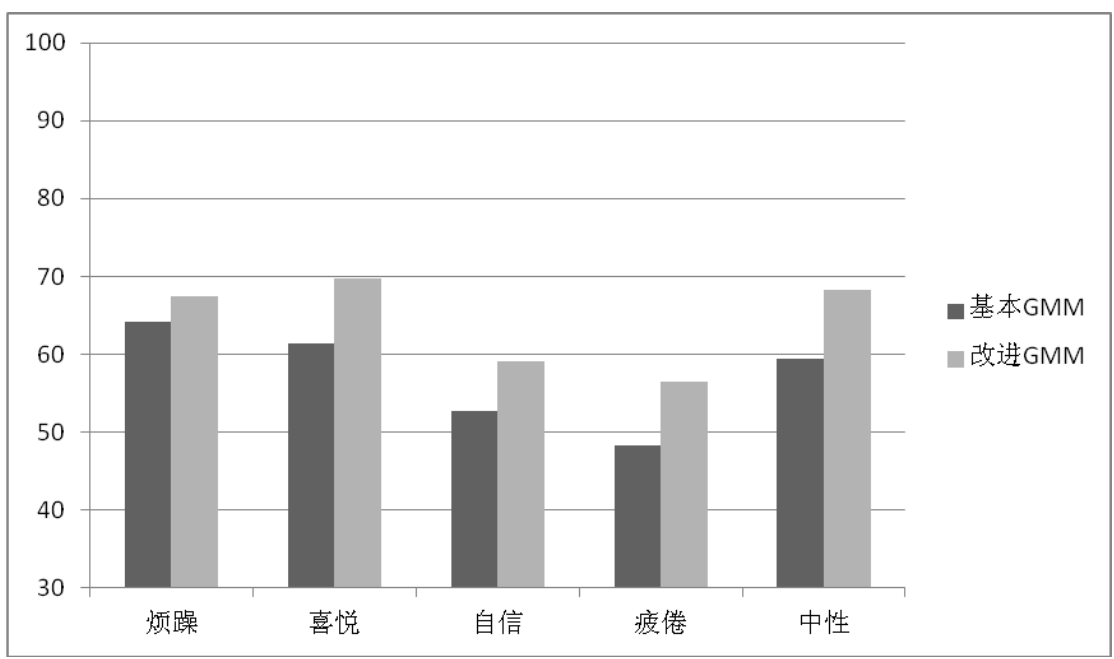

图 4.7 识别率提升比较图

采用本节中的结合两类分类器组的方法后，识别率如表 4.7 所示。我们可以看到平均识别率提高到了 64.2%，说明在多类分类问题分解为两类分类问题后，训练数据能够满足 GMM 训练的要求。基本 GMM 算法中，特征空间维数为 4，的混合度设置为 12 时，在小样本条件下获得较好的性能表现。本节中与两类分类器组结合后，特征空间维数为 1，GMM 混合度设置为 24，在小样本条件下获得了明显的性能提升。同时，将五种情感分别组成十个情感类别对后，能够分别优化各自的特征空间，提高了识别性能。图 4.7 中比较了基本的 GMM 与改进后的 GMM 分类器组的性能，可以直观的看到各类情感的识别率有明显的提升。

## 4.5 嵌入马尔科夫网络的高斯混合模型

本节中，我们研究了连续语音中的情感识别问题，在实际应用中系统接收到的往往是连续输入的语料，并且前后的情感变化具有连续性，因此研究连续语音信号中的情感识别具有重要的实际意义。

语音情感是包含在一段时长的语音信号中的，因此对于情感语料的起始点和结束点的界定有一定的困难，一个实际的措施是以语料中的自然停顿来分割。由于目前的情感识别系统中，对语料的分割是人为定义的，因此不难预见到在相邻的语料之间存在较强的情感相关性。我们对此进行了情感状态转移模型的研究，在此基础上进行了情感识别结果的纠错。

对于连续输入的语料，常用的做法是将语音信号进行人工的切分，制作成单词、短句、长句等数据样本，再对这些语音数据样本进行情感的标注，提供给训练模块和识别模块。实际当中，这个情感切分过程是困难的，一条情感语音的起点和终点往往难以判断。依靠自然停顿或者情感的变化来进行切分，往往带有很大的主观性和随机性。已有研究报告显示，不同的时长对语音情感识别有一定的影响[86]。在使用计算机技术对人类的行为模式进行的研究中，一段典型性行为样本的界定是研究的难点之一，不仅仅局限于语音情感识别之中。例如，在人体运动检测中，对于一个运动的起点和终点的切分，同样是研究的难点之一。

在本文中的高斯混合模型的基础上，我们采用了时域多尺度的情感识别方法，更加全面的分析连续语音中的情感信息，并且结合马尔科夫网络，对多尺度分析的结果进行融合判决。

### 4.5.1 马尔科夫网络

马尔科夫网络（Markov Network），是一类重要的统计机器学习方法，它是关于一组具有马尔科夫性质的随机变量的全联合概率分布模型。马尔科夫网络并不是纯数学专著中的研究对象，它是对马尔科夫特性的推广应用。最早将马尔科夫特性推广到二维图像研究中的是 Abend 与 Ahuja[87, 88]，由于当时的计算机运算性能的限制，这些算法并没有受到广泛的重视。到二十世纪八十年代中期，随着 Hammersley-Clifford 定理的提出[89]，建立了局部图像特征与全局图像特征的联系，使得马尔科夫网络在计算机视觉中的应用受到了研究者们的广泛重视。近年来随着最优化算法的发展，对高阶马尔科夫网络的优化问题吸引了很多研究者的关注。文献[90]提出了一种高效的优化算法（Quadratic Pseudo-Boolean Optimizer, QPBO），对计算机视觉研究领域做出了较大的贡献，使得很多视觉系统中能够快速有效的处理高阶的形状信息。

对于随机场的定义可以进行如下的表述：$s = \{s_1, s_2, \cdots, s_N\}$ 表示基于指标集 $S$ 的一组随机变量，其中每一个随机变量 $s_i$ 可以取集合 $\mathbf{L}$ 中的一个值 $s_i$ 则随机变量组 $S$ 被称为随机场[91]。

当且仅当一个随机场满足如下两个条件是，称为马尔科夫网络[89]：

（1）非负性，$P(s) > 0, \forall s \in S$                                     （4-39）

（2）马尔科夫性，$P(s_i | s_1, s_2, \cdots s_{i-1}, s_{i+1}, \cdots, s_N) = P(s_i | \{s_j\}), \forall j \in \eta_i)$    （4-40）

其中，$\eta_i$ 为 $s_i$ 的邻域。

马尔科夫网络与吉布斯随机场具有对应关系[89]。吉布斯随机场具有如下的形式：

$$P(S = s) = \frac{1}{Z} e^{-U(s)/T}$$         （4-41）

其中，$U(s)$ 为能量函数，由具体问题定义，$T$ 为温度，来自于吉布斯随机场的物理背景，$Z$ 是一个标准化常数：

$$Z = \sum_s e^{-U(s)/T}$$         （4-42）

由于两者的等价性，通过吉布斯分布可以计算出马尔科夫网络的条件概率，因此可以根据实际应用来选择合适的能量函数，据此表达出马尔科夫网络的联合概率，为马尔科夫网络在图像领域的应用提供了很大的便利。

### 4.5.2 多层次融合识别

对于实际中的连续语音，本文分别从两个层次进行语音的情感识别：短句层，以及相邻短句构成的长句层，如图 4.8 所示。这里"长句"的含义是指包含两个被识别的子语句对象，而并不是严格从时间长短来区分长句和短句。多尺度分析的优点在于，对于每一段数据同时在两个时间尺度上进行情感识别，有利于提高系统的可靠性。

对于每一个被识别的语句对象 $s$，情感识别的过程可以被看作是一个多标签的标记问题（Labeling Problem）。每种情感对应于一个标签 $l \in \mathbf{L}$，$\mathbf{L}$ 为标签的集合，一条语句对象 $s$ 即是标记问题中的一个位置（Site），$s \in \mathbf{S}$，$\mathbf{S}$ 为位置的集合。因此情感识别过程可以表达为映射：$\mathrm{F}: \mathbf{S} \rightarrow \mathbf{L}$。

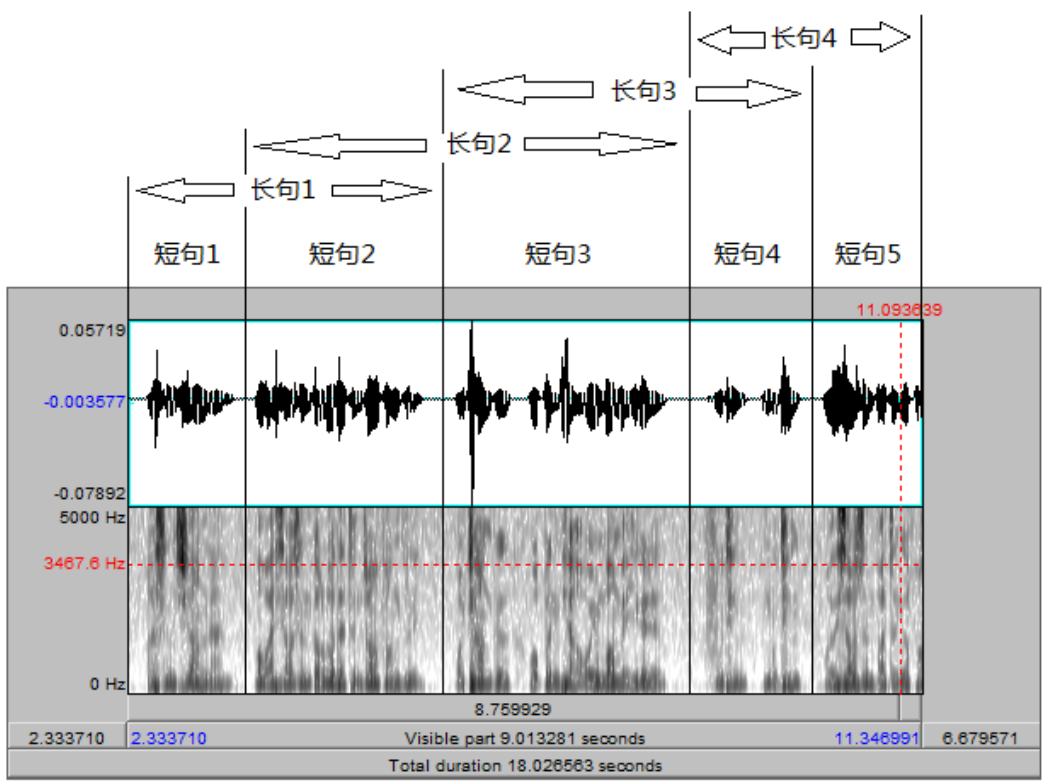

图 4.8 连续语音的时域多尺度情感分析

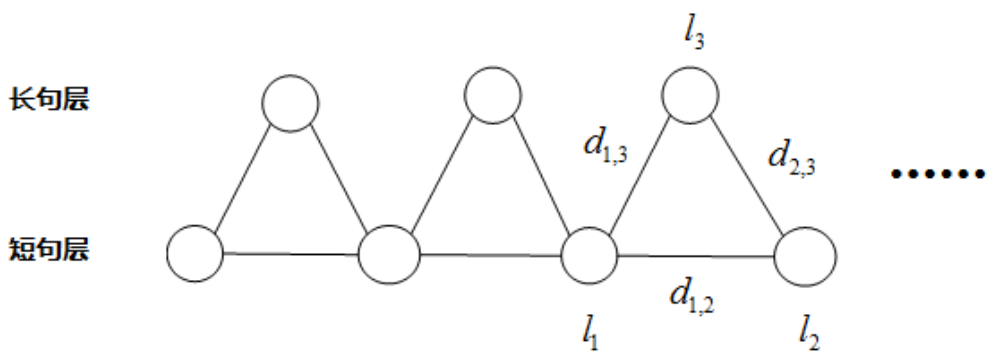

图 4.9 多尺度情感识别中的马尔科夫网络

　　根据情绪心理学的维度空间论，我们认为情感可以被看作为一个二维或者三维空间中的连续变量，并且，情感的变化在时间上具有前后的连续性。因此，相邻短句的情感，以及它们组成的长句的情感应该具有较高的一致性。本文中，我们假设这种相邻情感语句之间的关系具有马尔科夫性质，即在团内的节点（待识别的语句对象）之间具有依赖关系，团之间不具有依赖关

系。短句或者长句即为一个节点（即标记问题中的位置），相邻的两条短句，以及它们构成的一条长句共同构成一个团（clique），如图 4.9 所示。为了对连续耳语音中的这种相关性进行建模，我们在短句层和长句层的相邻语句上定义一个形状模型。选取相邻的两个短句和它们构成的长句，即标记问题中的三个位置，作为三角形的三个顶点。三角形的边长定义为对应的两个顶点的情感标签 $l_1$ 和 $l_2$ 之间的距离 $d_{1,2}$。

两个标签之间的距离，或者说情感标签的相似度，可以在"唤醒度-效价度"二维情感空间当中定义：

$$d_{1,2} = \left\| (a_1, v_1) - (a_2, v_2) \right\|_1 \tag{4-43}$$

或者，

$$d_{1,2} = \left\| (a_1, v_1) - (a_2, v_2) \right\|_2 \tag{4-44}$$

其中，$a_1$ 为唤醒度坐标，$v_1$ 为效价度坐标，$\| \ \|_1$ 为 L1 距离，$\| \ \|_2$ 为 L2 距离。

### 4.5.3 能量函数的定义

下面我们首先来考察三角形（即"团"的）势能，定义形状模型的形变惩罚函数。我们将三角形的三条边替换为三根弹簧，即得到弹簧模型（Spring Model），如图 4.10 所示。记三个顶点为 $s_1$，$s_2$，$s_3$，对应的三条边的距离为 $d_{1,2}$，$d_{1,3}$，$d_{2,3}$，设三根弹簧的弹性系数为，$\sigma_{1,2}$，$\sigma_{1,3}$，$\sigma_{2,3}$。在这里我们做第二个假设，一条语句的前半部分和后半部分对于情感的表达同样重要。因此，根据依据对称性原理，在长句与两条短句之间的弹性系数相等，即 $\sigma_{2,3} = \sigma_{1,3}$。归一化后有：

$$\sigma_{2,3} = \sigma_{1,3} = 1 \tag{4-45}$$

$$\sigma_{1,2} = \sigma_0 \tag{4-46}$$

惩罚函数即为（"团"的势能）可以直接定义为弹性势能的形式：

$$v = \frac{1}{2} \sigma_0 \sum_{\substack{i,j=1,2,3 \\ i<j}} d_{i,j}{}^2 \tag{4-47}$$

代入二维情感空间中的距离，采用 L1 距离有：

$$v = \frac{1}{2} \sigma_0 \sum_{\substack{i,j=1,2,3 \\ i<j}} \left\| (a_i, v_i) - (a_j, v_j) \right\|_1^2 \tag{4-48}$$

采用 L2 距离有：

$$v = \frac{1}{2}\sigma_0 \sum_{\substack{i,j=1,2,3 \\ i<j}} \left\| (a_i, v_i) - (a_j, v_j) \right\|_2^2 \tag{4-49}$$

其中，$v$ 为一个三角形的势能，整个马尔科夫网络的总能量为（L1 距离形式）：

$$V = U + \sum_{k=1}^{N} \frac{1}{2}\sigma_0 \sum_{\substack{i,j=1,2,3 \\ i<j}} \left\| (a_i^k, v_i^k) - (a_j^k, v_j^k) \right\|_1^2 \tag{4-50}$$

或者 L2 距离形式：

$$V = U + \sum_{k=1}^{N} \frac{1}{2}\sigma_0 \sum_{\substack{i,j=1,2,3 \\ i<j}} \left\| (a_i^k, v_i^k) - (a_j^k, v_j^k) \right\|_2^2 \tag{4-51}$$

其中，$N$ 为语句数量，$V$ 为马尔科夫网络的总能量，$U$ 为每个节点的一阶能量总和。高阶马尔科夫网络的能量定义一般具有如下形式：$V = U + \omega \sum_{k=1}^{N} v_k$，此处 $\omega$ 为权重系数，我们注意到可以将 $\omega$ 合并到弹性系数 $\sigma_0$ 中，因此在式（4-49）和式（4-50）中将 $\omega$ 略去。在本文中为了简化模型，我们不考虑二阶的能量函数，两个节点之间的相互依赖关系已经包含在了三阶的弹簧模型中。

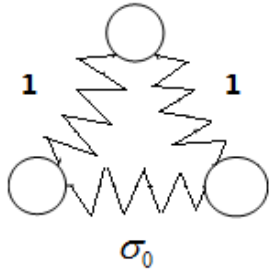

**图 4.10 惩罚函数的弹簧模型**

马尔科夫网络中的每个节点代表一次分类器的识别结果，其一阶能量可以通过分类器输出的置信度来定义。本文中我们将采用 GMM 似然概率的模糊熵形式来进行一阶能量的定义。

对于马尔科夫网络中一个节点的 GMM 判决输出，分别对应 $C$ 种情感的 GMM 似然概率密度值，以 GMM 似然概率密度值映射到 0 到 1 之间作为当前样本归属于第 $c$ 个情感类别的隶属度 $\mu_c(\boldsymbol{X})$：

$$\mu_c(\boldsymbol{X}) = \frac{\arctan\left(p(\boldsymbol{X}|\lambda_c)/10\right)}{\pi/2} \tag{4-52}$$

对于第 $c$ 个情感类别的所有可能的样本构成的模糊集 $E_c = \{\boldsymbol{X}_1, \boldsymbol{X}_2, ..., \boldsymbol{X}_n\}$，其隶属度分别为 $\mu_c(\boldsymbol{X}_1), \mu_c(\boldsymbol{X}_2), ..., \mu_c(\boldsymbol{X}_n)$，其模糊熵具有如下形式 [29, 69]：

$$e(\mu_c(\boldsymbol{X}_i)) = -K \ln \mu_c(\boldsymbol{X}_i) \tag{4-53}$$

其中 $K$ 是大于 0 的数。将式 (4-52) 代入得，第 i 个样本归属于第 $c$ 个情感类别的模糊熵为，

$$e(\mu_c(\boldsymbol{X}_i)) = -K\left(\ln\arctan\left(p(\boldsymbol{X}_i|\lambda_c)/10\right) - \ln\left(\pi/2\right)\right) \tag{4-54}$$

由于模糊熵 $e(\mu_c(\boldsymbol{X}_i))$ 反映了第 $c$ 个情感标签 $l_i$ 的置信度，将其作为马尔科夫网络的一阶能量定义形式，注意到总能量是一阶能量与三阶能量的加权形式，参数 $K$ 可以合并到弹性系数中，因此将其设为 1，可得到：

$$U_{l_c} = -\left(\ln\arctan\left(p(\boldsymbol{X}_i|\lambda_c)/10\right) - \ln\left(\pi/2\right)\right) \tag{4-55}$$

至此，通过高斯混合模型与马尔科夫网络的结合，我们将多尺度的情感识别问题转化为对应的马尔科夫网络的能量最小化问题。对于本文中的三阶马尔科夫网络的优化，我们采用了 Ishikawa 提出的通用框架[92]，以及文献[90]中提供的一种高效的优化工具（Quadratic Pseudo-Boolean Optimization, QPBO）。具体的多标签高阶马尔科夫网络的优化求解过程，在文献[92]与文献[31]中有详细的描述，这里不再赘述。

### 4.5.4 实验结果

为了验证嵌入马尔科夫网络的高斯混合模型识别算法，选用连续情感语料进行实验。实验数据中包含了烦躁、喜悦、疲倦、自信和中性状态等五种情感状态的样本，语句样本之间保留了录音采集现场的顺序编号信息，相邻语句可以通过相邻的顺序编号得到。情感语料来自认知作业中的诱发语音，详细采集情况参见第二章。每种情感语料包含 1000 条短句样本，总计 5000 条，其中 4000 条样本用于训练，1000 条样本用于测试，训练数据与测试数据的比例为 8：2。

在特征提取步骤中，采用了第三章中表 3.1 和表 3.2 中所示的特征。特征压缩采取第三章中 LDA 结合 PCA 的方法，PCA 压缩中截取前 20 维特征。GMM 的混合度设为 24，采用 K 均值聚类初始化，采用 EM 算法进行参数估计，其中 EM 算法迭代次数上限设为 50 次。在嵌入的马尔科夫

网络中，能量函数的形势采用了 L2 距离定义。弹性系数在实验中进行调节，此处设置为 0.5，弹性系数高代表了强化相邻语句间的情感延续性，弹性系数小代表了融合判决是更看重单个语句样本的高斯混合模型识别结果。

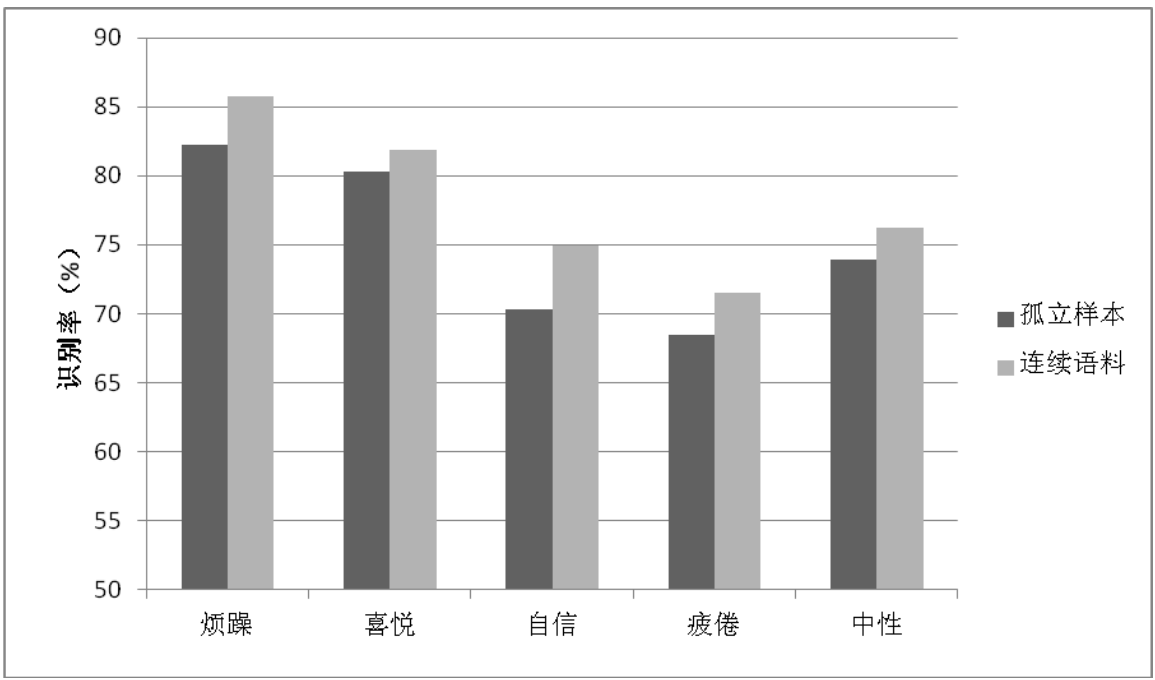

**图 4.11 考虑语料中情感连续性后的识别率提升**

**表 4.8 孤立短句语料的情感识别率（%）**

|  | 烦躁 | 喜悦 | 自信 | 疲倦 | 中性 |
|---|---|---|---|---|---|
| 烦躁 | **82.2** | 0.2 | 7.6 | 6.0 | 4.0 |
| 喜悦 | 0.6 | **80.3** | 5.7 | 3.1 | 10.3 |
| 自信 | 8.1 | 9.9 | **70.3** | 4.6 | 7.1 |
| 疲倦 | 8.3 | 2.9 | 5.2 | **68.5** | 15.1 |
| 中性 | 2.3 | 7.6 | 6.5 | 9.7 | **73.9** |

**表 4.9 马尔科夫网络连续语料识别结果（%）**

|  | 烦躁 | 喜悦 | 自信 | 疲倦 | 中性 |
|---|---|---|---|---|---|
| 烦躁 | **85.8** | 0.3 | 6.5 | 4.8 | 2.6 |
| 喜悦 | 0.4 | **81.9** | 6.2 | 3.7 | 7.8 |
| 自信 | 6.8 | 9.1 | **74.9** | 5.2 | 4.0 |
| 疲倦 | 9.2 | 2.3 | 4.5 | **71.5** | 12.5 |
| 中性 | 2.2 | 6.7 | 6.3 | 8.6 | **76.2** |

孤立样本短句的识别测试结果如表 4.8 所示，平均识别率为 75.0%。采用嵌入马尔科夫的高斯混合模型后，考虑了前后情感的连续性，识别率如表 4.9 所示，平均识别率提升到 78.1% 。疲倦和自信状态提升的比较明显，喜悦提升比较少，如图 4.11 所示。

## 4.6 关于系统性能的评价标准的讨论

本章中，我们讨论了适合于实用语音情感的识别算法，建立了基于 GMM 的识别系统。对于一个具有实用意义的系统，我们需要特别关注它的性能评价问题。一个合理的性能评价方法，有利于对系统进行分析和进一步的改进。最常用的性能评价指标是情感类别的平均识别率，通过平均识别率的高低可以反映出情感模型的可靠性和合理性。需要注意的是，平均识别率是在一定的数据集和一定的训练测试比例的前提下才有意义。在本文中，我们默认采用了这一性能指标进行算法的测试和研究。在实际的应用当中，我们可以考虑从另外一个角度来衡量系统的性能：判决错误的代价。

判决错误的类型可以分为两种，一种是"误识"，一种是"漏识"。这两种错误的代价在实际的应用系统中是非常不同的。一个典型的例子是目标检测，在水下声纳系统中，需要对鱼雷等军事目标进行识别。在这种情况下误识的代价不是很高，例如，水下的其它目标被误识为鱼雷。在这里误识也称为虚假警报，虚假警报带来的后果可能是浪费了一些应急响应的资源。然而，漏识的代价是非常高的，鱼雷等军事目标如果没有被及时的发现，带来的后果是可想而知的。因此，对于具有实际应用价值的识别系统，判决错误的代价是需要考虑的一个重要性能指标。我们需要尽可能的降低系统判决错误的代价。

对于实用语音情感识别，也存在判决错误的代价的问题。例如，在载人航天中，航天员在封闭的舱体内长时间的工作可能会引发烦躁等一系列负面的情感，对于烦躁情感的检测就具有重要的意义，可以防止任务执行中的失误。对烦躁情感的自动检测，可以给人工的心理干预提供一个客观的参考。如果发生误识，其带来的代价相对较小，如果发生漏识，带来的代价就较高。烦躁等负面情感没有被及时检测到的话可能会引发航天员情感和工作能力等方面的严重问题。

因此，我们需要根据实际应用的需求，来有效的降低系统的误判代价。有时需要牺牲误识率，来降低漏识率，在今后的研究中，我们可以考虑开放的调节两种错误的代价，让系统在实际需求中达到一个错误代价的平衡。

# 第五章 语音情感识别系统的鲁棒性研究

本章当中，我们从几个不同的角度研究了上一章中识别系统的鲁棒性：抗噪声性能、未知情感的兼容性与非特定说话人的鲁棒性。本章中还进一步研究了双模态融合情感识别，通过增加传感器通道，来提高抗毁性能。鲁棒情感识别系统的功能模块及相互关系如图 5.1 所示。

## 5.1 问题的提出

研究语音情感系统的抗噪声鲁棒性，主要是针对实际应用场合中难以避免的噪声环境。本文从实用的角度来研究语音情感识别，如何处理噪声的干扰，是提高系统实用价值的一个重要研究方面。在定量的研究情感特征和模型受到噪声的影响有多大的过程中，我们仅考虑了白噪声的情况，作为研究的一个起点。目前国际上对带有噪声的语音情感识别的研究非常少，希望本章中的这一部分研究能够起到一个抛砖引玉的作用。

针对系统的鲁棒性问题，除了考虑信号的噪声干扰外，还需要考虑可识别对象的范围。在理想的实验室环境中，系统遇到的输入都是经过人为选择和设定的。然而，在现实世界里，一个实用的情感识别系统，需要能够处理复杂的、多样的情感种类。因此，拒绝判断是一个必要的功能模块，据判功能需要在误识和漏识之间达到一个平衡。当未知类别的情感出现时，由于其超出了识别器的情感模型的范畴，应当给出拒绝判断的结果，表明这是一个未知的情感类型。识别系统能够给出自身识别能力的界限，是其鲁棒性的一种体现，反之，如果不具备这种能力，在实际当中遇到多种未知的情感样本时，系统很可能会给出荒谬的识别结果。本章中我们将这个方面的性能称为未知情感的兼容性，一个鲁棒的系统应该对任意的情感样本的输入都具有兼容性。

语音信号中的两种信息（说话人信息和情感信息）在一个语音情感识别系统中，不可避免的会发生互相干扰。通过特征归一化技术，避免这种干扰可以提高系统的鲁棒性。在说话人识别当中，不同的情感状态会对识别结果有较大的影响[93，94]。反过来，在情感识别中，说话人数量的增加，会使得情感识别的问题变得困难。由于不同的说话人，有各自的情感表达习惯，以及不同的声音特性，因此，在大量非特定说话人的场合中，情感语料内部的差异会较大。通过某种特征归一化技术，则有可能降低大量说话人带来的差异度。在面向实际的应用中，我们往往不能限定用户的身份，因此非特定说话人鲁棒性是非常值得研究的一项技术。

最后，在情感识别领域多模态是研究的热点之一。本章我们考虑了在传统的语音信号通道之外，给系统增加一个心电信号的通道，实现双模态的情感识别。这样的系统具有较强的抗毁性能，当环境中噪声严重时，心电通道依然能够提供情感识别的依据，当接触式的心电传感器运行不稳定时，还可以依靠非接触式的语音通道进行情感。虽然本文主要研究如何从语音信号中进行情感识别，然而融合心电信号后，可以对原有的语音情感识别系统起到一个辅助和提升的作用。

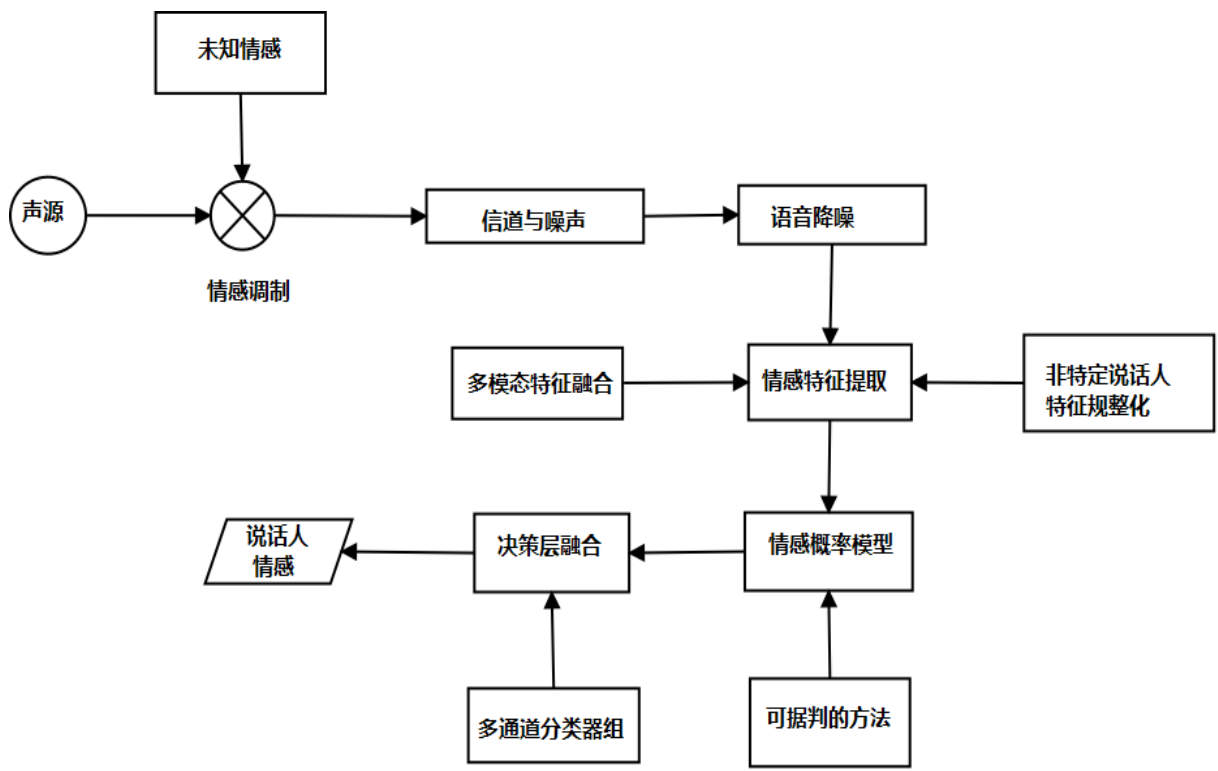

图5.1 鲁棒情感识别系统功能模块

## 5.2 系统的噪声鲁棒性的研究

### 5.2.1 应用环境对抗噪声性能的需求

在诸多语音情感的实际应用当中，噪声都是一个重要的干扰因素。例如，在话务中心的应用中有一定程度的背景噪声；在车载电子中，发动机和机械部件引起的噪声要更加严重；在一些军事环境的应用中，噪声环境也非常恶劣。因此，降噪模块是语音情感识别系统中必须的一个模块，对提高系统的鲁棒性具有关键的作用。

早在 1993 年，Vagar 与 Steeneken 深入研究了加性噪声对语音识别系统的性能的影响程度[95]。此后对抗噪声的鲁棒语音识别系统一直是研究的一个热点问题。在语音情感识别领域，对噪声干扰的研究起步较晚，目前这方面的研究成果非常少。Schuller 等人在 2006 年，首次研究了语音情感识别系统中的这个问题[16]。他们的研究工作中，采用了大量的声学特征，研究了噪声在支持向量机情感识别中的影响。Tawari 与 Trivedi 提出了一种适合于语音情感识别的降噪方法，他们采用了自适应阈值的小波换降噪方法，提高了情感识别率[96]。

## 5.2.2 研究方法

在本节中，我们考虑噪声对语音情感识别的影响。首先我们在纯净的语音信号中加入不同程度的白噪声，获得含噪情感语料。在噪声环境下研究语音情感分类器的训练和识别问题，可以分为两种测试方式。第一种是不匹配的训练与测试，将 GMM 分类器在纯净的数据上训练，在不同程度的含噪数据上进行识别测试。第二种是混合训练与测试，将不同信噪比的含噪数据进行混合，构成训练数据，训练获得的分类器能够反映出多种噪声条件下的情感模式，测试数据集也是含噪数据。

两种测试方式都能够客观的反映出系统的抗噪声性能。通常第一种方式的难度更大，数据不匹配问题更突出；第二种方式由于训练数据和测试数据差异小，识别测试相对容易。本文中采用了第一种测试方式，在纯净的语音数据上进行训练，测试时，在系统的前端加入语音降噪的模块，对含噪数据进行降噪后再进行情感识别。

系统的抗噪声鲁棒性，是通过信噪比和识别率之间的关系来定量的进行分析的。本章中我们主要针对以下两个问题进行了研究：（1）现有的情感特征和情感模型容忍噪声的程度如何，在噪声环境下其识别性能有何变化？（2）在严重的噪声环境下如何改善语音情感识别系统的性能，不同的降噪算法对系统性能的有何影响？

在算法的选择方面，我们选用了两种降噪算法，基本的减谱法和基于听觉掩蔽的降噪算法。减谱法是一种广泛应用的语音降噪算法，选用减谱法希望能够为语音情感识别系统提供一个抗噪声分析的性能基准。

基于人耳听觉掩蔽的降噪算法，是近年来受到特别重视并且在标准数据库上取得很好的效果的一类语音降噪算法。从感知觉心理学的角度来分析，情感的感知是一个需要高级意识参与的过程，对于人耳感知觉所不能捕获的声学特征，对情感听感知的作用是很有限的，如果不是

完全无关的话。因此，从情感识别的需求来看，基于听觉掩蔽的降噪算法具有先天的优势，它从听觉舒适度的角度出发，利用噪声掩蔽效应，去除噪声对人耳听辨的干扰。

### 5.2.3 减谱法降噪简介

减谱法是一种应用广泛的语音降噪算法。将不含有噪声的时域语音信号记为 $s(t)$，令符合零均值高斯分布的噪声信号为 $n(t)$。假设噪声是加性的，则含有噪声的语音信号 $y(t)$ 可以表示为：

$$y(t) = s(t) + n(t) \qquad (5\text{-}1)$$

对上式两边分别进行傅立叶变换，得到频域中的形式，并用 $Y(\omega)$、$S(\omega)$、$N(\omega)$ 表示 $y(t)$、$s(t)$、$n(t)$ 的傅里叶变换，则可得到：

$$Y(\omega) = S(\omega) + N(\omega) \qquad (5\text{-}2)$$

我们可以进一步得到其功率谱密度之间的关系为：

$$|Y(\omega)|^2 = |S(\omega)|^2 + |N(\omega)|^2 + S^*(\omega)N(\omega) + S(\omega)N^*(\omega) \qquad (5\text{-}3)$$

假设语音信号与加性噪声之间是相互独立的，那么上式后面的两项均为零，由此可以得到：

$$|Y(\omega)|^2 = |S(\omega)|^2 + |N(\omega)|^2 \qquad (5\text{-}4)$$

分别用 $P_y(\omega)$、$P_s(\omega)$、$P_n(\omega)$ 表示 $y(t)$、$s(t)$ 和 $n(t)$ 的功率谱密度，则有：

$$P_y(\omega) = P_s(\omega) + P_n(\omega) \qquad (5\text{-}5)$$

由于平稳噪声信号的功率谱在发声段之前和发声段之中基本无变化，因此可以根据发声前的"无声段"（假设这一时间段内无语音信号，而只包含噪声）来估计噪声的功率谱 $P_n(\omega)$，由此可得：

$$P_s(\omega) = P_y(\omega) - P_n(\omega) \qquad (5\text{-}6)$$

在实际中，为了防止出现负功率谱的情况，当 $P_y(\omega) < P_n(\omega)$ 时，令 $P_s(\omega) = 0$。完整的减谱法计算公式如下：

$$P_s(\omega) = \begin{cases} P_y(\omega) - P_n(\omega) & P_y(\omega) \geq P_n(\omega) \\ 0 & P_y(\omega) < P_n(\omega) \end{cases}$$ （5-7）

减谱法降噪的算法流程如图 5.2 中所示，可以看到减谱法的基本的算法原理和计算步骤。在频域处理过程中我们仅考虑了功率谱的变换，而在最后傅立叶反变换中需要借助相位谱来恢复降噪之后的时域语音信号。由于人耳对相位变化不敏感[97]，我们可以直接采用带噪语音的相位作为增强后语音的相位，即用 $y(t)$ 的相位谱来替代估计之后的语音信号的相位谱。

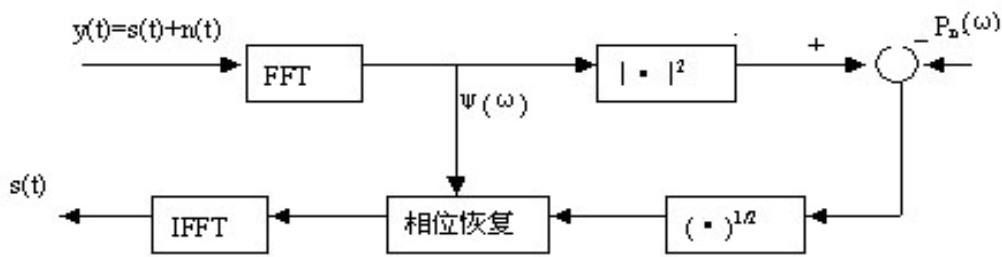

**图 5.2 减谱法降噪算法原理图**

### 5.2.4 基于听觉掩蔽的增强算法简介

本文中将文献[98]中提出的一种语音增强算法与实用语音情感识别算法相结合，系统的研究了白噪声下的语音情感识别问题。探讨了三种与认知有关的实用情感的识别，以及基本情感类别的识别。特别是比较了基本的减谱法和基于人耳听觉掩蔽效应的降噪算法之间的性能差异，以及不同信噪比下的识别率变化规律。

### 一、掩蔽阈值的计算

1961 年，Eberhard Zwicker 提出了 Bark 频带划分[99]，通常来说 24 个 Bark 频带就足以覆盖人耳的听觉感知频率范围，如表 5.1 所示。通过 Bark 刻度可以将线形频率变换到人耳听觉感知域，其间的函数关系可以用下式表示[99, 100]：

$$z = 13 \times \arctan(0.76 \times f(kHz)) + 3.5 \times \arctan(\frac{f(kHz)}{7.5})^2$$ （5-8）

## 表 5.1 临界频带的划分

| 临界频带 | 频率（Hz） | | |
| --- | --- | --- | --- |
| | 下限 | 上限 | 宽度 |
| 0 | 0 | 100 | 100 |
| 1 | 100 | 200 | 100 |
| 2 | 200 | 300 | 100 |
| 3 | 300 | 400 | 100 |
| 4 | 400 | 510 | 110 |
| 5 | 510 | 630 | 120 |
| 6 | 630 | 770 | 140 |
| 7 | 770 | 920 | 150 |
| 8 | 920 | 1080 | 160 |
| 9 | 1080 | 1270 | 190 |
| 10 | 1270 | 1480 | 210 |
| 11 | 1480 | 1720 | 240 |
| 12 | 1720 | 2000 | 280 |
| 13 | 2000 | 2320 | 320 |
| 14 | 2320 | 2700 | 380 |
| 15 | 2700 | 3150 | 450 |
| 16 | 3150 | 3700 | 550 |
| 17 | 3700 | 4400 | 700 |
| 18 | 4400 | 5300 | 900 |
| 19 | 5300 | 6400 | 1100 |
| 20 | 6400 | 7700 | 1300 |
| 21 | 7700 | 9500 | 1800 |
| 22 | 9500 | 12000 | 2500 |
| 23 | 12000 | 15500 | 3500 |
| 24 | 13300 | 22050 | 6550 |

在临界频带之间存在着相互掩蔽的效应，在任意临界频带处的扩展谱则是各临界频带处的 Bark 谱对扩展谱的贡献的总和，其影响的大小和临界频带之间的距离成反比。Schroeder 提出的扩展函数如下式[101]：

$$SF(k,j) = 15.81 + 7.5(|k-j| + 0.474) - 17.5\sqrt{1 + (|k-j| + 0.474)^2}\, dB \qquad (5\text{-}9)$$

根据频带之间的相互影响，扩展后的 Bark 频谱可以表示为：

$$C(k) = \sum_{j=1}^{j_{max}} SF(k,j) B_k \qquad （5\text{-}10）$$

其中 $B_k$ 为第 k 个频带的 Bark 能量，我们注意到将扩展函数进行傅立叶变换后的形式为 $-\lambda/\omega^2$，因此高频分量的衰减较大，更容易发生掩蔽。

对掩蔽阈值的计算注意到两种情况，一种情况是纯音掩蔽噪声，其偏移量为 $(14.5 + k) dB$。另一种情况是噪声掩蔽纯音，其偏移量为 $5.5 dB$。

因此我们可以定义一种称为谱平坦测度（Spectral Flatness Measure，SFM）的度量，来考察信号是更接近纯音还是更接近噪声：

$$SFM_{dB} = 10 \times \log_{10} \frac{G_m}{A_m} \qquad （5\text{-}11）$$

其中，$G_m$ 是功率谱密度的几何平均：

$$G_m = \sqrt[n]{x_1 \cdot x_2 \cdots x_n} \qquad （5\text{-}12）$$

$A_m$ 为功率谱密度的算术平均：

$$Am = \frac{1}{n} \sum_{i=1}^{n} x_i \qquad （5\text{-}13）$$

我们可以引入音调系数系数 $\alpha$ 来衡量信号接近噪声或是接近纯音的程度：

$$\alpha = \min(\frac{SFM_{db}}{SFM_{db\,max}}, 1) \qquad （5\text{-}14）$$

其中常量 $SFM_{db\,max} = -60 dB$。当 $SFM_{db} = 0 dB$ 时，$\alpha = 0$，则信号完全是噪声；当 $SFM_{db} = -60 dB$ 时，$\alpha = 1$，则信号完全是纯音。在实际当中的语音信号既不是完全的噪声，又

不是完全的纯音，其音调系数值介于零和一之间。根据 $\alpha$ 的数值，我们可以计算每个频带的掩蔽能量的偏移函数：

$$O(k) = \alpha \times (14.5 + k) + (1 - \alpha) \times 5.5 \qquad （5-15）$$

由此可以得到噪声掩蔽阈值为：

$$T(k) = 10^{\log_{10} C(k) - O(k)/10} \qquad （5-16）$$

由于我们计算的掩蔽阈值是针对扩展后的信号的，因此需要对其进行解卷积以获得扩展前原信号的掩蔽阈值。在这个过程中可能会出现负能量和零能量的情况，此时则应该采用能量归一化的方法，将得到的 Bark 带上的掩蔽阈值除以能量增益得到解卷积后的掩蔽阈值。

考虑到人耳的绝对听阈，当我们求出的原蔽阈值小于绝对听阈时，则取人耳的绝对听阈值。绝对听阈的定义为：

$$T_{abs}(k) = 3.64 f^{-0.8} - 6.5 \exp(f - 3.3)^2 + 10^{-3} f^4 \qquad （5-17）$$

最终的掩蔽阈值则可以表示为：

$$T^{'}(k) = \max(T(k), T_{abx}(k)) \qquad （5-18）$$

**二、基于听觉掩蔽的语音增强**

在上一节中介绍的减谱法是一种基本的语音降噪算法，尽管它可以较显著的提高信噪比，但是可能会带来残留的音乐噪声，这样不仅对信号质量和听觉质量带来一定的影响，还对所提取的情感的声学特征有一定影响。

本节中我们采用文献[98]中提出的一种基于听觉掩蔽的语音增强算法，首次将之用于实用语音情感识别中，研究了其在情感识别中的作用。基于人耳掩蔽效应的增强方法近年来得到深入的研究[98, 102-105]。一般来说，语音信号是强信号，而噪声相对来说是弱信号，人耳的听觉系统会根据具体的语音信号来确定频域上的听觉掩蔽阈值，如果滤波之后的残留噪声被限制在人耳的掩蔽阈值之内，那么该噪声就不会被人耳听觉所感知，也就实现了对带噪声语音的增强。

由于语音信号通常是逐帧进行处理的，写成帧的形式：

$$y(m, n) = x(m, n) + d(m, n) \qquad （5-19）$$

其中，$m$ 为帧的序号，$m = 1, 2, 3; \cdots$，$n$ 为帧内数据点序号，$n = 0, 1, \cdots, N-1$，$N$ 为帧长。对等式两边进行傅里叶变换可得，

$$Y(m,k) = X(m,k) + D(m,k) \qquad (5-20)$$

其中，$k$ 是离散频率，$Y(m,k)$、$X(m,k)$ 和 $D(m,k)$ 分别是含噪语音 $y(m,n)$、纯净语音 $x(m,n)$ 和噪声 $d(m,n)$ 的傅立叶变换。

根据文献[106]，增强以后的语音幅度谱函数可以表达为：

$$\hat{X}(m,k) = \arg \min_{\hat{X}} E\left\{ d\left[ X(m,k), \hat{X} \right] \middle| y_0^{m'} \right\} \qquad (5-21)$$

其中 $y_0^{m'}$ 是 $m'$ 帧带噪语音的傅立叶变换，

$$y_0^{m'} = \left\{ Y_0(0,k), Y_1(1,k), \ldots, Y_m(m',k) \right\} \qquad (5-22)$$

$d\left[ X(m,k), \hat{X} \right]$ 是 $X(m,k)$ 和 $\hat{X}$ 的距离度量函数，用来度量语音增强前后语音谱的接近程度。

式（5-21）的目标是找到 $\hat{X}(m,k)$ 使得在条件期望之下的距离度量函数最小。如果 $m' \le m$，那么就是对 $\hat{X}(m,k)$ 的因果估计；如果 $m' > m$，那么就是对 $\hat{X}(m,k)$ 的非因果估计。

根据文献[106]，增强函数的形式可以表达为：

$$G_{SP}(\xi_{m|m'}, \gamma_m) = \sqrt{\frac{\xi_{m|m'}}{1 + \xi_{m|m'}}\left(\frac{1}{\gamma_m} + \frac{\xi_{m|m'}}{1 + \xi_{m|m'}}\right)} \qquad (5-23)$$

其中先验信噪比为 $\xi_{m|m'} \triangleq \dfrac{\lambda_{X_{m|m'}}}{\lambda_{D_m}}$，后验信噪比为 $\gamma_m \triangleq \dfrac{|Y_m|^2}{\lambda_{D_m}}$。由 $\xi_{m|m-1}$ 递推 $\xi_{m|m}$ 的迭代算法为：

$$\xi_{m|m} = \frac{\xi_{m|m-1}}{1 + \xi_{m|m-1}}(1 + \frac{\xi_{m|m-1}\gamma_m}{1 + \xi_{m|m-1}}) \qquad (5-24)$$

定义信号功率谱估计：

$$\hat{\lambda}_{X_m} = E\left\{ A_m^2(k) \middle| y_0^{m'} \right\} \triangleq \lambda_{X_{m|m'}} = G_{SP}^2(\xi_{m|m-1}, \gamma_m)|Y_m|^2 \qquad (5-25)$$

其中 $A_m(k)$ 为第 $m$ 帧估计语音谱的幅度，则有增强后的语音谱函数为：

$$\hat{X}_m = G_{SP}(\xi_{m|m}, \gamma_m)Y_m \tag{5-26}$$

文献[98]提出了如下形式的 $\hat{X}_{m,k}$ 的参数化估计公式：

$$\hat{X}(m,k) = \sqrt{\frac{\xi_{m|m-1}}{\mu(m,k)+\xi_{m|m-1}}(1+\frac{\xi_{m|m-1}\gamma_m}{\mu(m,k)+\xi_{m|m-1}})}Y(m,k) \tag{5-27}$$

式中 $\mu(m,k)$ 是时间和频率的函数，定义如下的误差函数：

$$\delta(m,k) = X^2(m,k) - \hat{X}^2(m,k) \tag{5-28}$$

根据可听闻阈的要求，令：

$$E\left[\left|\delta(m,k)\right|\right] \le T(m,k) \tag{5-29}$$

式(5-29)就是令畸变噪声的能量在掩蔽阈值以下，而不被人耳感知。为了推导方便，令

$$M = \frac{\xi_{m|m-1}}{\mu(m,k)+\xi_{m|m-1}} \tag{5-30}$$

则有[98]：

$$
\begin{aligned}
E\left\{\left|\delta(m,k)\right|\right\} &= E\left\{\left|X^2(m,k)-\hat{X}^2(m,k)\right|\right\} = E\left\{\left|X^2(m,k)-M(1+M\gamma)Y^2(m,k)\right|\right\} \\
&= E\left\{\left|X^2(m,k)-M(1+M\gamma)(X(m,k)+D(m,k))^2\right|\right\} \\
&= \left|E\left\{X^2(m,k)\right\}-M(1+M\gamma)E\left\{(X(m,k)+D(m,k))^2\right\}\right| \\
&\le T(m,k)
\end{aligned}
\tag{5-31}
$$

注意到 $E\left\{X^2(m,k)\right\}=\sigma_s^2$，$E\left\{D^2(m,k)\right\}=\sigma_d^2$，化简式(5-31)有：

$$\sigma_s^2 - T(m,k) \le M(1+M\gamma)(\sigma_s^2+\sigma_d^2) \le \sigma_s^2+T(m,k) \tag{5-32}$$

当 $\sigma_s^2 - T(m,k) \le 0$ 时，即语音信号功率小于掩蔽阈值时，我们采用式(5-23)的滤波函数，令 $\mu(m,k)=1$。

当 $\sigma_s^2 - T(m,k) \geq 0$ 时，即语音信号功率大于掩蔽阈值时，由于 $M > 0$，所以

$$\frac{\sigma_s^2 - T(m,k)}{\sigma_s^2 + \sigma_d^2} \leq M(1 + M\gamma) \leq \frac{\sigma_s^2 + T(m,k)}{\sigma_s^2 + \sigma_d^2} \tag{5-33}$$

可以看出不等号两边 $\dfrac{\sigma_s^2 \pm T(m,k)}{\sigma_s^2 + \sigma_d^2}$ 相当于在维纳滤波的基础上做了修正，令

$$B = \frac{\sigma_s^2 - T(m,k)}{\sigma_s^2 + \sigma_d^2} \tag{5-34}$$

$$C = \frac{\sigma_s^2 + T(m,k)}{\sigma_s^2 + \sigma_d^2} \tag{5-35}$$

化简后得到：

$$\frac{-1 + \sqrt{4B\gamma}}{2\gamma} \leq M \leq \frac{-1 + \sqrt{4C\gamma}}{2\gamma} \tag{5-36}$$

即：

$$\frac{2\gamma\xi}{-1 + \sqrt{4C\gamma}} - \xi \leq \mu(m,k) \leq \frac{2\gamma\xi}{-1 + \sqrt{4B\gamma}} - \xi \tag{5-37}$$

参数 $\mu(m,k)$ 由人耳的听觉掩蔽阈值、估计的信号功率谱、噪声功率谱、先验信噪比以及后验信噪比共同确定，它可以动态的改变传递函数形状，以达到对语音畸变和残留噪声两种情况下的最佳折中，从而改善语音的听觉质量。

## 5.2.5 实用语音情感识别的抗噪声实验

我们在实用语音情感数据库上选取纯净的语音数据用于实验，实验数据集为第二章 2.6 节中的实验数据集 B。实验数据集中包含了烦躁、自信、疲倦、喜悦、生气、悲伤、害怕、惊讶和中性等九种情感。将情感语料随机等量分成 10 组，采用交叉验证的方法进行轮换测试，训练与测试数据比例为 9：1，测试识别率结果取平均值。我们在训练数据集中保持纯净的语音，不加入噪声，而在测试的数据当中，加入不同程度的白噪声，构成三类含噪语料集，信噪比分别为 5dB、10dB、15dB。在识别测试前，进行了两种算法的降噪处理，相应的识别测试结果如图 5.3 和图 5.4 所示。实验中，特征降维部分采用 3.4 节中的方法，PCA 压缩中截取前 20 个维度的分

量用于 LDA 降维。高斯混合模型的混合度设置为 32，采用 K 均值聚类算法进行模型的初始化，通过 EM 算法进行均值向量、协方差矩阵和权重的参数估计，迭代次数上限设置为 50。

我们初步研究了加性的高斯白噪声对语音情感识别系统的影响。在噪声的影响下，几种情感之间的可区分度受到了影响。实验显示，降噪模块是实用语音情感识别系统中的一个重要的环节。从识别率的总体看，随着噪声的加入，负面情感的识别率下降的相对较少，正面情感的识别率下降的相对较多。虽然这一差别并不显著，但由于是多次实验的平均结果，因此仍然有一定的分析价值。我们认为，噪声可能使得情感模型发生了迁移，分类器倾向于将样本判定为负面情感。这一现象可能是噪声使得音质特征发生变化导致的，音质特征是判定样本效价度的重要特征，噪声的干扰可能导致音质特征的变化。

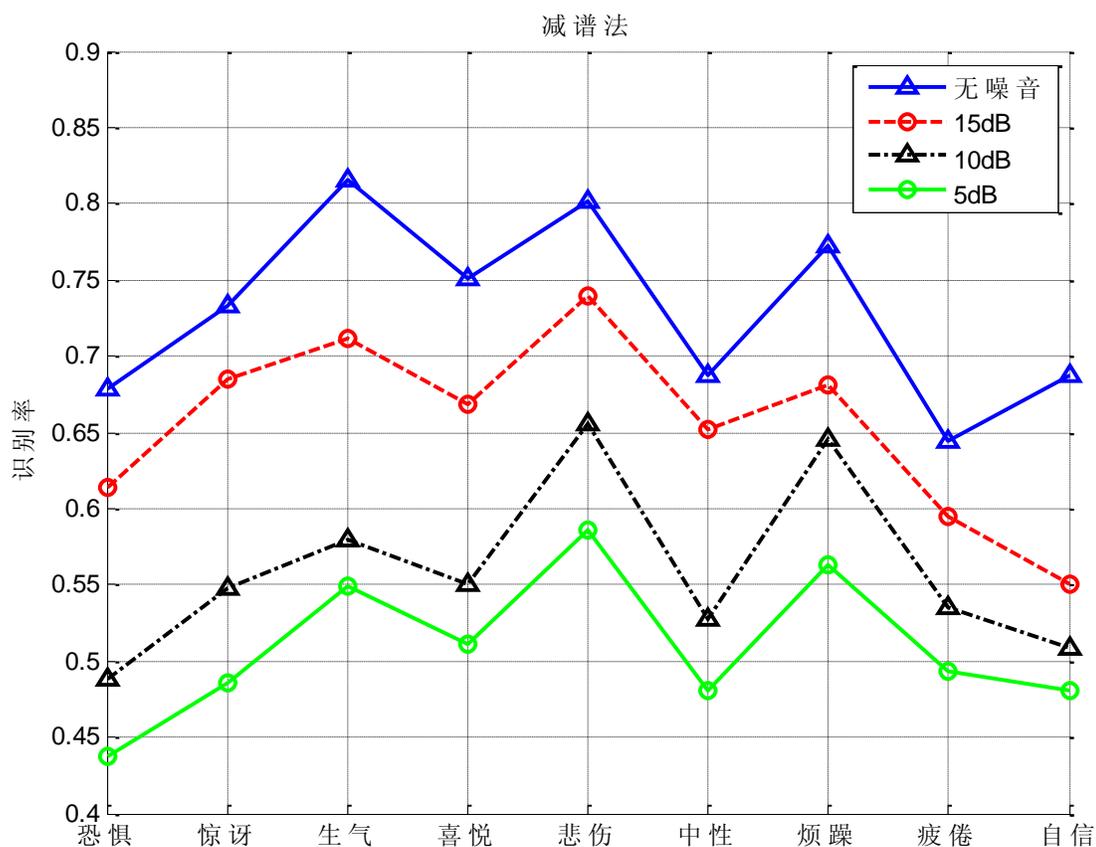

**图 5.3 减谱法降噪情感识别测试结果**

比较图 5.3 和图 5.4 中的识别率的高低可以看到，听觉掩蔽对情感识别的效果要优于减谱法。降噪算法可能会带来语音情感特征的畸变，基于听觉掩蔽的算法能够在语音畸变和噪声消除之间达到一个较好的平衡。因此在实用语音情感识别的应用中能够获得较好的效果。

　　本节中的实验结果可能与分类器算法有一定的联系，在今后的研究中我们可以研究不同的分类算法对噪声的容忍度，还需要进一步研究其它种类的噪声。在特征提取部分，我们沿用了第三章表 3.1 和表 3.2 中列出的特征。在今后的研究中，比较研究不同种类的噪声下情感特征的抗噪声性能，也将是一个有价值的研究方向。

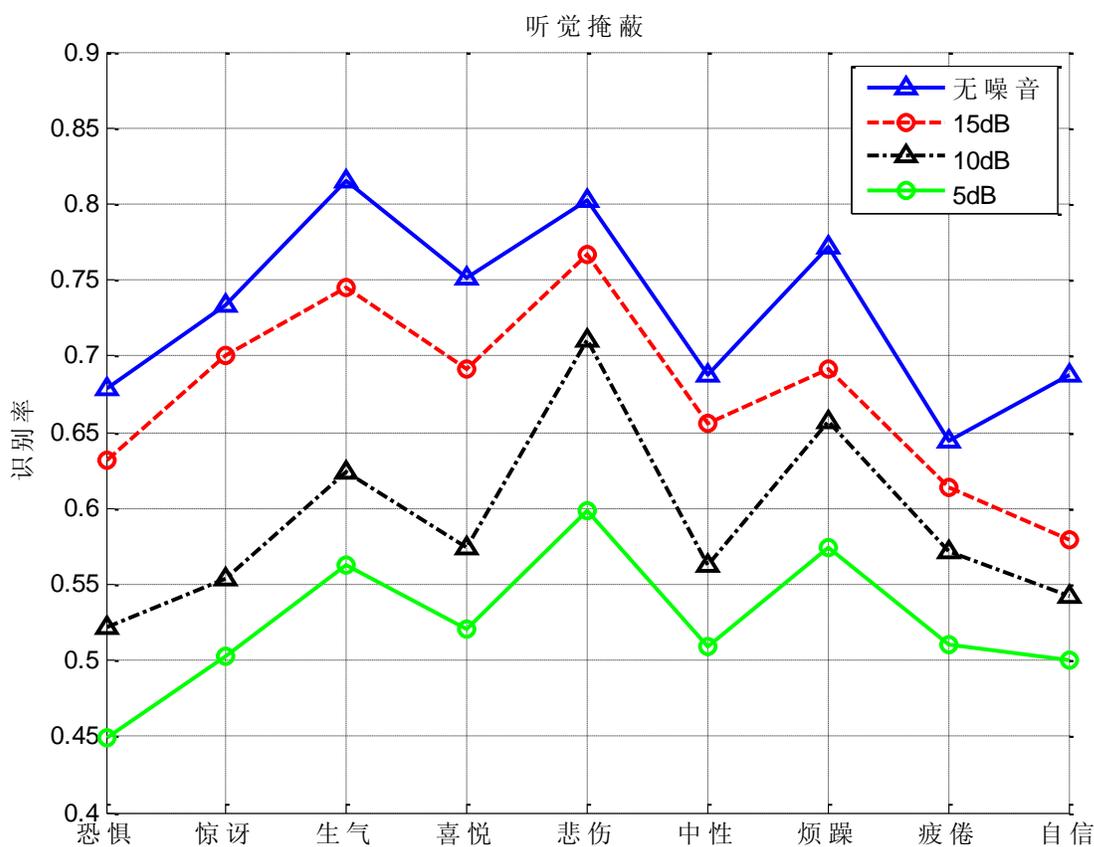

**图 5.4 基于听觉掩蔽的降噪算法测试结果**

## 5.3 未知情感类别的处理

　　上一节中我们考虑了系统对噪声的鲁棒性，这一节中我们来考虑鲁棒性的另一个方面——对未知输入类型的兼容性。在实际中我们无法保证输入的情感数据是何种类型，以往的情感识别系统仅在喜、怒、惊、悲等几种基本类别之间硬性划分，那么系统可能无法在实际环境中正常工作。

由于实际情感的模糊性和多样性，在实用语音情感的识别中，有必要考虑可拒判的识别方法。传统的识别方法，是将出现的样本硬性的划分为已知类别中的某一类，在实际中存在较多模糊不清的情感样本时，分类的可信度就较差，误判的概率就较高。因此本节采用可拒判的实用语音情感识别方法，对于不确定的或未知的情感样本，分类器给出拒绝判断的识别结果，即不属于需要检测的实用语音情感类别中的任何一类。

### 5.3.1 可拒判的实用语音情感识别方法

情感样本在特征空间里的分布，可以用多个高斯函数的叠加来描述。理论上来说，只要混合足够的高斯分量，高斯混合模型（GMM）能够拟合任意的概率密度分布函数。在本文中我们采用 GMM 对烦躁、喜悦和中性三种情感进行建模，每种情感对应一个 GMM 模型，通过最大后验概率准则判决。$\boldsymbol{x}_i$ 表示第 i 条语句样本，$\boldsymbol{\lambda}_j$ 表示情感类别 $j$ 对应的模型参数，最大后验概率可以表示为：

$$p(\boldsymbol{\lambda}_j | \boldsymbol{x}_i) = \frac{p(\boldsymbol{x}_i | \boldsymbol{\lambda}_j) \mathrm{P}(\boldsymbol{\lambda}_j)}{P(\boldsymbol{x}_i)} \qquad （5-38）$$

其中 $p(\boldsymbol{x}_i | \boldsymbol{\lambda}_j)$ 通过每个情感的 GMM 模型得到。对于给定的语句样本，特征矢量出现的概率是一个常量，假设每种情感等概率的出现，$C$ 为情感类别数。

$$P(\boldsymbol{\lambda}_j) = \frac{1}{C}, 1 \le i \le C \qquad （5-39）$$

那么，待识别的样本可以判决为，

$$j^* = \arg \max_j p(\boldsymbol{x}_i | \boldsymbol{\lambda}_j) \qquad （5-40）$$

其中，$j^*$ 表示样本所属的类别。

针对实际环境下情感的模糊与不确定性，实用语音情感种类的多样性，有必要研究可以拒判的识别实用语音情感方法。下面我们将采用一种基于似然概率模糊熵的拒判方法，采用模糊熵来对样本与情感类别之间的符合程度进行度量，从而实现对未知类别样本的拒判。待识别的样本到达时，分别通过 $C$ 种情感的 GMM 模型，得到 $C$ 个 GMM 似然概率密度值，以 GMM 似然概率密度值映射到 0 到 1 之间作为第 i 个样本归属于第 j 个情感类别的隶属度 $\mu_j(\boldsymbol{x}_i)$：

$$\mu_j(\boldsymbol{x}_i) = \frac{\arctan\left(p(\boldsymbol{x}_i | \lambda_j)/10\right)}{\pi/2} \tag{5-41}$$

其中采用的投影函数为，

$$y = \frac{\arctan\left(x/10\right)}{\pi/2} \tag{5-42}$$

函数图形如图 5.5 所示

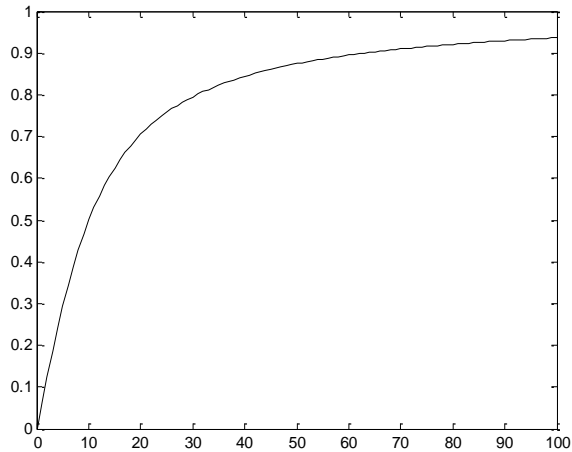

图 5.5 映射函数图形

对于第 j 个情感类别的所有可能的样本构成的模糊集 $\boldsymbol{E}_j = \{\boldsymbol{x}_1, \boldsymbol{x}_2, ..., \boldsymbol{x}_n\}$，其隶属度分别为 $\mu_j(\boldsymbol{x}_1), \mu_j(\boldsymbol{x}_2), ..., \mu_j(\boldsymbol{x}_n)$，令其模糊熵为 $e(\mu_j(\boldsymbol{x}_i))$，则 $e(\mu_j(\boldsymbol{x}_i))$ 应该满足[69]：

1) $e(\mu_j(\boldsymbol{x}_i))$ 随 $\mu_j(\boldsymbol{x}_i)$ 的增加而减少；

2) 当 $\mu_j(\boldsymbol{x}_i)$ 为 1 时，$e(\mu_j(\boldsymbol{x}_i))$ 为 0；

3) 两个独立的模糊集合的熵应该满足可加性。

这里可加性是一个严格的条件，只有满足可加性，模糊集合的熵才能唯一确定。对于独立的两个情感类别的模糊集合 $\boldsymbol{E}_j$、$\boldsymbol{E}_k$，其积为[69]：

$$E_j E_k : \mu_{jk}(\boldsymbol{x}_i) = \mu_j(\boldsymbol{x}_i)\mu_k(\boldsymbol{x}_i) \tag{5-43}$$

集合 $E_j E_k$ 的熵定义为：

$$e(\mu_{jk}(\boldsymbol{x}_i)) = e(\mu_j(\boldsymbol{x}_i)) + e(\mu_k(\boldsymbol{x}_i)) \tag{5-44}$$

类似于随机熵的证明，可以得到满足上面三个条件的模糊熵的表达式为[69]：

$$e(\mu_j(\boldsymbol{x}_i)) = -K \ln \mu_j(\boldsymbol{x}_i) \tag{5-45}$$

其中 $K$ 是大于 0 的数。将式(5-41)代入得，第 i 个样本归属于第 j 个情感类别的模糊熵为，

$$e(\mu_j(\boldsymbol{x}_i)) = -K\left(\ln \arctan\left(p(\boldsymbol{x}_i|\boldsymbol{\lambda}_j)/10\right) - \ln(\pi/2)\right) \tag{5-46}$$

对第 i 个待识别样本的 $C$ 个似然概率值构成的判决集合的平均模糊熵评价为，

$$S(\boldsymbol{x}_i) = \frac{1}{C}\sum_{j=1}^{C}\mu_j(\boldsymbol{x}_i)e(\mu_j(\boldsymbol{x}_i)) \tag{5-47}$$

将式（5-46）代入有，

$$S(\boldsymbol{x}_i) = -\frac{2K}{\pi C}\sum_{j=1}^{C}\arctan\left(p(\boldsymbol{x}_i|\boldsymbol{\lambda}_j)/10\right)\left(\ln \arctan\left(p(\boldsymbol{x}_i|\boldsymbol{\lambda}_j)/10\right) - \ln(\pi/2)\right) \tag{5-48}$$

对 N 个待识别情感的 GMM 模型，可以得到 N 个 GMM 似然概率密度值，分别代表样本与 N 个情感类别的符合程度。似然概率密度值构成的判决集合的模糊熵越高表示样本属于待识别情感的不确定程度越大，当模糊熵超过一定阈值 $Th$ 时则发生拒判，常数 $K$ 取 $\pi/2$。

$$S(\boldsymbol{x}_i) > Th \tag{5-49}$$

将式(5-48)代入即，

$$\frac{1}{C}\sum_{j=1}^{C}\arctan\left(p(\boldsymbol{x}_i|\boldsymbol{\lambda}_j)/10\right)\left(\ln(\pi/2) - \ln \arctan\left(p(\boldsymbol{x}_i|\boldsymbol{\lambda}_j)/10\right)\right) > Th \tag{5-50}$$

其中 $Th$ 为实验中确定的模糊熵阈值，在本实验中设为 0.11。阈值的选取既要保证待识别的目标情感类别得到正确的识别，又要兼顾未知的不确定的情感样本得到拒判。阈值设的过低则

会导致待识别的情感如烦躁、喜悦和中性等遭到拒判，识别率下降。阈值设的过高则会导致未知情感类别样本的误判，非烦躁、喜悦或中性的未知样本，被误判为其中的某一类情感。

### 5.3.2 未知情感类别的据判实验

模糊熵阈值的设置关系到样本的拒判，模糊阈值设定的过低，则对不确定样本的拒判效果不明显。模糊阈值设定的过高，则拒判的过多，会使得系统平均识别率降低。当部分样本离已知的情感模型距离较远时需要拒判，同时拒判也会使得某些测试样本不能得到正确识别。所以应该在保证烦躁、自信、疲劳、喜悦和中性等情感能够获得满意的识别率的前提下，调节模糊熵阈值。当平均识别率发生明显的下降时，此时的阈值为上限，在本实验中模糊熵设为 0.11。

实验中依然采用了实验数据集 A 中的数据，但是加入了大量的（500 条）未知情感类别的语料，测试与训练数据比例为 9:1。未知情感数据不需要进过 GMM 训练，直接可以进行识别测试实验。

特征降维部分采用 3.4 节中的方法，PCA 压缩中截取前 20 个维度的分量用于 LDA 降维。高斯混合模型的混合度设为 32，采用 K 均值聚类的方法进行初始化，通过 EM 算法迭代进行参数估计，其中的迭代次数上限设为 50 次。

**表 5.2 加入未知情感后的识别率（%）**

|  | 烦躁 | 喜悦 | 自信 | 疲倦 | 中性 | 拒判 |
|---|---|---|---|---|---|---|
| 烦躁 | **80.5** | 0.0 | 5.4 | 4.7 | 2.8 | 6.6 |
| 喜悦 | 0.0 | **78.9** | 3.6 | 2.2 | 7.3 | 8.0 |
| 自信 | 6.2 | 8.1 | **70.1** | 3.3 | 5.2 | 7.1 |
| 疲倦 | 7.1 | 2.2 | 3.7 | **66.5** | 11.8 | 8.7 |
| 中性 | 1.2 | 4.7 | 6.9 | 7.2 | **71.7** | 8.3 |
| 未知情感 | 6.1 | 8.6 | 7.4 | 6.0 | 6.1 | **65.8** |

## 5.4 说话人聚类与特征规整化技术

### 5.4.1 问题概述与算法思路

语音情感识别技术的研究正在从以往的实验室条件转向真实世界中的实际应用[107, 108]。以往的情感识别研究往往是依据表演方式采集的语料库，其中的情感类别数量较少，大

部分为基本情感类别，说话人的数量也相对较少。而在一些实际应用中，需要涉及到大量非特定说话人的情感语音，这就需要情感识别系统具有非特定说话人的鲁棒性。

以话务中心的语音数据处理为例，在银行、电信等大型服务行业领域，客户的满意度是一个重要的业务指标，因此在话务中心需要对客服通话进行录音，以便于分析和考核服务质量。然而对于大量的情感语音，很难进行人工听辨，通过自动识别的方式则可以快速的对录音数据进行筛选，识别出客户的情感信息。在这样的应用中，涉及到大量的非特定说话人，由于情感的个性化差异较大，会导致情感特征的复杂度增加，情感建模的困难加大，因此有必要研究非特定说话人的特征规整化技术。

很多规整化技术都可以用于提高一个识别系统的性能，在说话人识别、语种识别和自动语音识别中有很多的应用[109-111]。在说话人识别技术中，倒谱均值减法（Cepstral Mean Subtraction，CMS）经常用于镁尔倒谱系数，以降低信道方差。还有些研究者[110]在说话人识别中考虑了情感因素的影响，并且提出了情感规整化技术来提高说话人识别系统的性能。

基于性别差异的规整化[27，30，112]也是一类常见的降低说话人之间的特征差异的规整化技术，近年来不少文献将性别规整化用于语音情感识别系统中，获得了较好的性能提升效果。

Sethu 等人[111]较早的研究了语音情感识别系统中的说话人规整化问题，他们提出的特征规整化技术使得识别率平均提高了百分之六左右。然而他们的实验中涉及到的说话人数量较少，仅有七人。Vlasenko 等人[109]，在 SUSAS 情感语音库和柏林库上的实验中应用了说话人规整化技术（Speaker Normalization，SN）。SN 技术在柏林库上获得了百分之五的识别率提升。然而在很多实际场合中，说话人的身份是未知的，这就给 SN 技术的应用带来了困难。

本节中，我们提出了一种基于说话人聚类的情感特征规整化方法。我们通过说话人模糊聚类，自动获得情感语料中的说话人的身份信息，据此进行说话人规整化，以达到降低大量说话人带来的差异。首先，我们提取出能够反映说话人信息的特征空间，在此说话人特征空间内进行模糊聚类，获得"伪说话人"聚类。每一条样本按照其相似程度划分到不同的伪说话人分组中。其次，根据每条样本的伪说话人组别信息，进行情感特征的规整化处理，采用传统的说话人规整化方法，并且在规整化后的数据中加入相应组别的模糊隶属度信息。通过这样的说话人聚类规整化处理，情感特征空间中的样本分布更加清晰有效，降低了大量说话人带来的特征差异，适合应用于非特定说话人情感识别。算法的流程图如图 5.6 所示。

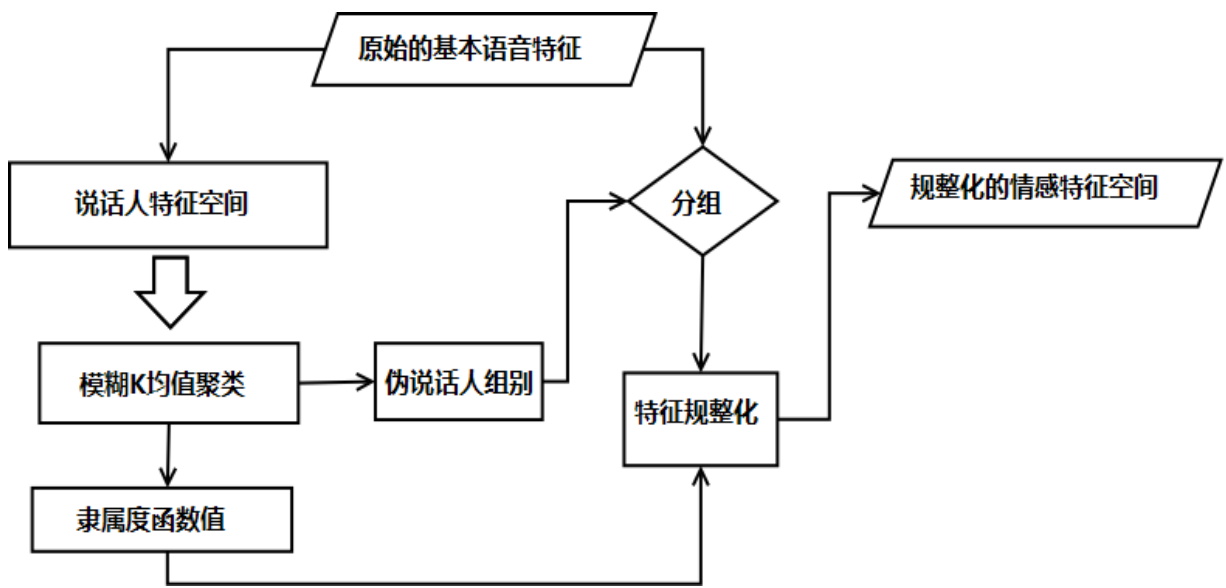

**图 5.6 基于说话人模糊聚类的情感特征规整化算法流程图**

### 5.4.2 特征规整化算法

说话人规整化可以提高语音情感识别系统的性能。语音中的情感信息与很多因素有关，如：性格、性别、年龄、文本内容等。如果将情感语料按照不同的因素归类，再进行规整化，则可以降低数据当中的差异，使得建模更为容易和可靠。

说话人规整化可以通过下面的公式进行[63]：

$$f_{u,id}^{'} = \frac{f_{u,id}(n) - \overline{f_{u,id}}}{\sqrt{\frac{1}{N_{u,id}-1}\sum_{m=1}^{N_{u,id}}\left(f_{u,id}(m) - \overline{f_{u,id}}\right)^2}} \tag{5-51}$$

其中 u 表示第 u 个特征值，id 代表说话人的身份，N 为同一个说话人的样本数量，$\overline{f_{u,id}}$ 则是给定说话人的数据样本的中心，由式（5-52）得到：

$$\overline{f_{u,id}} = \frac{1}{N_{u,id}-1}\sum_{n=1}^{N_{u,id}} f_{u,id}(n) \tag{5-52}$$

图 5.7 和图 5.8 所示，为第一共振峰均值在说话人规整化之前和之后的分布曲线。我们可以看到在进行说话人规整化之后，情感状态之间的可区分度得到了增加。

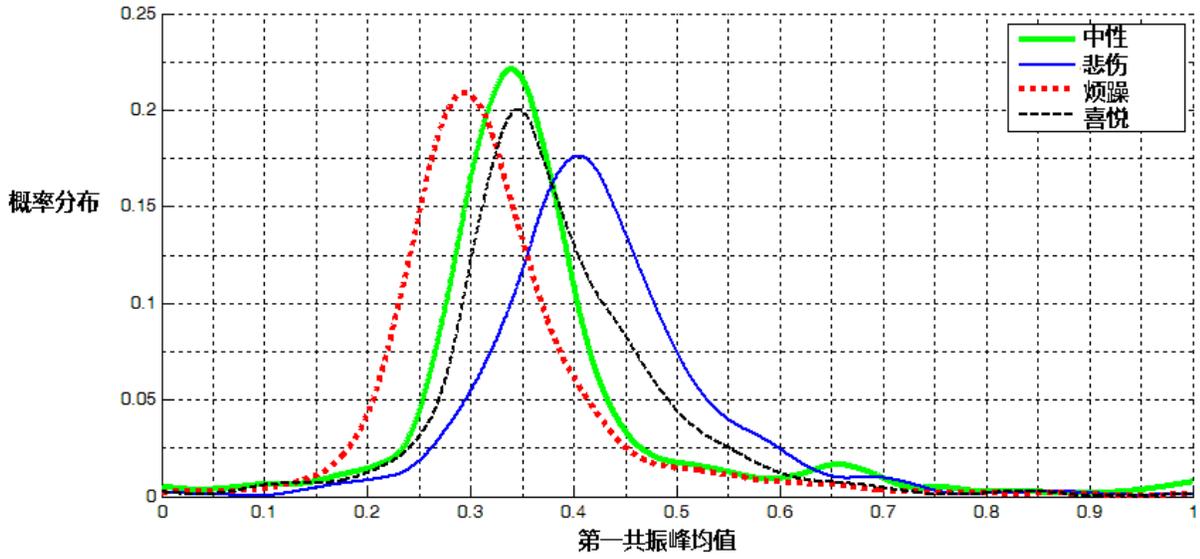

图 5.7 说话人规整化前的第一共振峰均值

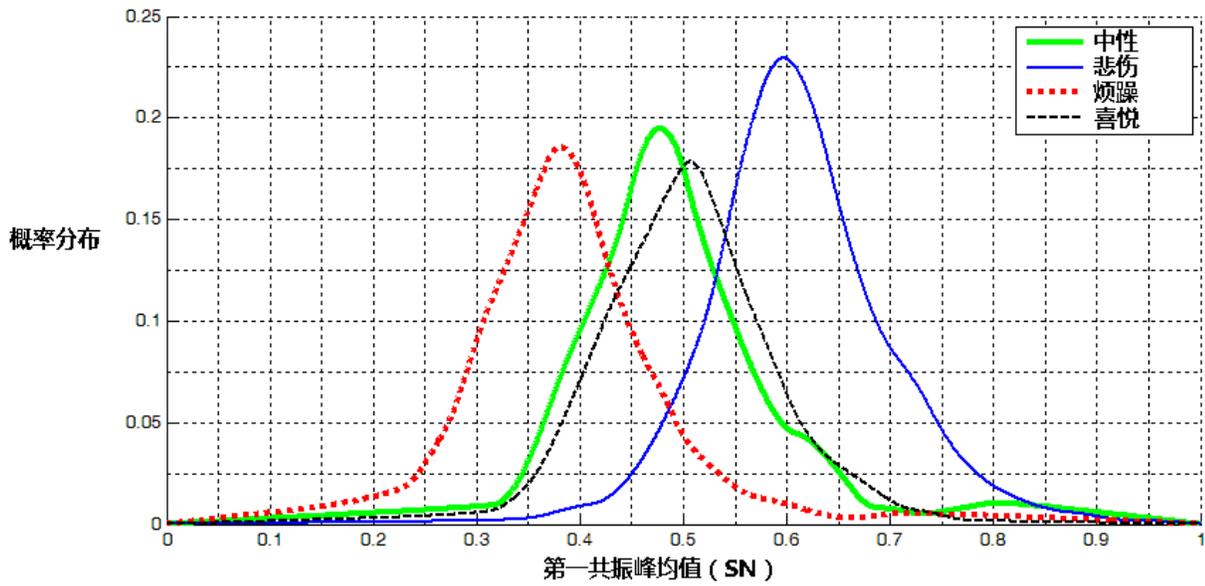

图 5.8 说话人规整化（SN）后的第一共振峰均值

### 5.4.3 模糊聚类算法

在非特定说话人的情感识别中，说话人的数量较大，说话人的身份是未知的变量。因此，将普通的说话人规整化算法直接用于非特定说话人中，就缺少了说话人身份的信息。本节中我

们考虑采用模糊说话人聚类的方法，以获得说话人聚类分组，以组别信息来代替规整化算法中的说话人身份。下面我们首先来介绍模糊聚类算法。

聚类问题可以表示为一个非线性的优化问题[113]：

$$\min J\left(\boldsymbol{W}, \boldsymbol{P}\right) = \sum_{t=1}^{k} \sum_{j=1}^{n} w_{ij} d^2\left(\boldsymbol{x}_j, \boldsymbol{p}_i\right)$$
$$s.t. \boldsymbol{W} \in \boldsymbol{M}_h \tag{5-53}$$

其中 $\boldsymbol{W}$ 是 k-划分矩阵，$w_{ij}$ 是类别标号，$\boldsymbol{P}$ 为 k 个聚类原型，$\boldsymbol{M}_h$ 为硬划分空间。

$$\boldsymbol{P} = \left(p_1, p_2, \cdots, p_k\right)^T \in \boldsymbol{R}^{kn} \tag{5-54}$$

样本与聚类原型之间的距离为，

$$d^2\left(x_j, p_i\right) = \left(x_j - p_i\right) \boldsymbol{A} \left(x_j - p_i\right)^T \tag{5-55}$$

在这里我们采用欧式距离，$\boldsymbol{A}$ 为单位阵。硬划分空间可以表示为：

$$\boldsymbol{M}_h = \left\{\boldsymbol{W} \in \boldsymbol{R}^{kn} \mid w_{ij} \in \{0,1\}, \forall i, j; \sum_{i=1}^{k} w_{ij} = 1; 0 < \sum_{j=1}^{n} w_{ij} < n, \forall i\right\} \tag{5-56}$$

Ruspini[114]将隶属度函数从离散的变量（0，1）扩展到连续的值域范围[0,1]中，得到了一个模糊划分空间：

$$\boldsymbol{M}_f = \left\{\boldsymbol{W} \in \boldsymbol{R}^{kn} \mid w_{ij} \in [0,1], \forall i, j; \sum_{i=1}^{k} w_{ij} = 1; 0 < \sum_{j=1}^{n} w_{ij} < n, \forall i\right\} \tag{5-57}$$

由此可得，模糊 K 均值聚类问题可以表示为[113]：

$$\min J_\alpha\left(\boldsymbol{W}, \boldsymbol{P}\right) = \sum_{t=1}^{k} \sum_{j=1}^{n} \left(w_{ij}\right)^\alpha d^2\left(\boldsymbol{x}_j, \boldsymbol{p}_i\right)$$
$$s.t. \boldsymbol{W} \in \boldsymbol{M}_f \tag{5-58}$$

其中，$\alpha \geq 0$，为平滑参数，通常在实验中设置为 1 到 5 之间。

### 5.4.4 情感特征上的伪说话人聚类与特征规整化算法

从情感语料中提取了 481 个基本声学特征参数，如第三章中的表 3.1 和表 3.2 所示。这里我们构造的特征参数，是针对语音情感识别的，并不是为说话人识别设计的。在整个语料上的统计特征，适合于语音情感识别，在每帧上提取的镁尔倒谱参数，适合于说话人识别。

我们提取的这些特征参数，会受到说话人因素的影响，说话人数量的增加，会使得情感特征中的方差增大。我们希望通过说话人聚类规整化，来减小说话人因素对这些特征的影响。因此，我们首先将这些原始的声学特征变换到一个对说话人敏感的特征空间中，将说话人因素的影响体现出来，再通过聚类算法对数据样本进行说话人分组，利用说话人组别信息，在原始的特征空间中进行规整化处理。

在原始的特征上通过 PCA 与 LDA 变换，使得每个说话人之间的可区分度最大化，我们可以得到说话人敏感的特征空间，如图 5.9 所示。在说话人特征空间中，情感数据样本的分布反映出其受到说话人因素影响的大小，样本聚合之处为同样的说话人的数据，样本分离之处代表了不同的说话人的数据。同过模糊聚类算法，我们进行说话人的聚类，获得说话人身份信息，从而可以进行规整化处理。但是通过聚类得到的说话人身份信息有一定的错误分类存在，因此可以称其为"伪说话人聚类"。伪说话人聚类是在语音情感特征上根据说话人因素的影响大小进行的情感样本自动聚类。

在完成了说话人的聚类后，将每条语料的聚类组别代替说话人身份，在原始的 481 维特征空间中进行规整化处理：

$$f_{u,v}^{'} = \frac{f_{u,v}(n) - \overline{f_{u,v}}}{\sqrt{\dfrac{1}{N_{u,v}-1}\sum\limits_{m=1}^{N_{u,v}}\left(f_{u,v}(m) - \overline{f_{u,v}}\right)^2}} \tag{5-59}$$

其中 u 表示第 u 个特征值，v 代表说话人聚类的组别，N 为同一个说话人的样本数量，$\overline{f_{u,v}}$ 则是给定说话人的数据样本的中心，由下式得到：

$$\overline{f_{u,v}} = \frac{1}{N_{u,v}-1}\sum\limits_{n=1}^{N_{u,v}} f_{u,v}(n) \tag{5-60}$$

完整的非特定说话人特征规整化算法步骤如表 5.3 所示。

### 表 5.3 非特定说话人特征规整化算法

**输入**：说话人敏感的特征 $\boldsymbol{x}^s$，481 维的原始特征 $\boldsymbol{x}^o$，$k=14, \alpha=2$

**输出**：规整化的情感特征 $\boldsymbol{x}^n$

Step 1: 模糊 K 均值聚类算法的初始化，$w_{ij} = \dfrac{w_{ij}}{\sum\limits_{i=1}^{k} w_{ij}}$

Step 2: 计算模糊聚类中心与隶属度函数

    **for** t=1 to T

$$p_i = \frac{\sum\limits_{j=1}^{n} \left(w_{ij}\right)^{\alpha} \boldsymbol{x}_j^s}{\sum\limits_{j=1}^{n} \left(w_{ij}\right)^{\alpha}}$$

$$w_{ij} = \frac{1}{\sum\limits_{m=1}^{c} \left(\dfrac{d_{ij}}{d_{mj}}\right)^{1/(\alpha-1)}}$$

    $t = t+1$

    **end for**

Step 3: 计算模糊聚类的分类标签集

$$\boldsymbol{L}_l = \left\{ j \mid \max\left(w_{ij}\right) = l \right\}, \quad l = 1, 2, \cdots, k$$

Step 4: 获得原始样本空间中的说话人聚类分组：

    $\forall j \in \boldsymbol{L}_l \longrightarrow \boldsymbol{x}_j^s \in \boldsymbol{\psi}_l \longrightarrow \boldsymbol{x}_j^o \in \boldsymbol{\psi}_l^{'}$，

    其中 $\boldsymbol{\psi}_l$ 为说话人特征空间中的聚类集合，$\boldsymbol{\psi}_l^{'}$ 为原始特征空间中的聚类集合

Step5: 扩展原始特征向量：$\boldsymbol{x}_j^o = \left[ x_{j,1}^o, x_{j,2}^o, \cdots x_{j,D}^o, w_{lj} \right]$

    其中 D=481

Step6: 计算说话人聚类的样本中心

$$\overline{x_{j,u,l}^o} = \frac{1}{N_l} \sum_{j=1}^{N_l} x_{j,u,l}^o$$

Step7: 特征规整化

$$x_{j,u,l}^n = \frac{\overset{\circ}{x}_{j,u,l} - \overline{x_{j,u,l}^o}}{\sqrt{\dfrac{1}{N_l-1} \sum\limits_{m=1}^{N_l} \left(\overset{\circ}{x}_{m,u,l} - \overline{x_{j,u,l}^o}\right)^2}}$$

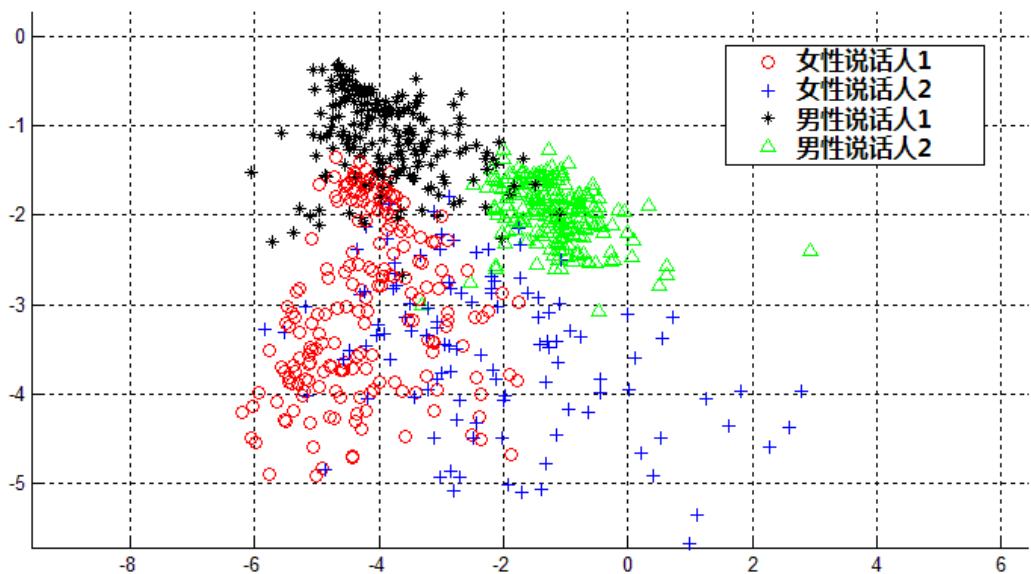

**图 5.9 从情感特征中获得的说话人特征空间**

### 5.4.5 实验结果

以往的情感数据库中所包含的说话人数量较少，既包含实用语音情感，又包含大量说话人的数据库更少。为了在大量的说话人中进行情感识别，实验数据集选择了富士通公司的非特定说话人语音情感数据库。该数据库包括了 51 名说话人（其中 28 名为女性），说话人的年龄段为二十至三十五岁之间。文本内容为无情感倾向性的短句。由于说话人的数量较大，因此数据中包含了更为丰富的情感表达模式，反映了不同说话人之间的性格差异、说话习惯等因素带来的情感变化。语音数据的录制环境为：采样频率为 48kHz，16 比特量化，单声道录制。

在本节的实验中，进行了烦躁、中性、悲伤和喜悦四种情感类型的识别，每种情感 2000 条样本，总计 8000 条样本，来自 51 个不同的说话人。实验中将数据集随机分成十份，训练数据集与测试数据集比例为 9：1，轮换测试后求取平均值做为识别测试的结果，实验结果如表 5.4 和表 5.5 所示。

为了验证非特定说话人的鲁棒性，进行了"Leave-one-speaker-out"的测试，实验结果如表 5.6 和表 5.7 所示。训练中，将被测试的说话人的数据剔除，将其余数据用于训练。在测试时，被测试的说话人没有在训练集中出现过，以达到非特定说话人的测试目的。训练数据集与测试数据集比例大约为 50：1。

表 5.4 五十一个说话人的交叉验证的测试结果

|  | 烦躁 | 喜悦 | 悲伤 | 中性 |
|---|---|---|---|---|
| 烦躁 | **83.0** | 2.6 | 13.8 | 0.6 |
| 喜悦 | 0.6 | **79.9** | 8.3 | 11.2 |
| 悲伤 | 7.1 | 4.2 | **78.7** | 10.0 |
| 中性 | 0.0 | 11.5 | 7.3 | **81.2** |

表 5.5 五十一个说话人的交叉验证的测试结果（特征规整化后）

|  | 烦躁 | 喜悦 | 悲伤 | 中性 |
|---|---|---|---|---|
| 烦躁 | **87.1** | 0.7 | 12.1 | 0.1 |
| 喜悦 | 1.1 | **83.9** | 6.8 | 8.2 |
| 悲伤 | 3.5 | 5.5 | **86.1** | 4.9 |
| 中性 | 1.2 | 9.5 | 4.6 | **84.7** |

表 5.6 非特定说话人的测试结果

|  | 烦躁 | 喜悦 | 悲伤 | 中性 |
|---|---|---|---|---|
| 烦躁 | **84.2** | 2.0 | 13.3 | 0.5 |
| 喜悦 | 0.4 | **82.1** | 8.7 | 8.8 |
| 悲伤 | 8.2 | 4.2 | **79.5** | 8.1 |
| 中性 | 0 | 9.7 | 7.3 | **83.0** |

表 5.7 非特定说话人的测试结果（特征规整化后）

|  | 烦躁 | 喜悦 | 悲伤 | 中性 |
|---|---|---|---|---|
| 烦躁 | **87.4** | 0.7 | 11.7 | 0.2 |
| 喜悦 | 2.3 | **85.2** | 6.5 | 6.0 |
| 悲伤 | 4.6 | 6.8 | **81.4** | 7.2 |
| 中性 | 1.0 | 9.1 | 4.6 | **85.3** |

从识别率实验的结果可以看到，采用本节中的特征规整化算法之后，识别率有了明显的提高。在特征空间中的样本分布中还可以进一步看到本节中算法的效果，图 5.10 和图 5.11 所示，为规整化前后的四种情感的样本分布。可以看到通过在说话人空间中聚类并规整化后，四种情感的可区分度得到了提高。

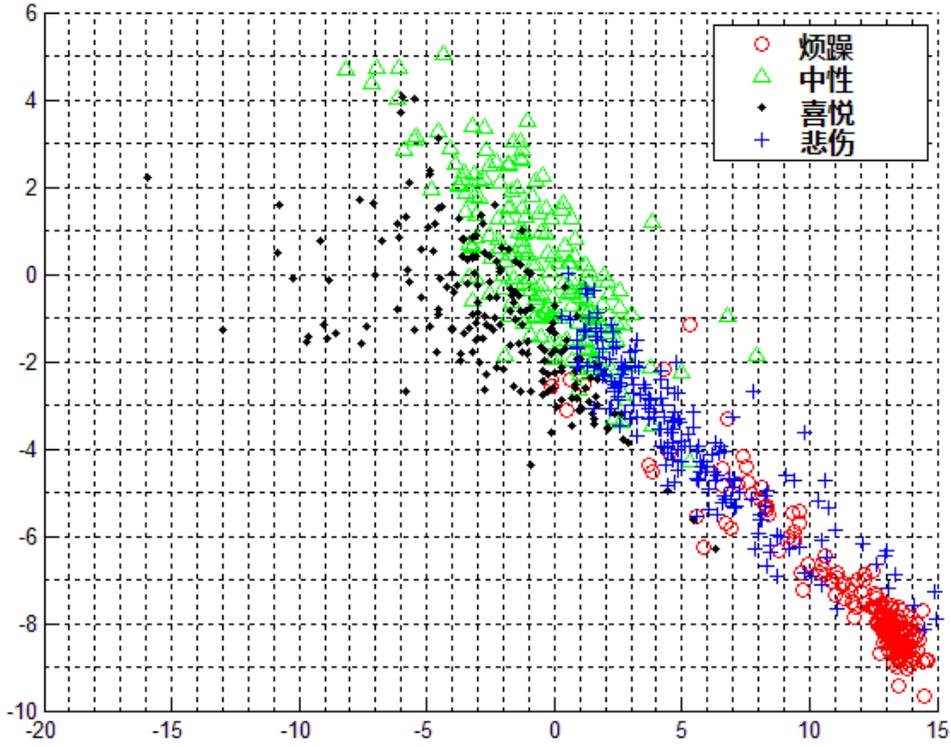

图 5.10 特征规整化前的情感样本分布

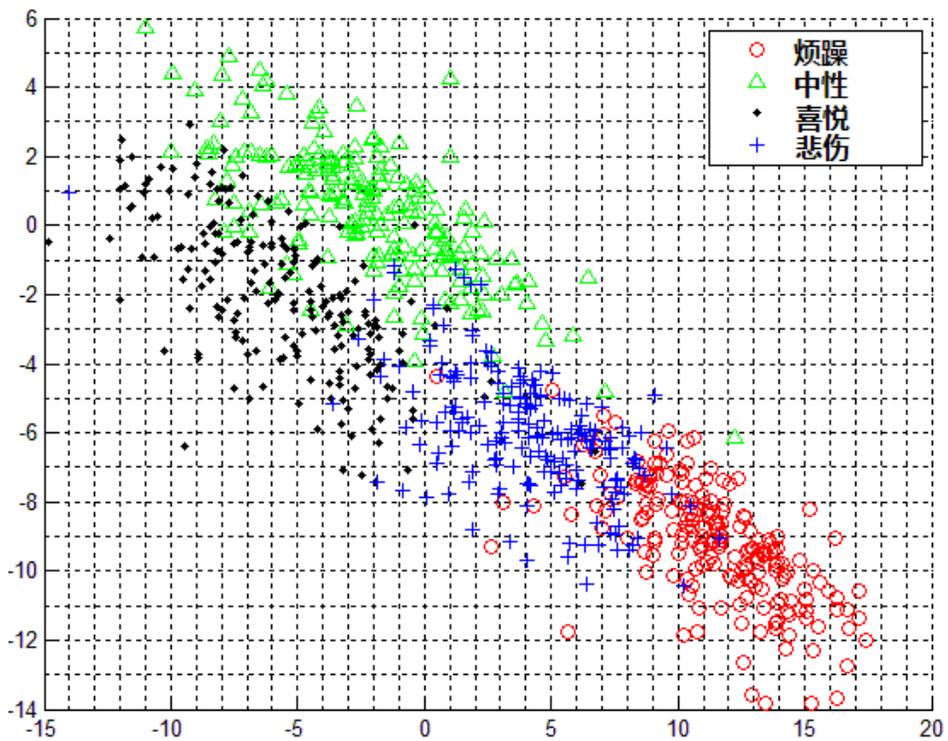

图 5.11 特征规整化后的情感样本分布

原始的特征首先进行了 PCA 变换，再进行了 LDA 变换，详细的步骤在第三章特征分析和降维中已经讨论过。可以看到烦躁情感被较好的分离开来，说明所提取的 481 维基本特征对烦躁的区分能力比较好。悲伤情感中的内部数据差异较大，中性情感和喜悦情感的样本靠得比较近。通过规整化算法处理后，不同说话人带来的情感表达之间的差异得到了削减，导致在 LDA 变换后的特征空间内各类样本之间的可区分度得到了提高。

### 5.4.6 关于参数设置的讨论

在特征降维部分，需要设置 PCA 的截断维数。我们选择截取 PCA 变换的前 20 个维度，以用于 LDA 变换。LDA 变换会受到训练数据量的制约，如果维度过高，而训练样本过少，则会造成计算的协方差矩阵为奇异矩阵，影响算法性能。我们采用高斯混合模型进行情感建模，在某些说话人识别的应用中，GMM 的混合度可能高达上千个，而在情感识别中，则低很多。高斯混合模型的混合度，我们设置成 4 至 128，其中，当 GMM 的混合度为 32 时，获得最佳的效果。高斯混合模型的其它参数，权值、均值向量、协方差矩阵等，通过 EM 算法进行估计，EM 算法的迭代次数设置为 50。模糊 K 均值聚类算法中，平滑参数 $\alpha$ 设置为 2，将样本聚类到 14 个伪说话人类别中时，所获得的识别效果最好。

## 5.5 双模态融合情感识别

在本节当中我们采用双模态的情感识别方法，首先通过情感诱发实验进行心电信号和语音信号的双通道数据采集。采用噪声刺激和观看喜剧影视片段的诱发手段，分别诱发被试人员的烦躁情感状态和喜悦情感状态。通过接触式的心电信号传感器和非接触式的人声麦克风进行双模态数据的同步采集，根据同一时段的两个通道的数据制作成双模态的情感数据，并进行了相应的情感标注，包括烦躁、喜悦、中性三个情感状态。

在双模态的情感数据组上，研究了特征层次的融合算法和决策层次的融合算法，比较了两者的识别率差异。实验表明，通过合适的融合算法，双模态情感识别系统的性能要优于单模态的系统。双模态的系统在一个通道信号缺失的情况下仍然能够正常工作，说明双模态情感识别具有较好的鲁棒性。

### 5.5.1 多模态情感识别的研究现状

目前，情感识别向多模态方向发展，单一的依靠表情、语音或者生理参数来进行情感识别的研究取得了一定的成果，但是如何将这些不同性质的情感信号融合，达到信息上的互补，从而建立一个鲁棒性强、识别率高的系统还需要进一步深入研究。

利用人脸表情、语音、眼动、姿态和生理信号等多个通道的情感信息之间的互补性来提高分类器的识别性能，除了能够使分类器的识别率提高以外，在实际噪声环境下，当某一个通道的特征受到干扰或缺失时，还能使分类器具有良好的鲁棒性。Hoch 等[115]通过融合语音与表情信息在车载环境下进行了正面（喜悦）、负面（生气）与中性等三类情感状态的识别。Busso 等[116]分析了单一的语音情感识别与人脸表情识别在识别性能上的互补性，分别通过特征层融合与决策层融合进行基于多模态信息的情感识别。Wagner 等[117]通过融合肌动电流、心电、皮肤电阻、呼吸等四个通道的生理参数，获得了 92%的融合识别率。

本节以语音信号与心电信号为基础，对烦躁、喜悦和中性等三种情感状态进行了识别，研究了心电信号与语音信号的融合情感识别及相应的融合算法和情感特征。通过在特征层面和判决层面进行融合，比较了基于语音信号与心电信号两种单模态分类器的识别率及其之间的互补性，并建立了基于多模态信息的分类器以提高情感识别性能。这种基于多模态信息的分类器在实际应用中具有重要意义，例如，在噪声等环境干扰下，当语音信号的采集受到影响时，生理信号为情感识别提供了重要的依据。此外，目前基于心电信号等生理参数的情感识别能分辨的情感种类较少，识别率相对较低，与语音特征融合后，可使识别性能得到了较大提高。

### 5.5.2 双模态的情感数据的采集

本节中介绍双通道的情感数据的采集，心电数据与语音数据在一个诱发实验中同时进行采集。双模态情感数据的获取可以用于建立更高鲁棒性的情感识别系统，当语音通道或者心电通道中的一个无法正常工作时，整个系统依然可以进行可靠的情感识别。例如，当环境噪声较大时，语音信号可能无法正常采集，而心电信号则不会受到噪音的干扰。在采集儿童的情感数据时，接触式的心电传感器可能不适合安置在儿童身上，则可以选择非接触式的录音方式。融合了两个通道的情感数据后，还可以提高情感识别系统的整体识别率。

针对心电数据采集的特点，没有延用第二章中计算机游戏诱发的方式，因为在手腕部位佩戴心电传感器电极后，不便于计算机游戏的操作。在采集心电信号的过程中尽量减少身体活动，以减少干扰。实验中，通过让被试人员在噪声环境下进行四则运算来诱发烦躁情感，通过观看喜剧片段诱发喜悦情感，并通过充分休息后采集平静状态下的数据。

实验过程如图 5.12 所示。参与实验的被试为 5 名男性和 5 名女性，年龄范围为 20 到 40 岁，健康状况良好，近期无药物服用。实验中要求被试人员读出指定的文本语句，录制烦躁、中性和喜悦等三种情感状态下的语音数据。在实验全过程中记录心电数据，并截取每条语音数

据开始前 30s 到结束后 30s 时间段内的心电数据，与相对应的语音数据绑定存储。由于情感的一次表现一般持续 1 分钟至 2 分钟，而 HRV 频谱等心电特征的提取一般需要至少 1 分钟的数据，因此在实验中截取 1 分钟至 2 分钟的心电数据作为一条样本。

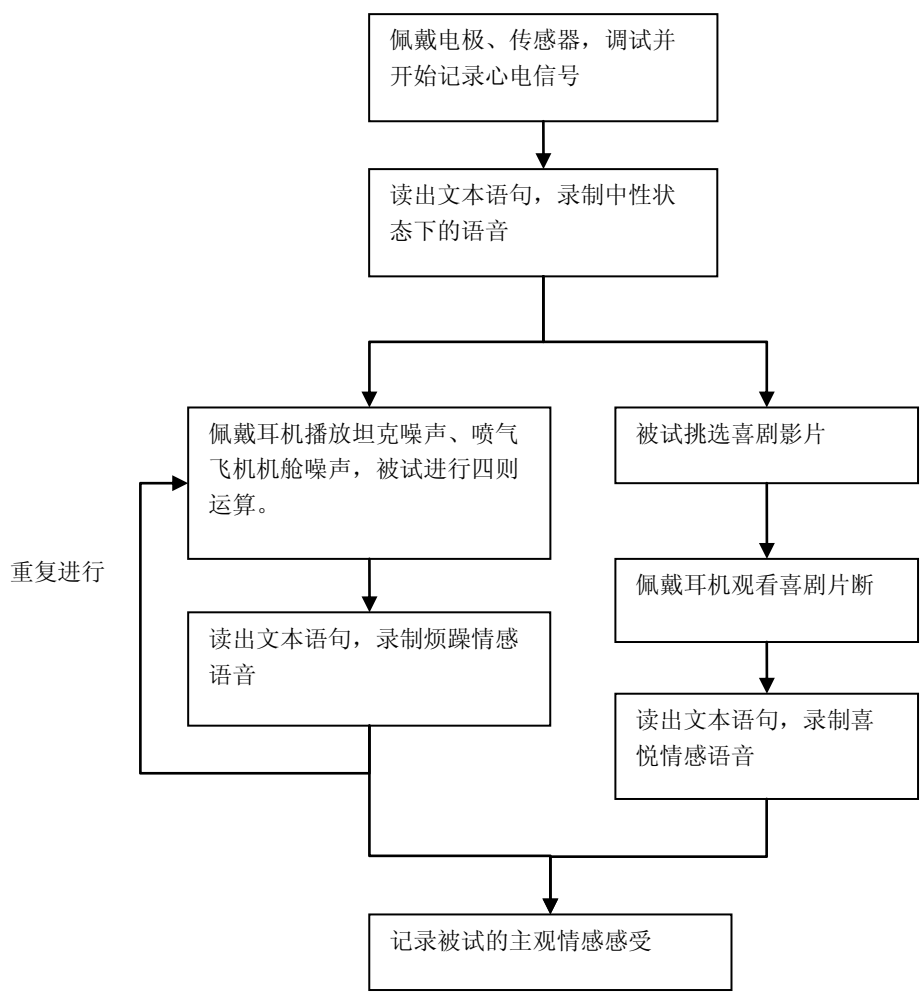

**图 5.12 双模态情感数据采集流程**

**表 5.8 双模态样本数量表**

| 传感器通道 | 情感类别 | 样本数量 | 被试人数 |
|---|---|---|---|
| 人声麦克风 | 喜悦 | 400 | 男性 5 人 |
| ECG 传感器 | 烦躁 | 400 | 女性 5 人 |
| | 中性 | 400 | |

### 5.5.3 情感特征提取

情感特征的优劣以及情感特征提取是否全面，直接影响到情感识别的性能。本节从语音信号与心电信号两个方面提取并构造了用于识别烦躁、喜悦与中性状态的特征。基于语音信号的情感识别研究相对较多，基音、能量、共振峰以及语速等参数是受到广泛认同的有效的语音情感特征。除了这些基本的语音情感特征外，还提取了谱能量分布、谐波噪声比（HNR）等方面的音质特征参数，用于加强对效价维度的区分能力（烦躁与喜悦在效价维度上差异较大）。本节中采用的语音特征包括韵律特征和音质特征等，与第三章中的相同，共 481 维。

目前，通过心电信号（ECG, Electrocardiogram）来进行情感识别的研究还较为缺乏，常用的心电情感特征有心率异常性（HRV, Heart rate variability）方面的时域、频域特征，更多有效的心电情感特征有待发掘，心电情感特征中的年龄差异等因素有待研究。本文除了提取常见的 HRV 特征外，还提取了心电信号的若干混沌特征，用于进行烦躁、喜悦与中性等三种情感的研究。

大多内脏器官都是受交感神经和副交感神经双重支配的，心电信号（ECG）也不例外。心脏的每次跳动都是由窦房结的起搏引起，窦房结内起搏细胞固有的节律性受自主神经调节，交感神经增快其自发激动，副交感神经减慢激动。研究表明，情绪的变化对心电信号有一定影响。心率变异性等指标被越来越多的研究者用于情绪的生理心理学研究中。

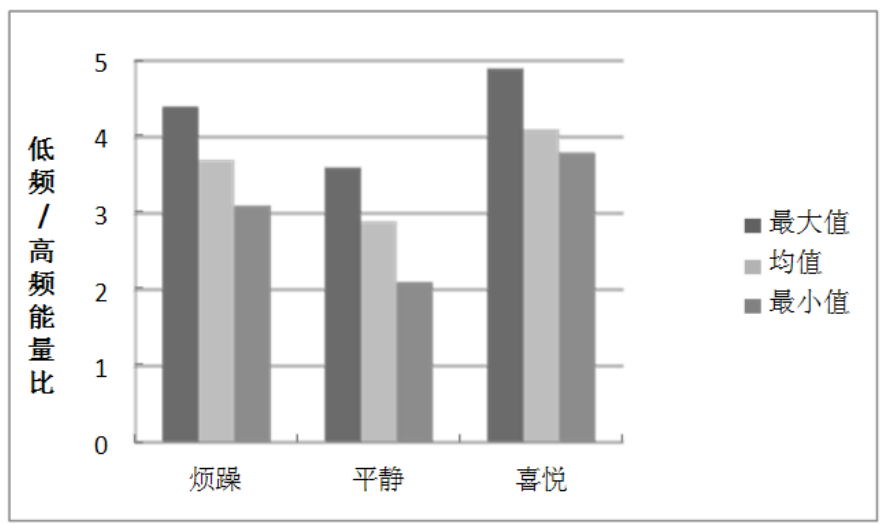

图 5.13 三种情感状态下的 HRV 特征分析

在 HRV 的频域分析中，短时 HRV 功率谱的高频成分（HF）认为是与呼吸同步的，可定量估计呼吸性心律失常，代表副交感神经活动指数，并可作为监测心脏迷走神经活动水平的定量指标；低频成分（LF）被认为是代表了交感神经活动的指数，随交感神经活动的增强而增加。低频/高频能量比则可作为评价心脏迷走－交感神经均衡性的定量指标，在一定程度上反映出情感状态的变化。三种情感状态下 HRV 的低频/高频能量比如图 5.13 所示。

本文中除了采用常见的 HRV、R 波、T 波等心电特征参数进行情感状态的分析外，还研究了心电的混沌特征，提取了心电的关联维数作为反映情绪变化的生理指标之一。ECG 信号不是一个单纯的周期信号，ECG 信号的非线性研究主要集中在非线性动力学参数的计算上，如分形维数、Lyapunov 指数等。

这里采用关联维数来描述心电信号的混沌特征。关联特征是从单变量时间序列中提取维数信息，表示系统在多维空间中的疏密程度，反映了系统点与点之间的关联程度。实际计算中心电信号的嵌入维数设定为 8 较为合理，采用 Grassberger-Procaccia 算法[118]（G-P 算法）得到的三种情感状态下的关联维数如表 5.9 所示。

对烦躁、喜悦和中性 3 种情感状态的识别，本文一共采用了以下 23 个心电特征参数：

特征 1-8：关联维数、Lyapunov 指数的最大值、最小值、均值、方差；

特征 9-12：RR 间期的最大值、最小值、均值、方差；

特征 13-15：HRV 的低频能量、高频能量、低频/高频能量比；

特征 16-23：T 波及 R 波能量的最大值、最小值、均值、方差。

**表 5.9 心电的关联维特征分析**

| 关联维特征 | 烦躁 | 中性 | 喜悦 |
|:---:|:---:|:---:|:---:|
| 均值 | 2.553 | 2.875 | 2.621 |
| 最大值 | 3.041 | 3.433 | 3.142 |
| 最小值 | 2.163 | 2.359 | 2.310 |

心电参数的提取算法可以参照相应的文献[119]。为了充分利用基于语音信号与心电信号的情感特征，下面将分别通过判决层融合算法和特征层融合算法来进行语音与心电的双模态数据的融合识别。

### 5.5.4 判决层融合算法

在判决层融合算法中，首先分别设计出基于语音信号的情感分类器和基于心电信号的情感分类器，将两个分类器依据一定的准则进行判决融合，得到最终的识别结果（见图 5.14）。

本节中待识别的情感类别包括烦躁、喜悦和中性三个类别。对于两种分类器，均采用高斯混合模型来进行每种情感类别的概率模型训练。

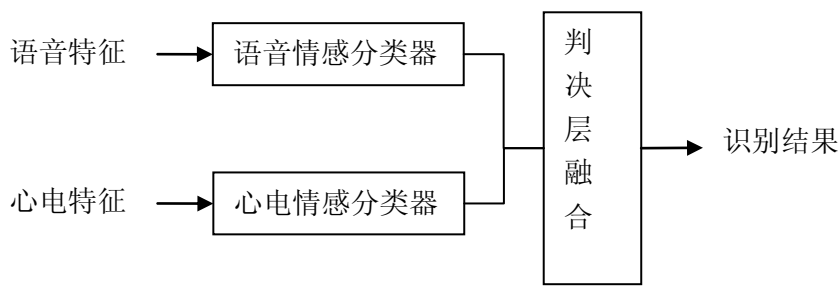

**图 5.14 判决层融合算法**

当存在噪声干扰时，语音分类器的性能会发生下降；当心电电极受到抖动、碰撞或者仪器内部的基线漂移干扰时，心电分类器的性能会发生下降。这就需要在选择判决层融合算法时，考虑评价各个子分类器在某一时刻的置信度，并根据分类器的输出置信度来进行融合判决。这里采用一种样本自适应的方法来衡量分类器对当前样本的判决是否可靠，对置信度高的分类器给予较高的融合权值，对于置信度低的分类器赋予较低的融合权值[56]。子分类器（语音分类器与心电分类器）给出的三类情感类别的 GMM 似然度分别记为 $P(X|\lambda_k)$，$k=1,2,3$ 时分别对应了三种情感类别。如果属于各个类别的 GMM 似然度基本相等或差别不大时，则认为该样本很可能处于概率分布模型的重叠区域，该子分类器的判决置信度较低；当分类器给出的似然度值较为分散时，则认为样本处于概率分布模型的非重叠区域，该子分类器的判决置信度较高。因此，每个子分类器的融合权值 $w_j$ 可表示为

$$w_j = \frac{\displaystyle\sum_{1<m<n<3}\left|\ln(P(X|\lambda_m))-\ln(P(X|\lambda_n))\right|}{\left|\displaystyle\sum_{k=1}^{3}\ln(P(X|\lambda_k))\right|} \tag{5-61}$$

式中，$j$ 为子分类器编号，且 $j=1,2$ 。

当分类器判决越可靠时，差值越大，$w_j$ 越大；反之当 $w_j$ 越小时，说明样本距离重叠区域越近，分类可靠性越差。定义了子分类器的融合权值后，对每个子分类器的判决进行加权融合，则最终的分类器融合判决输出为：

$$i^* = \arg\max \left\{ \sum_{j=1}^{2} w_j \, \mathrm{P}^j(\boldsymbol{X} \mid \boldsymbol{\lambda}_k) \right\} \tag{5-62}$$

### 5.5.5 特征层融合算法

在特征层融合算法中，并不设计多个单模态的情感分类器，而是将来自多个通道的大量情感特征通过特征选择算法进行优化选取，使用单个分类器对由语音数据与心电数据共同构成的最佳特征组进行分类识别，如图 5.15 所示。对双模态数据组成的特征，采用高斯混合模型进行训练与识别。

用于情感识别的原始语音特征包括韵律特征和音质特征等，与第三章中的相同，共 481 维；与情感状态有关的心电特征，包括非线性特征、时域特征、频域特征等 23 维。特征层融合的关键是对这些原始的情感特征进行优化选取，使得语音和心电双模态数据高效组合，提高特征与情感的相关程度。进行特征优化的方法通常有两种：包装法（wrapper）和滤波法（filter）。其中包装法与情感识别系统后端所采用的识别器相关性较大，采用不同的模式识别方法，会对特征选择的结果产生较大的影响[120]；滤波法能够在一定的准则下找出最佳意义上的情感特征，因而其通用性较好。采用 PCA 方法进行语音特征与心电特征的融合与降维。Ververidis 等[121]认为，4 到 5 个特征已经可以描述情感的分布。本节中的情感类别数量较少，仅为三种，因此实验中截取 PCA 变换的前 10 个维度来构成识别用的特征矢量。

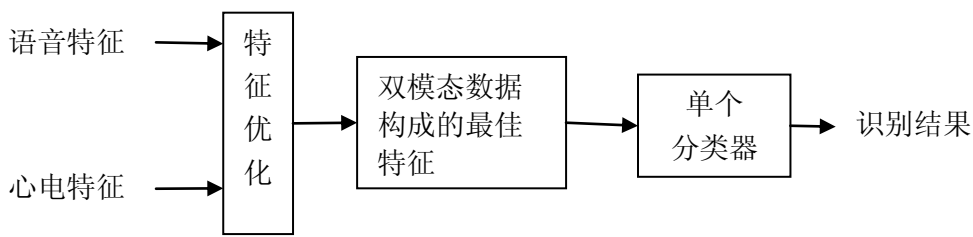

**图 5.15 特征层融合算法**

### 5.5.6 单模态与双模态实验结果

实验中采用 GMM 模型来拟合各类情感的概率分布结构。本节中的情感数据样本量较少，因此 GMM 混合度的设定也较小：混合度为 6，即高斯分量的个数为 6。训练样本集包含每种情感的 300 条语音样本与 300 条心电样本，测试集包含每种情感的 100 条语音样本与 100 条心电样本。实验中采用相同的训练集与测试集对单模态、多模态分类器进行测试，结果见表 5.10 和表 5.11。

**表 5.10 基于语音信号的单模态分类器的识别率（%）**

| 测试样本类别 | 烦躁 | 中性 | 喜悦 |
|---|---|---|---|
| 烦躁样本 | **83** | 8 | 9 |
| 中性样本 | 13 | **78** | 9 |
| 喜悦样本 | 6 | 12 | **82** |

**表 5.11 基于心电信号的单模态分类器的识别率（%）**

| 测试样本类别 | 烦躁 | 中性 | 喜悦 |
|---|---|---|---|
| 烦躁样本 | **71** | 15 | 14 |
| 中性样本 | 19 | **69** | 12 |
| 喜悦样本 | 11 | 15 | **74** |

由表 5.10 与表 5.11 可知，基于语音信号的单模态分类器的平均识别率达到 81%；烦躁情感在实际中具有重要的应用价值，其识别率为 83%，说明本文中采用的语音情感特征与烦躁情感的相关性较高，能够用于烦躁情感的识别。基于心电信号的单模态分类器对三种情感状态的区分能力较弱于，平均识别率略高于 71%，因此单纯依靠心电数据的情感识别在实际应用中会遇到一定的困难，需要同其他类型的情感数据相结合来进行多模态的情感识别，以提高识别率与可靠性。

在多模态的情感识别中，采用图 5.14 与图 5.15 所示的识别系统，通过判决层融合与特征层融合两种算法来进行情感识别。实验结果显示，相比单模态分类器，多模态分类器的识别性能有了显著提高。其中，判决层融合算法的平均识别率达到 89%，基于特征层融合的平均识别率则达到 91%。虽然基于心电信号的单模态分类器对情感的识别能力有限，但是心电信号提供了一

部分语音数据所不能替代的生理信息。通过加入心电数据后情感识别系统的性能得到了明显提高，相比传统的语音单模态识别系统，其平均识别率提高了10%。

两种融合算法的识别结果见表5.12和表5.13。可以看出，判决层融合算法在平均识别率上略低于特征层融合算法，后者对喜悦状态的识别率达到94%。对于烦躁状态的识别，两种融合算法的性能均较高，说明多模态融合算法获得了预期的效果。判决层融合算法的优势在于，每个分类器都是相互独立的，当某一通道的情感数据无法获取或质量较低时，判决层仍然能够进行情感识别，鲁棒性较高。特征层融合算法能够在一定条件下获得最佳的特征压缩与优化性能，在识别测试中识别率略高于判决层融合算法。

**表5.12 判决层融合算法的识别率（%）**

| 测试样本类别 | 烦躁 | 中性 | 喜悦 |
|---|---|---|---|
| 烦躁样本 | 89 | 9 | 2 |
| 中性样本 | 5 | 88 | 7 |
| 喜悦样本 | 8 | 2 | 90 |

**表5.13特征层融合算法的识别率（%）**

| 测试样本类别 | 烦躁 | 中性 | 喜悦 |
|---|---|---|---|
| 烦躁样本 | 90 | 6 | 4 |
| 中性样本 | 4 | 89 | 7 |
| 喜悦样本 | 4 | 2 | 94 |

### 5.5.7 结论

本节中我们过诱发手段采集了语音与心电的情感数据，建立了双模态的情感识别系统，并对语音情感特征和心电情感特征进行了分析与优化选择。以往的情感识别研究大多依靠单通道的数据来进行，如表情识别、语音情感识别等。通过将多种不同性质和不同来源的情感特征进行融合，实现多模态识别，从而提高系统的识别率和鲁棒性。将语音信号与心电信号融合，建立情感识别系统，能够有效地利用这两种不同的生物信号进行情感识别，实验结果表明，融合后系统的识别率得到了明显提升。判决层融合算法和特征层融合算法在实验中显示了不同的识别特性，其中特征层融合算法的平均识别率较高。何种语音特征和心电特征能够有效地反映出

烦躁等具有实际意义的情感，还需要进一步深入研究；心电情感特征与情感维度，（如激活维、效价维、控制维等）之间的关系也是今后值得探讨的问题。

选择心电信号进行双模态研究的原因，主要是希望心电信号能够从另一个角度提供情感信息。我们知道人类情绪的产生有其内在的一些生理机制。心电信号，作为一种内在的生理信号能够提供与语音、表情等外在行为不同的一个研究角度。语音和表情等外显的行为特征，可能较容易进行掩饰和控制，而心电等生理信号则不容易进行伪装和控制。从这方面来看，增加心电信号进行双模态情感识别，也同时增加了识别结果的可信度。

值得注意的是人脸表情识别的研究，近年来取得了不少进展[122-125]。在今后的研究中，可以进一步将人脸表情、语音情感与心电信号融合，进行多模态的情感识别。

# 第六章 跨语言与耳语音中的情感识别技术

本文建立了基于 GMM 的情感识别系统，将其用于三种与认知有关的实用情感类型的识别中。本文中实用语音情感识别的含义主要是指研究具有实际应用价值的、与认知过程相关的特殊情感类别的自动识别。以往的语音情感识别研究集中在一些基本的情感类别上，而这些具有特殊意义的情感类别，由于其在实际应用当中的价值，更加需要我们进行系统的研究。

本文中实用语音情感识别的另一层含义，是研究和解决实际条件下可能出现的技术难点，例如上一章中的噪声鲁棒性分析、未知情感类别的处理等等。在实际中可能出现各种干扰因素，导致系统无法正常工作，因此在实用条件下研究如何扩展系统的使用范围是一个有意义的课题。

本章主要讨论语音情感识别在实际应用中遇到的两个特殊的问题，希望能够将基于高斯混合模型的语音情感识别系统应用到跨语言的环境中，以及应用到耳语音这种特殊的畸变语音信号中：（1）以德语和汉语为例，研究跨语言的情感识别；（2）以耳语音信号为例，研究畸变语音下的情感识别。

通过本章的研究，我们分析并指出了将基于 GMM 的识别系统应用到这两个实际领域中所遇到的主要难点，以及可能的解决途径。

在 6.1 节中我们介绍跨语言语音情感识别的初步研究成果，6.2 节中我们讨论如何从耳语音信号中识别说话人的情感状态。由于与认知过程相关的情感数据的采集较为困难，目前烦躁、自信、疲倦等实用语音情感数据相对缺乏。耳语音数据库和柏林数据库上，只能够提供几种基本的情感类别数据。因此在本章中，我们只针对几种基本情感类别进行识别测试。

## 6.1 跨语言语音情感识别

### 6.1.1 问题的概述

情感的表达方式与很多因素有关，如年龄、性别、语境和文化背景等。因此语音情感识别比起其它的模式识别问题有特殊的困难。环境感知的方法可能是语音情感识别进入实际应用的一个关键途径。Tawari [28]分析了语音情感识别中的上下文信息，采用了基于性别差异化的情

感识别器。James A. Russell [25]从不同文化背景的角度研究了人脸表情识别的普遍性。Elfenbein 与 Ambady[126]研究了情感识别中的文化差异。

跨语言和跨文化的研究，一直是行为科学中的重要的研究内容。在情感识别研究领域，Smith 曾经研究过来自不同文化的被试人员对情感的识别能力[26]。在语音情感识别领域，目前还没有一个成熟的跨语言的识别系统。

本文中讨论的跨语言语音情感识别问题，是指在情感数据的语言差异性下的情感模型的同一性问题。我们对跨语言的语音情感识别问题的定义，是从训练数据和测试数据的同一性上来进行的。所谓同一性，就是不同对象之间具有相同的本质。训练数据和测试数据，必须是同一种情感本质在不同的数据样本中的表现，这是识别系统有可能正常工作的逻辑前提。以非特定说话人的情感识别为例，如果训练数据和测试数据来自同一个说话人同一种语言，则情感识别较为容易。再以噪声鲁棒性为例，如果训练数据是纯净的语音，而测试数据含有噪声，则会带来情感模型的不匹配，需要通过降噪等方法处理，使其显示出情感本质的同一性。而跨语言的语音情感识别，就是要研究在不同语言中情感的同一性本质。

本文对跨语言语音情感识别问题进行了初步的研究，分析了跨语言问题中的主要挑战和困难。主要贡献有： (a) 研究了在汉语和德语数据库上通用的语音情感特征； (b) 针对生气情感，在德语和汉语数据库上获得了一个通用的生气情感模型，在下文中的识别测试中获得了较好的效果。

跨语言语音情感识别中会遇到情感声学特征的差异的问题，在一种语言的数据库上研究获得的情感特征，在另一种语言的数据库上效果往往会变差。这是由于不同的语言发音特点差异导致的。我们知道在情感特征的研究和评选中，较大的困难在于排除音位因素的干扰，也就是需要选择出对情感变化敏感而对语义变化不敏感的声学特征。因此不同语言中发音特点的变化会对情感的声学特征有很大的影响，导致在不同语言的数据库上情感特征的不一致。

跨语言语音情感识别中的另一个重要因素，是不同文化背景带来的情感表达与理解的差异。在实际中伴随着跨语言情感数据出现的往往是来自不同地域和文化背景的说话人，因而跨语言情感识别问题本身就部分的包含了跨文化的情感识别问题。虽然有研究文献指出[25, 126]，人类对情感的理解是有一定的全球共通性的，来自不同语系的说话人对情感大致有着共同的理解，但是文化背景差异依然会导致语音情感表达的差异。这种差异带来的识别困难在下文中的跨语言测试实验中有所表现。

　　本文中进行的跨语言情感测试，是指训练数据来自一种语言，而测试数据来自另一种语言。我们称之为跨语言的"不匹配测试"。这对情感识别系统带来了很大的挑战，需要一个与语言无关的情感模型，也就是需要寻找不同语言之间情感语音上的共同点。这种共同点是存在的，例如提高说话的声音一般表示激动，可能与生气有关。再如基音轨迹的上升和下降等变化，与喜悦情感的关系密切。这些发音上的特性，在近年来的语音情感识别研究中都有具体的报道[21, 22, 42, 60, 69]。

### 6.1.2 汉语与德语数据库上的跨语言特征分析

　　我们在汉语和德语数据库上研究了跨语言语音情感识别。研究语音情感的普遍特征对于理解人类的情感行为具有重要的意义。我们重点研究不同语言中的语音情感特征，以及对识别效果的影响，提取了跨语言的语音情感特征。训练了基于高斯混合模型的语音情感模型，并进行了跨语言的测试。生气情感在汉语和德语上得到了普遍的识别，实验结果显示了语音情感在不同语言上的普遍性。

　　中文情感语音库[56]包含六名男性和六名女性的表演语音，包括了六种基本的情感：生气、恐惧、喜悦、中性、悲伤和惊讶。广泛使用的柏林语音库[23]中包含了五名男性和五名女性的七种情感：生气、中性、恐惧、枯燥、喜悦、悲伤和厌恶。在跨语言的情感特征选择中，我们选择了五种共同的情感：生气、恐惧、喜悦、中性和悲伤。实验数据集如表6.1所示，我们选取了两个数据库中的共同的情感类型，并且在中文情感语料中选取了情感表达效果较好的样本。由于柏林库中的语料数量较少，为了保证两种语言中语料的平衡，从中文语料中选出的实验语料不宜过多，每种情感在两百条左右。

**表6.1 跨语言实验数据集**

| 数据集1 | | 数据集2 | |
|---|---|---|---|
| 柏林情感语音库的数据集 | | 中文情感语音库的数据集 | |
| 情感类别 | 样本数量 | 情感类别 | 样本数量 |
| 生气 | 127 | 生气 | 219 |
| 恐惧 | 69 | 恐惧 | 189 |
| 喜悦 | 71 | 喜悦 | 189 |
| 中性 | 79 | 中性 | 214 |
| 悲伤 | 62 | 悲伤 | 217 |

　　本节中提取了基本的声学特征，包括：基音、短时能量、共振峰，和MFCC等。在此基础上构造了二阶差分和三阶差分作为进一步的特征。构造了最大值、最小值、均值、方差、和范围等统计特征。一共采用了 375 个特征用来进行特征选择和识别，如表6.2所示。

表 6.2 声学特征列表

| 序号 | 特征 |
|---|---|
| 1–15： | 基音、基音一阶差分、二阶差分的均值、最大值、最小值、范围和标准差 |
| 16–90： | 第一至第五共振峰频率、及其一阶、二阶差分的均值、最大值、最小值、范围和标准差 |
| 91–165： | 第一至第五共振峰带宽、及其一阶、二阶差分的均值、最大值、最小值、范围和标准差 |
| 166–180： | 短时能量、短时能量一阶差分、二阶差分的均值、最大值、最小值、范围和标准差 |
| 181–375： | 镁尔倒谱系数 MFCC-0 至 MFCC-12*、及其一阶、二阶差分的均值、最大值、最小值、范围和标准差 |

*其中 MFCC-n 代表第 n 阶镁尔倒谱系数

表 6.3 跨语言特征选择结果

| 数据库与权重 | 前 20 个最佳特征选择结果（特征编号） |
|---|---|
| 数据集 1（德语） | 3，1，213，183，184，211，256，345，214，330，215，260，340，185，360，241，315，300，233，286 |
| 数据集 2（汉语） | 3，1，176，168，172，179，178，169，167，174，327，171，101，329，173，8，131，96，185 |
| 跨数据集评选（权重 0.1） | 3，1，176，168，177，172，179，178，169，167，174，327，171，101，329，8，173，131，185，184 |
| 跨数据集评选（权重 0.3） | 3，1，176，168，177，172，179，178，169，167，174，327，184，171，8，213，185，211，183，101 |
| 跨数据集评选（权重 0.5） | 3，1，176，168，177，213，184，183，172，211，179，185，171，178，169，167，256，8，174，214 |
| 跨数据集评选（权重 0.7） | 3，1，213，183，184，211，176，256，345，214，168，185，330，260，215，177，340，360，241，171 |
| 跨数据集评选（权重 0.9） | 3，1，213，183，184，211，256，345，214，330，215，260，185，340，360，241，315，300，233，286 |

采用 Fisher 判别系数（FDR）进行特征选择。 采用 FDR 可以对每一个原始特征获得一个特征评价的分数。我们分别在数据集1（德语）和数据集2（汉语）上进行特征评价。由于两个数据

集的数据量有很大的不同，德语数据集的数据量要少很多，因此不能简单的将两个数据集合并进行特征的选择。

在特征评价之后，我们采用加权融合的方法获得跨语言的特征评价分数：

$$FDR = w \times FDR1 + (1-w) \times FDR2 \tag{6-1}$$

此处 FDR 是在汉语和德语上的综合特征评价结果，FDR1 是德语数据上的评价结果，FDR2 是汉语库上的结果。w 是融合权重，在实验中，我们尝试了五个不同的权重取值。如表 6.3 所示。

### 6.1.3 跨语言语音情感识别实验

在基于GMM的情感识别实验中，我们测试了表6.3中的各组跨语言特征。测试结果如表6.4和表6.5所示。GMM的混合度在每个实验中调节到最佳的识别效果。为了研究跨语言特性，训练集和测试集是不同的语种，在德语上训练，在汉语上测试，或者在汉语上训练，在德语上测试。由于数据量不充足，特征选择所采用的数据集和识别测试的数据集是相同的。

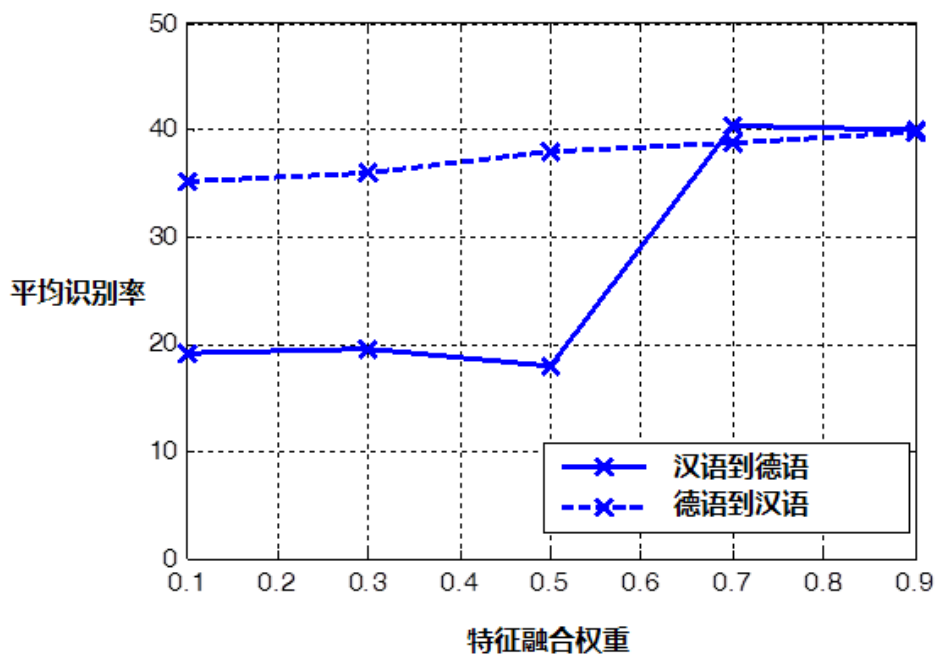

图6.1 跨语言识别测试的平均结果与特征融合权重的关系曲线

　　随着融合权重的升高，平均识别率也随之变化，如图6.1所示。从平均识别率的角度来看，当特征选择的融合权重为0.7时，系统的性能最佳。如表6.4所示，除了中性情感之外，调节权重和GMM混合度，其它的情感识别率都能分别到达70%以上。例如，喜悦的识别率在融合权重0.9，GMM混合度20时达到71.83%。虽然在情感之间的误识率比较高，对一个或两个目标情感的高识别率，仍然显示了在这些声学模型在德语和汉语上具有相同的情感模式。

### 表6.4 跨语言的不匹配训练与测试结果

#### (a) 在汉语上训练，在德语上测试 （%）

| 特征选择权重 | GMM 混合度 | 生气 | 恐惧 | 喜悦 | 中性 | 悲伤 |
|---|---|---|---|---|---|---|
| 0.1 | 16 | 63.779 | 0 | 0 | 0 | 32.258 |
| 0.3 | 20 | 63.779 | 0 | 0 | 0 | 33.871 |
| 0.5 | 32 | 63.779 | 0 | 0 | 0 | 25.806 |
| 0.7 | 32 | 98.425 | 82.608 | 0 | 15.189 | 6.451 |
| 0.9 | 20 | 100 | 0 | 71.831 | 0 | 1.612 |
| 0.9 | 16 | 100 | 100 | 0 | 0 | 0 |

#### (b) 在德语上训练，在汉语上测试(%)

| 特征选择权重 | GMM 混合度 | 生气 | 恐惧 | 喜悦 | 中性 | 悲伤 |
|---|---|---|---|---|---|---|
| 0.1 | 16 | 94.063 | 0 | 0 | 0 | 82.027 |
| 0.3 | 12 | 94.063 | 0 | 0 | 0 | 85.714 |
| 0.5 | 4 | 94.063 | 0 | 0 | 0 | 95.852 |
| 0.7 | 4 | 94.063 | 0 | 0 | 0 | 100 |
| 0.9 | 4 | 100 | 0 | 0 | 0 | 99.078 |

　　同时在汉语和德语上，生气情感获得了成功的识别 。对生气的跨语言识别测试，获得了94%以上的识别率，如表6.5所示。虽然在有些测试中，生气情感的识别率达到了100%，但是其它情感被误判为生气的错误概率也较高。进一步调节GMM模型的参数可以获得更加均衡的识别性能。

　　在本节中我们首次进行了不匹配的跨语言（汉语和德语）语音情感识别的实验研究。Fisher判别系数结合加权特征融合的方法被用来进行跨语言特征选择。采用高斯混合模型进行情感的建模。实验结果显示，在汉语和德语上，存在一个通用的声学模型来描述同一种语音情感行为。生气情感的跨语言识别测试获得了成功，而中性情感的错误率较高。

　　对基于GMM的识别系统在多语种环境下进行了扩展性的研究，希望能够扩展语音情感识别系统在实际中的应用范围。特别是针对不匹配条件下的识别测试，更加能够反映出情感特征组是否具有较好的通用性。在下一节中我们将进一步探讨GMM情感识别系统在耳语音情感识别中的使

用。

## 表 6.5 生气情感的跨语言识别

### (a) 生气 vs. 恐惧 (%)

| 训练 （汉语数据集） | 测试 （德语数据集） | | |
|---|---|---|---|
| | 生气 | 恐惧 | 中性 |
| 生气 | 100 | 0 | 0 |
| 恐惧 | 2.89855 | 73.913 | 23.1884 |
| 中性 | 0 | 67.0886 | 32.9114 |

GMM 混合度为 12，特征选择权重为 0.7

### (b) 生气 vs. 喜悦 (%)

| 训练 （汉语数据集） | 测试 （德语数据集） | | |
|---|---|---|---|
| | 生气 | 喜悦 | 中性 |
| 生气 | 100 | 0 | 0 |
| 喜悦 | 5.6338 | 91.5493 | 2.8169 |
| 中性 | 30.3797 | 58.2278 | 11.3924 |

GMM 混合度为 12，特征选择权重为 0.9

### (c) 生气 vs. 悲伤 (%)

| 训练 （汉语数据集） | 测试 （德语数据集） | | |
|---|---|---|---|
| | 生气 | 悲伤 | 中性 |
| 生气 | 94.0639 | 0 | 0 |
| 悲伤 | 6.07477 | 0 | 90.6542 |
| 中性 | 0 | 0 | 100 |

GMM 混合度为 4，特征选择权重为 0.7

## 6.2 耳语语音中的情感识别

### 6.2.1 耳语音情感识别简介

目前世界上仍然有不少残疾人士需要通过辅助仪器来进行正常的语音交流[127]，对耳语音信号处理技术的研究，对失音者的语音恢复具有重要的意义。耳语音是一种特殊的语音信号，它在人们的日常生活中广泛的出现，在会场、音乐厅、图书馆等禁止大声喧哗的场所人们往往采用耳语音进行交谈。耳语音的一个主要特点是基音的缺失，从而对语音信号的处理带来了挑战。对耳语音情感的研究涉及到信号处理、模式识别、发音生理学和情绪心理学等多个相关学科领域，是交叉性十分强的一个课题。目前对耳语音情感识别的研究国内外还处于起步阶段，虽然受到了研究者们的广泛关注[128-135]，但是对耳语音的研究深度和研究成果还比较缺乏。

　　研究耳语音具有重要的理论意义和实际价值。Schwartz 早在 1968 年研究了从耳语音当中辨别说话人的性别的问题[136]，Tartte 等人研究了耳语音中元音的听辨，研究表明某些语言中的元音[a]和[o]在耳语音中有较为严重的混淆[134, 137]。早期的研究主要从发音学的角度去研究耳语音的特性，采用的研究方法集中在人耳听辨上。2002 年乔治亚理工的 Morris 从信号处理的角度较为系统的研究了耳语音的增强和识别问题[138, 139]，从语音转换，语音编码和语音识别角度研究了耳语音的处理。根据 Morris 的研究发现，在正常音上训练的语音系统在耳语音信号上发生了严重的性能衰落，通过模型的自适应能够获得较明显的改善效果。维多利亚大学语言学系的研究者们研究了汉语耳语音的音调[140]，Gao 通过感知实验研究了在基音缺失的情况下耳语音音调信息的表达。赵鹤鸣等人研究了耳语音的端点检测与共振峰估计[131]，并且从语音信号处理的角度研究了耳语音音调的自动识别[130]。金赟等人初步研究了耳语音情感识别的问题，建立了耳语音情感数据库[141]。赵艳采用共振峰、语速等耳语音特征，结合量子遗传神经网络进行了耳语音的喜、怒、悲三种情感的识别[142]。龚呈卉等人[132]对汉语耳语音的情感特征进行了研究，研究表明时长和短时能量等耳语音韵律特征对生气和悲伤能够进行较好的区分。

　　情感数据库的建立是情感识别研究的先决条件，目前正常音的情感数据库已经比较丰富，耳语音情感数据库还相对缺乏。耳语音情感数据库的建立过程如图 6.2 所示，主要包括情感数据采集方案的制定、耳语音的录制、录音数据的检验与补录、情感耳语音的切分和标注以及对耳语音情感数据的人耳听辨检验等步骤。表 6.6 中总结了常用的中文耳语音情感数据库，并且与正常音的情感数据库进行了比较。在识别率方面，耳语音数据库明显低于正常音数据库；在采集方式上，正常音情感数据库已有情感诱发等自然度较高的采集方式，耳语音情感数据库还集中在表演方式上；在包含情感类别方面，正常音情感的研究已经比较广泛，包含烦躁等实用情感数据，耳语音情感数据库中包含的情感类别还比较少。随着研究的展开，耳语音情感数据库还有待进一步的扩充。

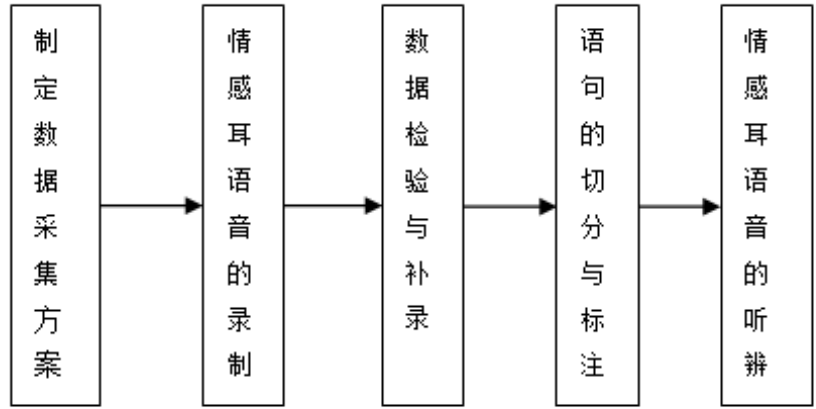

图 6.2　耳语音情感数据库的制作过程

表6.6 耳语音情感数据库比较

| 发音状态 | 情感类别 | 数据库 | 识别率 | 采集方式 | 应用 | 研究算法 |
|---|---|---|---|---|---|---|
| 耳语音 | 平静 | 南京大学汉语耳语音库[143] | 59.7%（声调识别） | 表演 | 语音识别和转换 | 概率神经网络 |
| 耳语音 | 平静、生气、惊讶、恐惧、喜悦、悲伤、厌倦 | 苏州大学耳语音数据库[133] | N/A | 表演 | 说话人识别、情感识别 | 类别可分性分析 |
| 耳语音 | 喜悦、生气、悲伤、恐惧 | 东南大学耳语音情感数据库[141] | 49% | 表演 | 情感识别 | 量子神经网络 |
| 正常音 | 喜悦、平静、烦躁 | 东南大学实用语音情感数据库[144] | 76.6% | 诱发 | 情感识别 | 高斯混合模型 |

## 6.2.2 耳语音的情感数据采集与分析

耳语音情感数据的采集，是后续情感识别研究的基础，具有非常重要的意义。目前正常音的情感数据库已经比较丰富，数据的采集环境也多种多样，一般可以在实验室的安静环境下采集。由于耳语音的特殊性，录制耳语音信号需要高质量的麦克风和声卡，并且在静音室里完成

情感数据的录制，以减小噪声的干扰。

在采集耳语音情感数据的过程中，我们一般同时采集相应的正常音数据，以便于后续的对比研究。在平静状态下录制的耳语音和正常音，其时域波形图和语谱图如图 6.3 所示，正常音的基音轮廓如图 6.4 所示，耳语音的共振峰提取结果如图 6.5 所示。由于耳语音基音特征的缺失，共振峰特征就格外重要。因此在耳语音情感数据的采集阶段，要保证语音的质量以便于后续研究中的参数分析。由于耳语音的短时能量要比正常音弱很多，噪声干扰的问题对于耳语音情感识别就更为突出。Schuller 等人在 2006 年首次研究了噪声对于情感识别的影响[16]，Tawari 等人在车载环境下进一步研究了情感识别系统中的降噪方法[96]。他们的研究表明，基于小波变换的语音增强方法能够对抗噪声对正常音的情感识别的影响，但是噪声依然是制约实用情感识别系统性能的一个重要因素。因此我们在采集耳语音情感数据的时候，需要尽可能的降低噪声干扰，为后续的识别研究提供便利。

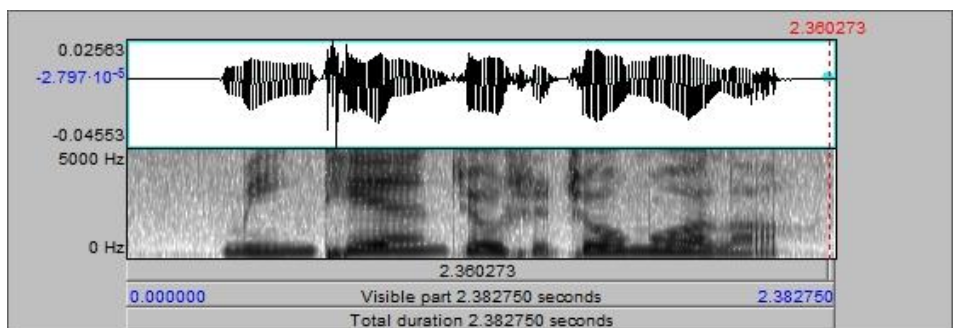
（a）正常音的波形与语谱图

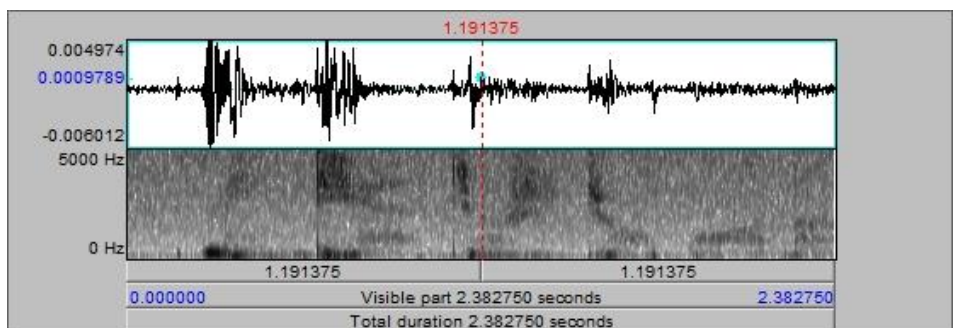
（b）耳语音的波形与语谱图

**图 6.3 中性状态下的耳语音与正常音**

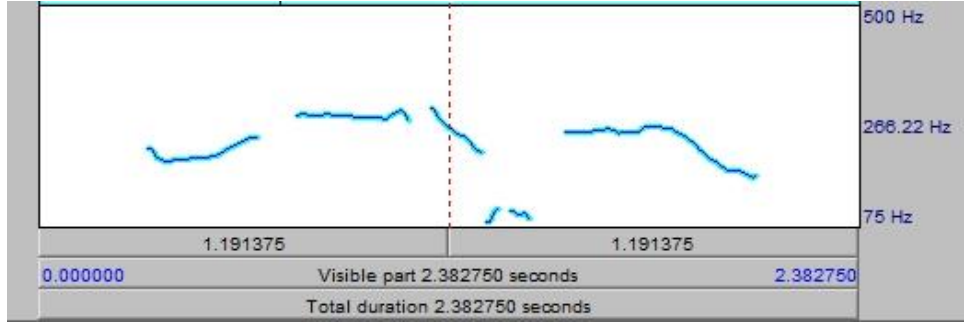

图 6.4 正常音的基音轮廓

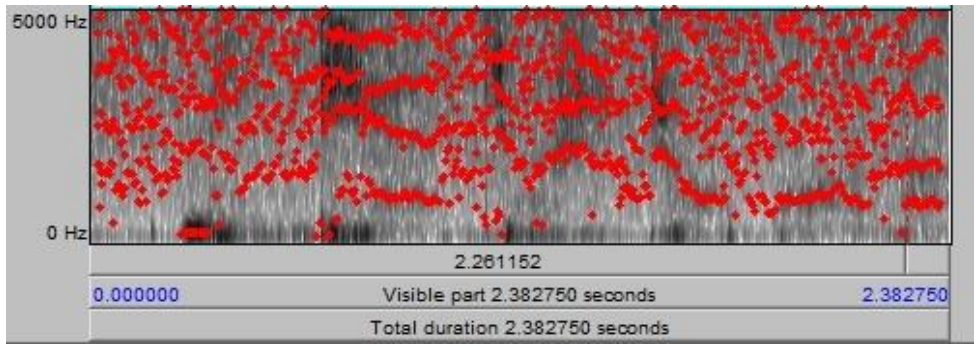

图 6.5 耳语音的共振峰提取

　　情感数据的获取，按照自然度来分，可以分为自然语音、诱发语音和表演语音三类。对于耳语音情感的诱发方法，可以延用正常音的情感诱发方法。例如，Johnstone 等人利用计算机游戏进行情感的诱发[54]，建立了自然度较高的诱发语音库。对于烦躁等实用情感类别，可以利用睡眠剥夺、噪声刺激，以及重复性的认知作业来进行诱发[74]。

　　在不同的情感状态下，我们可以看到耳语音信号会有相应的变化，通过表演方式采集到的生气、喜悦和中性状态下的耳语音信号如图 6.6 所示。我们可以直观的看到不同情感状态下的耳语音的时长变化较为明显。

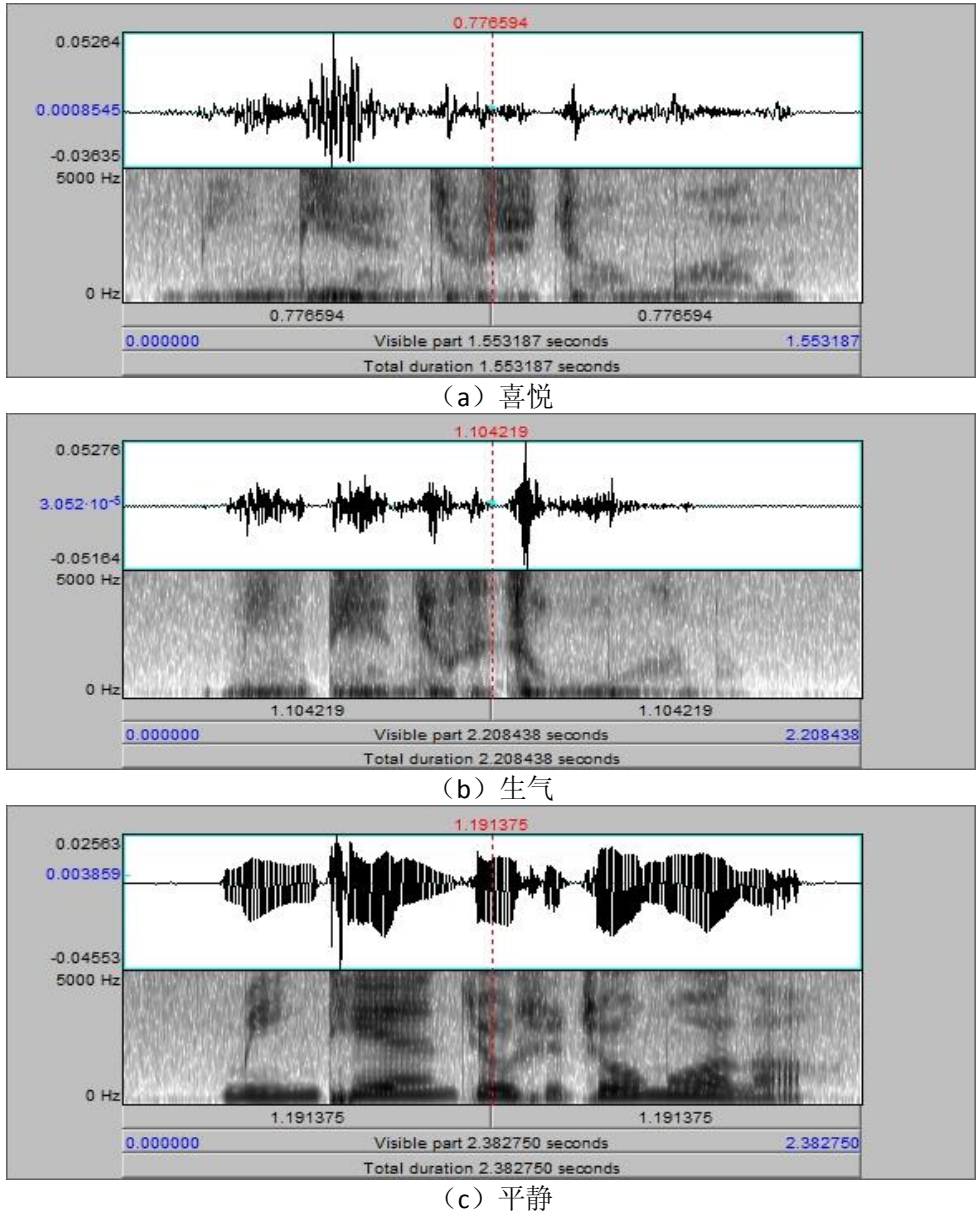

（a）喜悦

（b）生气

（c）平静

**图 6.6 经过情感渲染的耳语音信号示例**

### 6.2.3 耳语音的情感特征研究

语音信号中的何种特征能有效的反映出情感的变化，一直是语音情感识别中的关键问题之一。在过去的几十年中，研究者从语音学、心理学等角度，作了大量的研究。语音情感特征，可以分为韵律特征与音质特征两类。韵律特征又称为超音段特征，是指大于一个音位的语音单位如音节（syllable）或比音节更大的单位所表现出来的音强（intensity）、音长

（length or duration）、音高（pitch）、重音（accent）、声调（tone）和语调（intonation）等语音特征参数。在早期的语音情感识别中研究，韵律特征的应用最为广泛，例如基音特征、语速及其衍生参数等。音质特征主要指语音的音色和语谱方面的特征，也被称作是音段特征，反映发音时声门波形状的变化。韵律特征和音质特征共同反映了情感的变化。

情感特征的分析是进行情感识别的先决条件，本节中首先介绍耳语音的发音特点，然后对耳语音的情感特征进行了初步的总结。正常音的声母部分发音时，声道某处有一定阻碍，韵母部分发音时，声道没有阻碍，声带相对靠拢，形成窄缝声门，气流从窄缝挤出时引起声带振动，形成准周期的脉冲串声源。耳语音的清擦音、塞擦音和塞音声母部分与正常音的发音方式没有大的差异，而韵母部分发音时，声门保持半开状态，声门前部完全靠拢，后部的气声门有一个宽三角裂隙，声带不振动，从肺部出来的气流通过开放区产生摩擦噪声，故声源为噪声[142, 145]。由于发耳语音时，伪声带区域变窄，声门保持半开状态，使得声道增加了气管和肺部分，产生附加的零极点，改变了声道传输函数，所以耳语音的韵母部分与正常音的韵母部分有较大的差异[142, 145]。

### 表6.7 耳语音情感声学特征总结

| 特征类型 | 关联的耳语音情感 | 所属情感维度 | 是否可用于耳语音 | 参考文献 |
|---|---|---|---|---|
| 基音 | N/A | 唤醒维 | 否 | Dellaert等[7, 8, 13, 15, 146] |
| 共振峰 | 生气、悲伤、喜悦 | 效价维 | 是 | Tartter等[129, 132, 134] |
| MFCC | 生气、悲伤、喜悦 | 效价维 | 是 | 林玮等[142, 147] |
| LPC | N/A | 效价维 | 是 | Nwe等[78] |
| 短时能量 | 生气、悲伤 | 唤醒维 | 是 | 龚呈卉等[132] |
| 语速 | 生气、悲伤 | 唤醒维 | 是 | 龚呈卉等[132] |
| 时长 | N/A | 唤醒维 | 是 | 龚呈卉等[132] |
| TEO算子 | 生气、悲伤、喜悦 | 效价维 | 是 | 赵艳等[64, 142] |

表 6.8 耳语音情感特征选择结果

| 最佳特征排序 | 耳语音情感特征 |
|---|---|
| 1 | 第三共振峰最大值 |
| 2 | 第三共振峰均值 |
| 3 | MFCC4 一阶差分的标准差 |
| 4 | MFCC2 一阶差分的标准差 |
| 5 | MFCC6 一阶差分的标准差 |
| 6 | MFCC5 一阶差分的标准差 |
| 7 | MFCC10 一阶差分的标准差 |
| 8 | MFCC11 一阶差分的标准差 |
| 9 | MFCC8 一阶差分的标准差 |
| 10 | MFCC3 一阶差分的标准差 |
| 11 | 第二共振峰的最大值 |
| 12 | MFCC9 一阶差分的标准差 |
| 13 | 第二共振峰的均值 |
| 14 | MFCC7 一阶差分的标准差 |
| 15 | MFCC11 的最大值 |
| 16 | MFCC12 的最大值 |
| 17 | 短时能量的抖动 |
| 18 | MFCC6 的最大值 |
| 19 | 语速 |
| 20 | MFCC12 一阶差分的标准差 |

*注 1：表中 MFCCn 表示第 n 阶梅尔倒谱系数

*注 2：基本声学特征集不同：耳语音的特征提取，与正常音的特征提取有所差别，最典型的是基音缺失，有一部分声学特征是可以用于正常音的情感识别，而没有用于耳语音的情感识别。因此在本节所用的基本声学特征集和前几章中所用的不同。

我们对常用在语音情感识别中的声学特征进行了总结，如表 6.7 所示。其中基音等声源参数是区分正常音中情感信息的有效特征，然而耳语音的激励源是噪声不具备基音特征，故基音特征无法用于耳语音情感识别中。共振峰、镁尔倒谱参数（MFCC）、线性预测参数（LPC）等声道特征参数，可用于耳语音的情感特征分析。这一类特征参数常作为正常音的效价维特征，反映了说话人情绪中的愉悦度，对耳语音的情感特征分析同样具有重要的意义。短时能量、语速、

时长等韵律参数，常用语正常音情感的唤醒度分析，研究表明能量和语速也是区分耳语音情感的重要特征[132]。文献[142]对耳语音TEO(Teager Energy Operator)算子特征进行了研究，并用于耳语音的喜、怒、悲三种情感的识别。

在耳语音的基本声学参数之上，包括短时能量、语速、共振峰、MFCC等，可以构造出适合用于识别的特征，一般分为静态特征（全局统计特征）和动态特征（短时特征）两类。由于动态特征对音位信息的依赖性太强，使用差分、均值、最大值、最小值、方差等全局统计特征有利于建立与文本无关的情感识别系统。静态特征可以同GMM分类器结合进行情感识别，动态特征一般与HMM分类器结合。

耳语音的特征选择是一个重要的研究问题，在本文中的第三章里我们探讨了特征压缩和特征选择两类方法。 为了研究何种情感特征能够对耳语音的情感识别有更大的贡献，本节中采用基于FDR的特征评价和选择的方法来获取识别用的特征组。前几章中都采用了PCA加LDA的方法来进行特征的压缩，以获得较高的识别率，然而这样的特征压缩算法不能直观的给出何种声学特征对耳语音情感识别是有价值的。我们通过本节中选取的耳语音情感特征组，可以看到哪些常用的语音情感特征能够区分耳语音情感，如表6.8所示。

### 6.2.4 耳语音的情感识别问题

Hultsch与Tartter等人的研究中指出，耳语音中对"喜悦"情感和"恐惧"情感的表达比较困难[128]。随后，Cirillo等人对这个问题再次进行了听辨实验研究[129]，得到了相似的结论。在他们的听辨研究中，耳语音中的"喜悦"，容易被混淆为"恐惧"或者"中性"，进一步的频谱分析显示这几种情感之间的混淆可能是由于音调质量的下降造成的。在语音质量较差的条件下的听辨实验显示，通过电话线路传送的耳语音信号中，喜悦被听辨为"悲伤"或者"中性"。Cirillo的实验结果显示，耳语音中的"悲伤"和"生气"可能较容易通过听辨区分，文献[132]的声学特征分析中也支持了这一结论。

在一组区分度较高的情感特征集上，常用的模式识别算法都可以用来研究语音情感的分类问题。目前对于耳语音情感识别算法的研究还比较少，何种分类算法适合耳语音情感的特点还尚未有定论。文献[132]和文献[141]从模式分类的角度对耳语音情感特征进行了分析，文献[142]将量子遗传神经网络应用于耳语音情感识别。量子遗传神经网络是采用量子遗传算法来进行神经网络的连接权值优化。采用量子遗传算法优化BP(Back Propagation)神经网络的连接权值，既有神经网络的学习能力和鲁棒性，又有遗传算法的强全局搜索能力。BP网络基于梯度下

降方法，因而对网络的初始权值非常敏感，不同的初始权值会导致完全不同的结果。因此基于量子遗传神经网络算法的基本思想就是先利用量子遗传算法训练随机产生的初始数据，较快地搜索到最优解，定位出一个较好的搜索空间，并将该最优解作为BP网络的初始权值和阈值。然后利用BP算法在这个小的搜索空间内搜索出最优参数，最快地实现网络的收敛。

### 6.2.5 耳语音的唤醒维度与效价维度的识别

一般对情感维度的研究认为，在正常语音信号中，韵律特征对应唤醒维，音质特征对应效价维。需要引起注意的是，对于耳语音信号来说，由于发音特点的改变，何种声学特征对应于何种情感维度的问题还没有进行系统的研究。一种合理的假设是耳语音特征与正常音特征一致，韵律特征对应与唤醒维，音质特征对应于效价维。我们就此假设，进行了识别实验的验证。

在我们建立的耳语音情感数据库中[141]，"悲伤"和"生气"都处于效价维的负向，"喜悦"处于效价维的正向。在唤醒维上，"生气"、"喜悦"和"惊讶"位于正向区域，而"悲伤"处于负向区域。基于高斯混合模型的识别实验中，每种情感选用200条样本，训练与测试比例为3：1，单独采用音质特征（共振峰）进行识别与单独采用韵律特征（短时能量、语速、时长）进行轮换测试的结果如表6.9所示。

惊讶与悲伤的区分度较低，生气的识别率较高。单独使用共振峰对耳语音情感识别的效果不佳，而短时能量等韵律特征的区分效果较好。从识别结果可以看到，音质特征对唤醒维度的贡献较低，而韵律特征与唤醒维和效价维都有关联。

**表6.9 耳语音情感的唤醒维和效价维识别率研究**

| | 音质特征 | 韵律特征 | 对应的情感维度差异 |
|---|---|---|---|
| 惊讶vs悲伤 | 62%(惊讶)，66%（悲伤） | 58%(惊讶)，48%（悲伤） | 唤醒维差异 |
| 喜悦vs生气 | 60%(喜悦)，64%（生气） | 70%(喜悦)，72%（生气） | 效价维差异 |
| 生气vs悲伤 | 70%(生气)，66%（悲伤） | 72%(生气)，76%（悲伤） | 唤醒维差异 |
| 平均识别率 | 64.6% | 66.0% | |

### 6.2.6 嵌入马尔科夫网络的高斯混合模型在耳语音情感识别中的应用

在第四章 4.5 节中，我们探讨了嵌入马尔科夫网络的高斯混合模型，在这里我们将其应用与连续耳语音语料段的情感识别中。根据情感的维度空间论，耳语音信号中的情感信息具有时间上的连续性，因此可以利用三阶的马尔科夫网络对多尺度的耳语音情感分析进行上下文的情感依赖关系的建模。在第四章 4.5 节中提出的方法中，我们采用了一种弹簧模型来定义二维情感维度空间中的高阶形变，并且利用模糊熵评价将高斯混合模型的似然度转化为马尔科夫网络中的一阶能量。

实验中，采用了 2000 条情感短句作为短句的高斯混合模型的实验数据集，1000 条长句作为长句的高斯混合模型的实验数据集。短句训练的高斯混合模型的混合度设置为 32，长句训练的高斯混合模型的混合度设置为 64，分别训练了五种情感的概率模型。测试数据集采用连续的耳语音段落，包含五种情感类型，每种情感的耳语音语料约为 15 分钟左右。根据人耳对连续耳语音段落中的情感进行听辨标注，由于连续的耳语音段中，情感包含的成分较为复杂，我们将每个短句的情感标注为"喜悦"、"生气"、"惊讶"、"悲伤"、"中性"和"其它"，六个标签中的一个。在识别测试中，忽略对"其它"情感耳语音的识别结果，不将其包括在识别率的统计中。识别结果如表 6.10 所示。

**表 6.10 耳语音情感识别结果**

| 测试样本 | 识别结果 | | | | |
|---|---|---|---|---|---|
| | 喜 | 怒 | 惊 | 悲 | 中性 |
| 喜 | **53.1%** | 7.0% | 9.6% | 14.2% | 16.1% |
| 怒 | 8.7% | **64.3%** | 11.8% | 4.6% | 10.6% |
| 惊 | 10.1% | 12.3% | **56.7%** | 9.4% | 11.5% |
| 悲 | 13.9% | 5.2% | 9.1% | **58.2%** | 13.6% |
| 中性 | 14.1% | 8.4% | 9.2% | 11.7% | **56.6%** |

**表 6.11 弹性系数对耳语音情感识别结果的影响**

| 弹性系数 | 识别结果（L1 距离/L2 距离） | | | | |
|---|---|---|---|---|---|
| | 喜 | 怒 | 惊 | 悲 | 中性 |
| 0.1 | 39.7%/41.8% | 47.2%/49.7% | 38.6%/40.2% | 41.7%/42.9% | 37.9%/40.1% |
| 0.2 | 44.5%/46.3% | 51.1%/53.2% | 40.3%/42.4% | 44.9%/46.8% | 43.1%/44.3% |
| 0.3 | 49.0%/51.4% | 57.2%/60.9% | 47.1%/50.3% | 49.8%/53.1% | 48.2%/50.5% |
| 0.4 | **55.3%/53.1%** | **62.1%/64.3%** | **54.3%/56.7%** | **56.9%/58.2%** | **55.1%/56.6%** |
| 0.5 | 48.1%/52.6% | 56.8%/61.4% | 46.2%/51.7% | 49.7%/54.4% | 49.3%/51.4% |

实验结果显示，本文提出的情感识别算法在连续耳语音数据上获得了较好的识别结果，对愤怒的识别率达到了 64.3%。实验结果进一步显示，与正常音的研究结论不同，耳语音中的喜悦情感的识别相对困难，而愤怒与悲伤之间的区分度较高，与 Cirillo 等人[129]进行的人耳听辨研究结果一致。

在不同的弹性系数设置下的识别率如表 6.11 所示，在嵌入的马尔科夫网络中，L2 距离要优于 L1 距离。弹性系数高代表了强化相邻语句间的情感延续性，弹性系数小代表了融合判决是更看重单个语句样本的高斯混合模型识别结果。

耳语音情感识别的特点，是基音缺失。在基音缺失的情况下，我们应该充分发掘共振峰和频谱特征等其它重要的语音情感特征。并且充分利用上下文信息、性别差异、说话人规整化等方法，来提高耳语音情感识别系统的性能。

### 6.2.7 耳语音情感分析的意义

对耳语音情感识别的研究，有助于建立一个和谐的人机交互环境，通过情感识别，能够使计算技术更加成熟的融入于人类社会的各个领域中，更好的理解用户的需求，提供智能化、个性化的服务。在监听和安保领域，耳语音是一个重要的研究对象。通过对耳语音情感状态的识别，有助于推断说话人所处的环境和危险状况。在消费电子当中，越来越多的电子设备采用语音交互接口。在一些公共场合，采用耳语音进行菜单操作和文字输入等，更适宜保护隐私，使语音操作更容易为用户接受。目前的语音情感识别系统正在往多模态方向发展，未来的耳语音情感识别系统可以利用其它通道的信息来辅助识别情感，或者作为辅助信息提高其它识别系统的性能。例如耳语音情感特征可以同文本特征、表情特征、生理特征等进行多通道的信息进行融合，以提升情感识别系统的可靠性。

# 第七章 应用与展望

## 7.1 载人航天中的应用的设想

本文中涉及的实用语音情感，包括烦躁、疲倦和自信等，在第二章中探讨了这些特殊类型的情感所具有的实用价值。本文中研究的几种语音情感类型有两个特点，第一个特点是与认知过程关系密切，第二个特点是具有特殊的应用前景。

烦躁情感具有特殊的应用背景，在某些严酷的工作环境中，烦躁是较为常见的、威胁性较大的一种负面情感。保障工作人员的心理状态健康是非常重要的环节。本文中设想在未来可能的长期的载人任务中，对航天员情感和心理状态的监控与干预是一个重要的研究课题。

在某些特殊的实际应用项目中，工作人员的心理素质是选拔和训练的一个关键环节，这是由于特殊的环境中会出现诸多的刺激因素，引发负面的心理状态。例如，狭小隔绝的舱体内环境、严重的环境噪声、长时间的睡眠剥夺等因素，都会增加工作人员的心理压力，进而影响任务的顺利完成。

因此，本文设想在天地的通信过程中，有必要对航天员的心理健康状况进行检测，在发现潜在的负面情绪威胁的情况下，应该及时的进行心理干预和疏导。在心理学领域，进行心理状态评估的方法，主要是依靠专业心理医师的观察和诊断。而近年来的情感计算技术，则为这个领域提供了客观测量的可能。我们设想，语音情感识别技术可以用于分析载人航天任务中的语音通话，对说话人的情感状态进行自动的、实时的监测。一旦发现烦躁状态出现的迹象，可以及时的进行心理疏导。

图 7.1 所示的系统，是本文中设想的实用语音情感识别在载人航天中的一种可能的应用方式。我们设想在未来可能的长期的载人航行中，对烦躁等负面心理状态的监测将发挥重大作用。

本文设想在载人航天的应用中，需要考虑几个特殊的问题。

第一，我们识别的对象群体是特定的。一般来说，一个国家的航天员队伍是相对稳定的一个群体，在人数上不会过多，人员变动也相对较少。因此，在载人航天中的应用与电信话务中

心等大量说话人的场合不同，我们面对的是特定说话人的情感识别问题。在识别技术上，可以为每一个说话人定制所需的声学特征和识别模型，以提高识别的准确度。

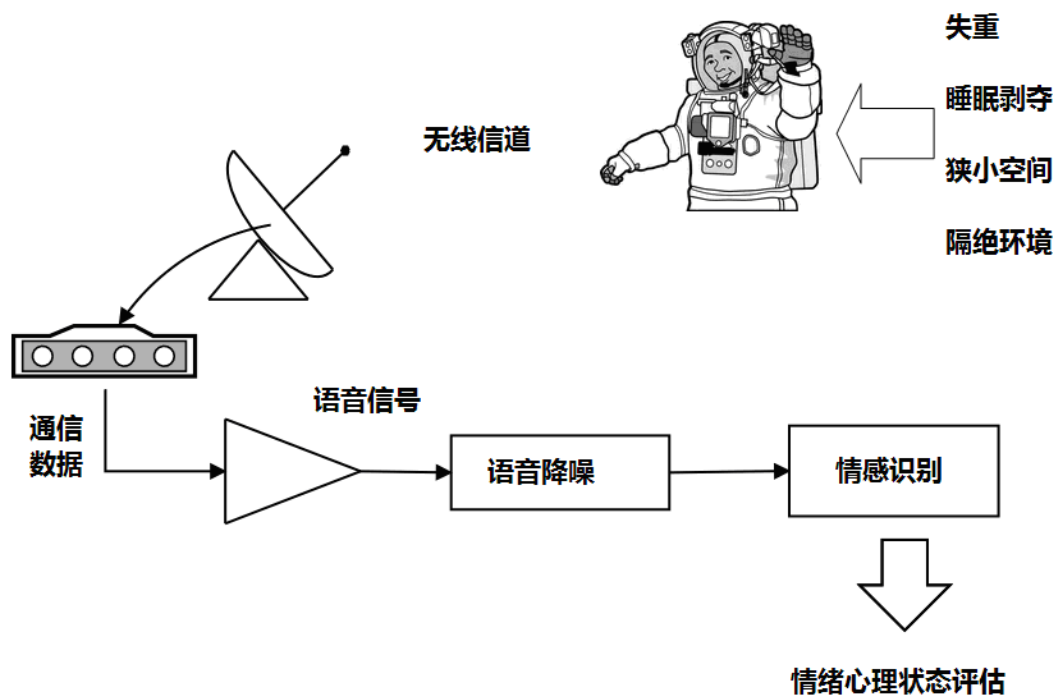

**图 7.1 载人航天中的语音情感分析应用的设想图**

第二，我们假设航天员在工作状态下的说话习惯与普通人不同，具有一定的特点。在情感建模中，说话人个性的差异会带来很大的干扰，例如在鲁棒性分析章节中对大量非特定说话人的研究中就发现了这个问题。因此，我们有必要考虑针对不同的性格与不同的说话方式，调整已有的情感模型。模拟载人航天环境中的研究显示，在特殊的工作环境中，被试人员倾向于隐藏负面情绪的流露，语音中的唤醒度比正常条件下高。

第三，预计环境噪声非常恶劣，需要有效的抗噪声解决方法。我们知道在车载电子中，语音识别主要受到发动机等噪声源的干扰。在载人飞船内的噪音干扰可能会直接制约语音情感识别技术在其中的应用。然而，目前在情感识别领域，对噪声因素的研究尚处于起步阶段，在今后的实际应用中，对降噪技术的需求会越来越显著。

## 7.2 情感多媒体搜索

语音情感识别技术的另一个重要应用，是在基于内容的多媒体检索中。传统的搜索引擎，一般是进行文本的检索，对网络上的多媒体数据的内容无法进行识别和搜索。目前，基于内容的检索技术已经带来了一些有趣的应用。MIT 的研究者们提供了一个实验平台，他们的搜索引擎，能够对教学视频中的语义内容进行关键字的检索。首先在视频数据中分离出音频部分，对其中的语音信号进行自动语音识别，再对识别出的语义关键词进行匹配和搜索。基于图像内容的检索也带来了一些特殊的搜索引擎。例如，"Picitup"、"Exalead"、"Face Search"等网站，都可以通过用户上传的图片的内容，寻找到网络上类似的图片和出处。如果上传名人的人脸图像，可以直接搜索到网络上该人的其它相关图片。

情感识别技术可能会给多媒体检索领域带来更多、更有趣的应用。多媒体数据中蕴含了大量的情感信息，例如摄影作品、音乐歌曲、影视作品等，都是丰富的情感信息源。如果我们可以对多媒体数据进行情感检索，也就是根据指定的情感类型，找寻出对应的多媒体数据，那么能够给网络用户提供的，将是一个广阔的情感多媒体搜索平台。情感信息的检索技术在娱乐产业中会有很大的应用前景。例如，在优酷网、土豆网等视频资源网站，目前仅能够根据人工标注的方式，对海量的视频资源进行分类和检索。这样的多媒体资源管理模式是低效的，急需一种基于情感信息内容的检索技术来对网络视频进行自动的分类和管理。

在用户进行网络视频搜索时，可以指定一些特殊的视频类型进行检索，例如"喜剧片"、"真实"、"清新"等与情感有关的描述词。这样的检索方式会给用户提供一个比现有的语义搜索平台更加广阔的情感信息搜索平台。然而，目前商用的搜索引擎还停留在对视频文件名和文件描述进行检索的阶段，有待将实用语音情感识别技术融合到音频内容的检索中。

图 7.2 是一个检索系统的系统模块设计，可以对网络视频进行基于音频内容的情感信息分析与检索。

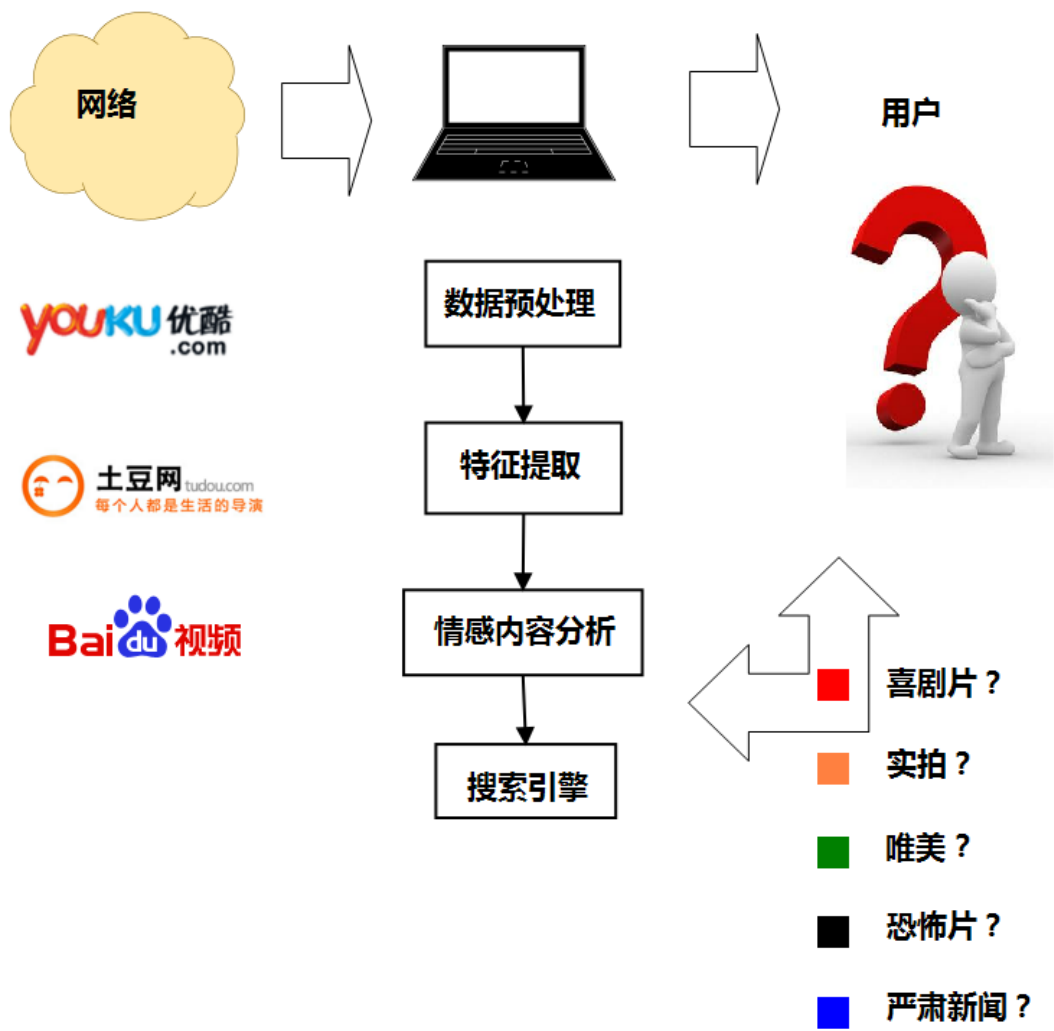

**图 7.2 网络情感多媒体搜索**

## 7.3 个人机器人中的应用

在本文的第二章中，介绍了现有的一些语音情感数据库，其中值得我们注意的是 AIBO 情感数据库。它是德国柏林的研究者们，通过 51 名儿童与索尼的智能机器狗 AIBO 之间的交流，采集建立的情感数据库。在他们的实验中，采用了儿童与机器狗之间交互的这样一种环境来研究语音情感识别技术。这方面的研究的成功，使得语音情感识别技术在人与智能机器人之间的交流中显得日益重要。

　　智能机器人技术是一个具有良好发展前景的领域，在我们的日常生活中，智能家居可以给我们带来自动化的便捷的生活方式，荷兰飞利浦公司的"House of the Future"研究项目就是一个典型的例子。相比之下，更加人性化的、更深层次融入人们生活中的智能机器人，应该会给我们的生活方式带来更大的冲击。目前，除了索尼的智能机器狗外，还有一些有趣的智能机器人的产品，例如可以表达七种情绪的机器人 KOBIAN、陪伴孤寡老人的 Telenoid R1 机器人等。

　　语音是人类交流与沟通的最自然、最便捷的方式，在人与机器人的交互中，语音交互亦是首选的技术之一。情感识别技术与智能机器人技术的结合，可以使得冰冷的机器能够识别用户的情感，是机器人情感智能的基础技术。在智能机器人拥有了情感识别能力之后，才有可能进行同用户的情感交流，才有可能成为"个人机器人"，更加深入的融合到人们的社会生活和生产劳动中。

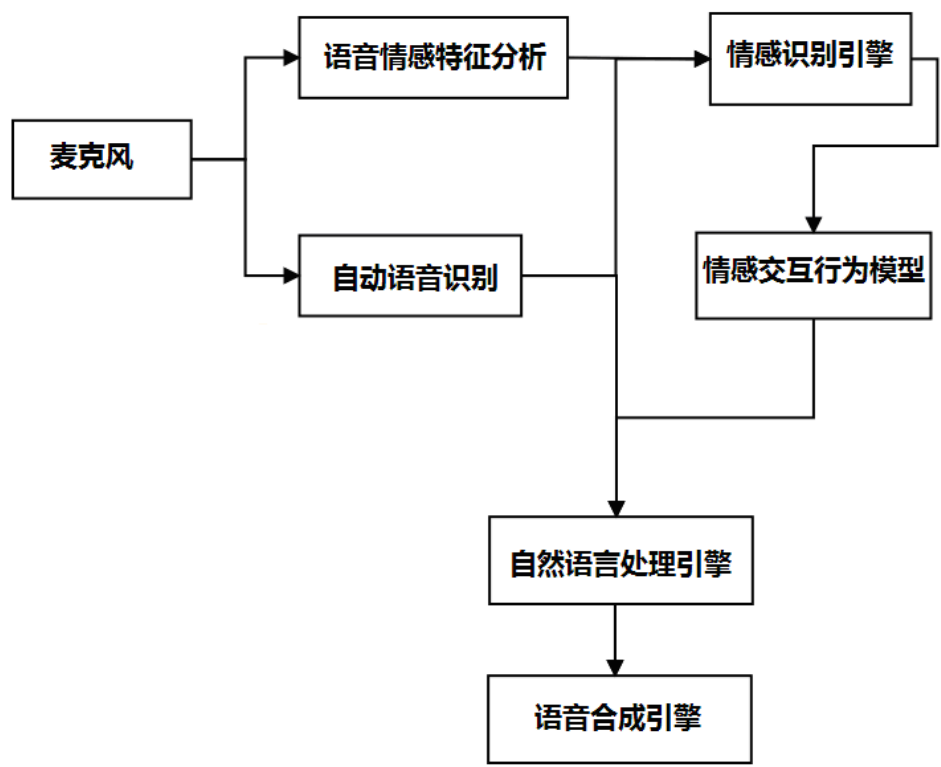

**图 7.3 智能机器人中的语音情感交互模块**

　　机器人的情感，是一个有趣的话题，在很多文艺作品中有生动的讨论。从语音情感识别技术的角度看，具备一定情感智能的机器人，有可能进入人们生活的一个途径，可能是在儿童的

智能玩具中。情感语音的识别与合成技术可能带来一系列具备虚拟情感对话能力的玩具,在模拟情感交流的环境中,培养儿童的沟通与情感能力。语音情感识别技术在儿童的发展与教育科学中的应用,亦是值得我们探讨的一个课题。

图 7.3 是智能机器人的语音情感交互模块,能够通过用户的语音和行为,识别出说话人的身份、性格和情感状态,进而对短期和中长期的人机交互方式进行自适应的调节。

## 7.4 全文总结与展望

本文探讨了实用语音情感识别技术,它是情感计算的核心领域之一。从源头上讲,情感计算是起源于人们对制造智能机器的热情。当"Thinking Machine"这个词语首次出现在人们的讨论中时,它既体现了人们对自身智能的好奇,也体现了对人造物的崇拜。当明斯基指出,情感能力是智能的重要部分,当 Picard 质疑"没有情感能力的人工智能是否能实现"时,情感计算领域的研究被开启了。研究者们对赋予机器以人类的情感,表现出了非常大的兴趣。

在以往的诸多语音情感识别的研究中,主要围绕基本情感类别进行研究。这些基本情感类别并不能满足实际应用的需求。本文中关注了几种实用语音情感类型,它们是烦躁、自信和疲倦等,这几种情感类型与认知过程有密切的关系,既受到认知过程的激发,又对人员的认知能力产生影响。因此对它们在语音信号中的表达和自动识别的研究具有重要的实用意义。

高自然度的情感数据是研究的基础和保障,我们通过多种诱发方式采集了可靠地情感语料,建立了实用语音情感研究的基础数据库。通过噪声刺激可以诱发被试人员的负面情感,通过重复的四则运算可以诱发被试的烦躁情感,通过喜剧影视片段可以诱发喜悦情感,通过长时间的认知作业和睡眠剥夺容易激发疲倦状态,通过认知作业中的难易度的调节能诱发被试的自信情感的出现。对情感语料的标注采取了多人听辨并计算一致度的方法,以筛选出可靠地情感数据。

我们将语种识别和说话人识别中获得成功应用的高斯混合模型(GMM),应用于实用语音情感的建模与自动识别中。GMM 对训练数据有较大的依赖性,而在情感识别领域,情感数据较难获取。在训练数据充足的条件下,基于 GMM 的情感识别系统获得了理想的效果。在小样本条件下,我们将 GMM 与两类分类器组结合,对情感配对特征优化和建模进行了研究,改善了小样本条件下 GMM 识别系统的性能。

从样本数量的角度分析了基于 GMM 的识别系统的性能之后，我们从样本的语料连续性角度进行研究。连续语音信号中的情感状态应该具有一定的连贯性，因此我们采用马尔科夫网络对上下文信息进行建模，提出了嵌入马尔科夫网络的高斯混合模型。实验结果显示，考虑情感状态的上下文连贯性，能够提升连续语音中的情感识别性能。

本文在建立了基本的 GMM 实用语音情感识别系统后，进一步考虑了系统的鲁棒性问题。首先，实际环境中的噪声问题是不可避免的，我们将基于听觉掩蔽的降噪方法与情感识别系统结合，分析了不同性噪比和不同降噪算法下实用语音情感的识别性能。其次，对未知情感类别提出了可据判的方法，提高了识别系统的稳定性。目前的情感识别系统，仅能够识别出数量很有限的情感类型，而人类的情感非常丰富，真实条件下的复杂的未知的情感类型如果没有得到适当的据判，就会导致大量的错判。再次，我们考虑到情感的表达非常具有个性特点，受到说话人的影响很大，大量的非特定的说话人会增加语音情感识别的困难。针对这个问题，我们提出了一种基于说话人特征聚类的情感特征规整化方法，提高了系统在大量非特定说话人条件下的性能。最后，为了提高系统的抗毁性，我们引入了心电信号通道，研究了双模态情感识别。本文中比较了两种融合心电和语音信号的算法，特征层融合与判决层融合。实验结果显示，两种融合方法都能提高系统的性能，特征层融合算法在识别率上略有优势。

我们在情感语音的诱发和数据的录制过程中，筛选并标注了两个实用语音情感实验数据集（见第二章 2.6 节）。为了便于本文中的对比研究和系统的分析问题，在上面提到的大多数分析和实验中，都使用了这两个实验数据集，我们可以在不同的章节之间交叉对比和验证实验结果。本文中研究的基于 GMM 的情感识别系统，还能够扩展应用跨语言和耳语音等特殊的问题上。

语音情感受到很多因素的影响，在不同语言中，发音和情感表达的方式也不同。在跨语言和跨文化的条件下，能否建立通用性较高的情感识别系统，是一个尚待解决的问题。本文中，我们研究了跨语言的情感识别测试，在一种语言上（如汉语语音数据）进行情感建模，在另一种语言上（如德语语音数据）进行识别测试。这样的跨语言测试难度较高，但是可以反映出情感特征在不同语言中的共通性。我们在 FDR 特征选择的基础上采用了权重融合的方法，获得了通用性较好的情感特征，在跨语言条件下生气情感具有较高的辨识度。

基于 GMM 的情感识别系统可以扩展应用到耳语音情感识别领域。我们将嵌入马尔科夫网络的高斯混合模型在耳语音情感数据上进行了验证，获得了较为理想的性能提升效果。耳语音信号中的情感表达与正常音不同，耳语音中对"喜悦"情感和"恐惧"情感的表达比较困难。

今后的研究工作可能在情感模型和情感特征方面有较大的发展空间。首先，情感维度空间模型在语音情感识别中的应用才刚刚开始，诸多算法可以与之结合，出现更为合理的情感识别方法。虽然心理学中的"唤醒度-效价度-控制度"三维模型比较流行，但是我们可以从语音信号的实际特点出发研究更加合适的情感模型。其次，情感特征还有待进一步的研究，从声学特征到心理状态的映射是非常困难的，如何构造可靠的情感特征一直是本领域的一个主题。特别是结合跨语言和跨数据库的研究，有利于发掘情感特征中的通用性。

虽然情感计算的研究已经进行了多年，然而情感的科学定义还并不明确。情感可以从进化论得到解释，认为情感是动物在生存斗争中获得的能力，使得动物能够趋利避害。情感还可以从社会心理学的角度得到解释，人类作为群居动物，成员个体之间需要进行有效的沟通，为劳动协作建立关系，而情感则是一种有效的交流手段，体现出个体的意图和心理状态。从这个角度来看，人工智能中是不可缺少情感识别技术的，它能够进行复杂意图信息的直接表达和有效传递。

虽然从哲学的角度看人工智能的可实现性具有争议，虽然对情感的科学定义还并不明确，然而从工程实际的角度看，对人类的情感行为进行分析和测量，是完全可行的。对人类的情感能力的部分的模拟，也已经获得了初步的进展。

人类语音当中包含的丰富多彩的情感信息，计算机能够理解到何种程度？语音情感识别技术是仅能够模仿一部分的人类情感感知能力，还是有可能超越人类的能力，捕获到人耳亦所无法感知的信息？这个问题值得我们深思。

从情感的含义上看，既然只有有人类和动物才具有情感，那么人类的情感也就通过人类自身得到了界定，人耳所不能感知到的信息，似乎不在语音情感的范畴内。然而，情感的感知通道，并不仅限于人耳听觉。通过内省知觉的方式，说话人自身能够体验到的情感是"体验情感"（Felt Emotion），通过人耳听觉感知到的他人的情感，是"听辨情感"（Perceived Emotion）。从这个角度考虑，语音情感识别技术，有可能超过人耳的听辨能力，获取到更多的说话人的体验情感的信息。人们在日常生活和工作中无意识的流露出的情感心理状态，能够通过情感计算技术得到准确的测量和分析，在此基础上发展出的技术应用有着广阔的前景。

# 参考文献

# 博士研究生阶段的科研成果

**第一作者论文:**

**第二作者论文：**

1, Yun Jin, Chengwei Huang, Li Zhao, A Semi-Supervised Learning Algorithm Based on Modified Self-training SVM. Journal of Computers, Vol 6, No 7, 2011, pp.1438-1443

2, 余华, 黄程韦, 金赟, 赵力, 基于改进的蛙跳算法的神经网络在语音情感识别中的研究, 信号处理, 2010, 26(9):1294-1299

3, 余华, 黄程韦, 金赟, 赵力, 基于粒子群优化神经网络的语音情感识别, 数据采集与处理, 2011, 第1期, pp.57-62

4, 余华, 黄程韦, 金赟, 赵力, 语音情感的维度特征提取与识别, 数据采集与处理, 2012 年, 第 3 期, pp.389-393

5, 余华, 黄程韦, 张潇丹, 金赟, 赵力, 混合蛙跳算法神经网络及其在语音情感识别中的应用, 南京理工大学学报, 2011, 35(5), pp.659-663

6, B. Efraty, Chengwei Huang, S.K. Shah, I.A. Kakadiaris, Facial landmark detection in uncontrolled conditions, International Joint Conference on Biometrics, Washington, DC, USA, Oct. 11-13, 2011.

7, Cairong Zou, Chengwei Huang, Dong Han, Li Zhao, Detecting Practical Speech Emotion in a Cognitive Task, International Conference on Computer Communications and Networks, Maui, HI, USA, July 31 2011-Aug. 4 2011

8, Hua Yu, Chengwei Huang, Yun Jin, Li Zhao, Automatic recognition of speech emotion using Shuffled Frog Leaping Algorithm. International Congress on Image and Signal Processing, Yantai, China, 16-18 Oct., 2010.

9, 余华, 黄程韦, 赵力, 邹采荣, 儿童情绪监测与情感电生理参数采集系统的研究, 电子器件, 2011, 33(4), pp.516-520

**项目技术报告**（第二作者）：

B. Efraty, Chengwei Huang, M. Papadakis, S. Shah, I. Kakadiaris, Agglomerate of cascaded fern regressors for point landmark detection, Technical Report, University of Houston, Houston, Texas, USA, 2012.

**发明专利申请**（第二发明人）：

1, 赵力, 黄程韦, 邹采荣等, 基于情感对特征优化的语音情感分类方法, 发明专利公开号: CN101894550

2, 赵力, 黄程韦, 邹采荣等, 一种针对烦躁情绪的可据判的自动语音情感识别方法, 发明专利公开号: CN101937678A

3, 赵力, 黄程韦, 邹采荣等, 一种基于心电信号与语音信号的双模态情感识别方法, 发明专利公开号: CN101887721A

4, 赵力, 黄程韦, 魏昕等, 基于特征空间自适应投影的语音情感识别方法, 发明专利公开号: CN102779510A